\documentclass[fleqn,usenatbib]{mnras}

\usepackage[T1]{fontenc}

\usepackage{subcaption}

\DeclareRobustCommand{\VAN}[3]{#2}
\let\VANthebibliography\thebibliography
\def\thebibliography{\DeclareRobustCommand{\VAN}[3]{##3}\VANthebibliography}

\usepackage{graphicx}	% Including figure files
\usepackage{amsmath}	% Advanced maths commands
\usepackage{amssymb}	% Extra maths symbols
\usepackage{newtxtext,newtxmath}
\usepackage{xcolor}     % Allows to write coloured text
\usepackage{url}        % Allows to put url in references
\usepackage{hyperref}

\title[Clustering effects on dark sirens]{Not all those missing are lost: leveraging galaxy clustering in incomplete catalogs to unleash dark sirens cosmology}

\author[Barbieri, Kalomenopoulos et al.]{
Riccardo Barbieri$^{1}$\thanks{E-mail: riccardo.barbieri@aei.mpg.de},
Marios Kalomenopoulos$^{2,3}$,
Matteo Tagliazucchi$^{4,5}$,
Jonathan Gair $^{1}$,
Sadegh Khochfar$^{3}$
\\
$^{1}$Max Planck Institute for Gravitational Physics, Potsdam Science Park, Am Muhlenberg 1, D-14476 Potsdam, Germany \\
$^{2}$Department of Physics and Astronomy, University of Nevada, Las Vegas, NV 89154, USA\\
$^{3}$Institute for Astronomy, University of Edinburgh, Royal Observatory, Edinburgh EH9 3HJ, UK \\
$^{4}$Dipartimento di Fisica e Astronomia ``Augusto Righi'' -- Universit\`{a} di Bologna, via Piero Gobetti 93/2, I-40129 Bologna, Italy\\
$^{5}$INFN - Sezione di Bologna, Viale Berti Pichat 6/2, I-40127 Bologna, Italy \\
}

\date{Accepted XXX. Received YYY; in original form ZZZ}

\pubyear{2026}

\begin{document}
\label{firstpage}
\pagerange{\pageref{firstpage}--\pageref{lastpage}}
\maketitle

% Abstract of the paper
\begin{abstract}
Gravitational waves offer a unique opportunity to solve the Hubble tension. In order to do so, we have to extract as much information as possible from cross correlating gravitational wave events (used as "dark sirens") with incomplete galaxy catalogs. Traditional methods assume a uniform in comoving volume distribution for galaxies missing from the catalogs, which neglects the fact that galaxies tend to cluster, leading to less precise and possibly biased posteriors. In this paper, we introduce a new method for accounting for galaxy clustering when dealing with an incomplete galaxy catalog, by adding back galaxies to the incomplete catalog and distributing them in pixels according to the correlation function $\xi(r)$. We find that our method drastically improves on traditional ones when the galaxy catalog is only $1\%-10\%$ complete, leading to posteriors that are between 2 and 4 times as precise, depending on the completeness fraction, without introducing any biases. 
Our method produces results comparable to traditional methods at extremely low catalog completeness fractions (<0.5\%) or for very high uncertainties in the recovered luminosity distance and sky localization of the gravitational wave events.
\end{abstract}

% Select between one and six entries from the list of approved keywords.
% Don't make up new ones.

%%%%%%%%%%%%%%%%%%%%%%%%%%%%%%%%%%%%%%%%%%%%%%%%%%

%%%%%%%%%%%%%%%%% BODY OF PAPER %%%%%%%%%%%%%%%%%%

\section{Introduction}

Cosmology has recently faced what many refer to as the "cosmological crisis": local, late-Universe measurements of the Hubble constant, $H_0$, (obtained for example by the SHOES experiment, that uses Type-Ia supernovae, \citep{SHOES}) seem to be in stark contrast with cosmic microwave background (CMB) measurements (obtained for example by the Planck experiment, \citep{Planck}), leading to a $>5\sigma$ discrepancy. Gravitational waves (GWs) offer another independent method of estimating the Hubble constant, with the potential to determine whether one of the two contrasting measurements is affected by a systematic bias, or if indeed new physics needs to be introduced.  

The idea that gravitational waves can be used in cosmology is now 40 years old, and was introduced for the first time by \cite{Schutz1986}. Electromagnetic waves are complementary to gravitational waves for this goal: the former ones give, folded into the signal, an estimate of the redshift of the source, without any direct measurement of its distance; the latter ones give, folded into the signal, an estimate of the distance of the source, without an independent measurement of the redshift, as it is degenerate with another signal parameter, the chirp mass. Using gravitational waves as standard sirens allows us to skip the cosmic distance ladder, the convoluted and multi step method that standard candle probes have to use to infer the distance of a source. Bernard Schutz's idea relies on combining the information: %this combination: 
if a host galaxy can be associated to a gravitational wave event, we get the best of both worlds.

In 2017, a very lucky event happened: GW170817, the detection of the merger of a binary neutron star (BNS) \citep{GW170817a} with the associated electromagnetic counterpart \citep{GW170817b}, leading to the first gravitational wave measurement of $H_0$. In this case, detecting the electromagnetic counterpart meant that we could unequivocally identify the host galaxy of the event, and therefore the redshift of the source. Unfortunately, GW170817 remains the only such event observed to date, with only one other BNS detected (GW190425, \cite{GW190425}) but without an associated electromagnetic signal. In the absence of a counterpart, we have to resort to alternative approaches to extract the redshift of the source. This can be done in three main ways:
\begin{itemize}
    \item The "spectral siren" method, which infers the redshift from the properties of the mass distribution of the GW sources \citep{Farr_2019, EquiagaSpectralSiren, SuvodipSpectral, Karathanasis_2023, leyde2022currentfutureconstraintscosmology}.
    \item The galaxy catalog method, where the redshift of the source is inferred from the redshift of the many possible host galaxies in the GW sky localization volume. This method has also a variant, the cross-correlation method, where the cosmological inference is obtained by comparing the spatial clustering of GW sources and of galaxies \citep{Suvodip_CrossCorrelation, SuvodipCrossCorr2, Bera_2020, DiazCrossCorr, MariosPaper}.

    \item The tidal siren method, where the tidal coupling contribution to the waveform is used to break the degeneracy between chirp mass and redshift, therefore obtaining a redshift measurement from the gravitational wave signal itself, without using any electromagnetic data \citep{Messenger_2012_TidalSiren}
\end{itemize}

Our paper will focus exclusively on the galaxy catalog method, and in particular on the statistical method \citep{DelPozzo, Chen_2018, Fishbach_2019_GW170817, Gray_2020, Gray_Pixelatedgwcosmo, Leandro} which tries to guess the host galaxy in the absence of a counterpart from the galaxies within the GW event localization area. 

In more recent times, this approach was used in the \texttt{gwcosmo} code \citep{Gray_Pixelatedgwcosmo, gray2023jointcosmologicalgravitationalwavepopulation}, a product of the LVK collaboration to measure the Hubble constant, leading to the most recent measurement of $H_0 = 71.0 ^{+9.0}_{-7.1} \ \text{km}\ \text{s}^{-1} \ \text{Mpc}^{-1}$ \citep{theligoscientificcollaboration2026gwtc50constraintscosmicexpansion}. In this code, built upon the Bayesian framework first introduced in \cite{DelPozzo}, each galaxy in a galaxy catalog is assigned a probability of being the host of a gravitational wave event. These probabilities are then used to statistically average over all possible host galaxies, with the possibility of introducing additional weights such as their luminosities. 
However, there might be the case in which the host is not part of the catalog. Usually, each galaxy catalog is complete only up to a certain distance, and gets less and less complete at higher redshifts. In the statistical method, the possibility of the source being outside the galaxy catalog is dealt with by introducing an incompleteness term, where the Hubble constant is estimated by averaging over the whole sky coverage, effectively assuming that every missing galaxies is distributed uniformly in comoving volume. 

While this assumption is serviceable and usually does not introduce biases, it might be losing a significant amount of information, that is encoded in galaxy clustering. Going back to Schutz's original idea, even if we do not see the GW event host galaxy, the chances are that it is located somewhere in the vicinity of galaxies we do see. Not all missing galaxies are therefore completely lost.

The idea of incorporating galaxy clustering, or more generally improving on the assumption of a simple uniform in comoving volume distribution for the missing galaxies, has been recently explored in various works: \cite{Dalang_Fiorini_Baker_2024} use a factor based on the variance of the distribution of galaxies in the catalog;  \cite{Cosmic_cartography_2025} used a Bayesian method for galaxy density and magnitude reconstruction based on large-scale structure information; \cite{VanWyngarden_et_al_2025_cat_completeness} and \cite{Borghi_2026} use an astrophysical approach, alleviating the effects of catalog incompleteness.

In this paper, we explore a  new method to include galaxy clustering in the usual Bayesian statistical method, by repopulating an incomplete galaxy catalog and performing a statistical cosmological analysis assuming a (now-) complete galaxy catalog. This introduces an inherent randomness to the process, which can be averaged out by repeating the completion several times. This method is equivalent to the traditional statistical method if one neglects galaxy clustering. We show, however, that it is possible to take into account galaxy clustering via the galaxy correlation function \citep{peebles1980} when repopulating the catalog. We use the MICE galaxy catalog as our mock galaxy catalog of reference \citep{MICECarretero_2014, MICECrocce_2015}, limited to redshift less than 0.1, generating mock gravitational wave events and exploring different "detector sensitivities", represented in our work by a luminosity distance detection threshold. We show how accounting for galaxy clustering results in a more precise, yet still unbiased, measurement of the Hubble constant, and we explore how several parameters impact this measurement.

Our paper is structured as follows: in section \ref{sec: Methods} we introduce our approach to generate mock gravitational wave events, our procedure to complete a galaxy catalog (with or without clustering), and how to use it to estimate the Hubble constant. In section \ref{sec: Results} we present our findings, comparing the "clustered" analysis with the "uniform" approach and studying the impact of various parameters. In section \ref{sec: Conclusions} we discuss our conclusions and possible follow-ups.

\section{Methods}
\label{sec: Methods}
In this section we present our complete procedure, starting from data generation to the cosmological analysis.
This section is divided into subsections, each detailing a piece of our pipeline:
\begin{itemize}
    \item General cosmology introduction.
    \item Data generation process: how we, starting from a complete galaxy catalog, generate an incomplete catalog and simple mock GW events.
    \item Given a galaxy catalog (whether complete or incomplete), how the Hubble constant posterior is obtained. This is the standard, Bayesian, uniform statistical method, previously discussed in literature and here reported for clarity and for the sake of completeness.
    \item How, given an incomplete galaxy catalog, we repopulate it with its missing galaxies, whether accounting or not for galaxy clustering. The power of our method relies on the ability to introduce galaxy clustering, but the case of uniform completion (meaning, adding back galaxies to the catalog assuming an identically zero correlation function) is here reported for comparison with the traditional uniform statistical method.
\end{itemize}

\subsection{Cosmology}
\label{sub: cosmology}
The cosmological expansion is described by the Hubble parameter, which can be expressed as a function of the redshift $z$ as:
\begin{equation}
    \label{Hubble parameter}
    H(z) = H_0 \sqrt{\Omega_m (1+z)^3 + \Omega_k (1+z)^2 + \Omega_\Lambda}
\end{equation}
where $H_0$ is the Hubble constant and $\Omega_m$,  $\Omega_k$, $\Omega_\Lambda$ are, respectively, the fractional matter density, curvature energy density and dark matter density, with $\Omega_m +\Omega_k + \Omega_\Lambda = 1 $. This expression can be used to establish a relationship between a source redshift and its luminosity distance 
\begin{equation}
    \label{eq: Nonlinear dl_zH0}
    d_L(z, H_0) = \dfrac{c}{H_0} (1+z) \int_0^z \dfrac{H_0}{H(z')} dz.
\end{equation}
This equation can be used, from observations of $d_L$ and $z$ (either EM sources or GW sources, or both) to simultaneously constrain the four parameters, or, with a prior assumption on the three densities, to constrain the Hubble constant $H_0$ itself. The latter is the approach we will follow in this paper. We assume as values for density the same values used for simulating the MICE galaxy catalog \footnote{The MICE catalog is publicly available at this page: https://cosmohub.pic.es/catalogs/1}  \citep{MICECarretero_2014, MICECrocce_2015}, as well as the same injected value of the Hubble constant

\begin{equation}
    {H_0}_{\rm true}=70 \ \text{km}\ \text{s}^{-1} \ \text{Mpc}^{-1}, \, \, 
    \Omega_k = 0, \, \, \Omega_m=0.25, \, \, \Omega_\Lambda=0.75
\end{equation}

As we limit the MICE catalog to redshift 0.1, equation \eqref{eq: Nonlinear dl_zH0} can be well approximated by
\begin{equation}
    \label{eq: Linear dl_zH0}
    d_L(z,H_0) = \dfrac{cz}{H_0},
\end{equation}
which we will use for the rest of this paper.

\subsection{Data Generation}
\label{sub: Data Generation}
For our analysis, we start from the MICE galaxy catalog, considering only the galaxies up to redshift 0.1. This will be our full catalog describing our Universe, where each galaxy will be identified by its coordinates (right ascension $\rm RA_{gal}$, declination $\rm dec_{gal}$ and redshift $z_{\rm gal}$) and by its apparent $r$-band luminosity $\rm m_{gal}$. We also assume perfect knowledge of the galaxy's redshift.

\subsubsection{Generating Incomplete Catalogs}
\label{subsub: generating incomplete catalogs}
Real galaxy catalogs are, however, not complete, being limited by the observational sensitivity of the survey. For example, the galaxy catalog GLADE+ is complete up to $d_L = 47^{+4}_{-2} \ \text{Mpc}$, and roughly 90\% complete up to $d_L \simeq 130 \ \text{Mpc} $ \citep{GLADE+}. 
In order to generate an incomplete galaxy catalog, we set an apparent magnitude threshold, $\rm m_{th}$, so that every galaxy with apparent magnitude lower than $\rm m_{th}$ is considered detected in our survey, while we discard every galaxy with apparent magnitude higher than $\rm m_{th}$. 

In this work, we will explore various completeness levels, always expressed as the overall fraction of galaxies in the incomplete catalog over the starting number of galaxies in the complete one.

\subsubsection{GW event generation}
\label{subsub: gw event generation}
From our complete galaxy catalog, we now need to create gravitational wave events. We assume, as is commonly done in the literature, that each gravitational wave event happens inside a galaxy. Furthermore, we use a toy model for our GW events, so that each event can be completely described by its coordinates $\rm RA_{gw}$, $\rm dec_{gw}$ and ${d_L}_{\rm gw}$, neglecting parameters such as chirp mass, inclination, eccentricity, etc. The event's $\rm RA$ and $\rm dec$ are identical to its host galaxy $\rm RA$ and $\rm dec$, while the luminosity distance $d_L$ can be obtained by the galaxy's redshift through equation \eqref{eq: Nonlinear dl_zH0} or \eqref{eq: Linear dl_zH0}. 

We also assume that every galaxy is equally likely to be the host of a gravitational wave event. It is possible to insert additional weights in the likelihood based on various properties of a galaxy, such as its luminosity in a certain band. For example, r-band and B-band luminosity can be used to trace stellar mass and star formation rate \citep{calzetti2012starformationrateindicators, Mobasher_2015}, which in turn can be used as indicators for the probability of a galaxy to be host of a GW event. Studies on effects of this additional weighting can be found in \cite{li2025usinggravitationalwavedark}, \cite{Gray_2020}, \cite{Borghi_2026}, \cite{HanselmanWeighting}, \cite{PernaHostingProbability}.

\subsubsection{GW event observation}
\label{subsub: event obs}

Following \cite{Gair_2023}, we use a simple model where each event is defined by three coordinates: right ascension (RA), declination (dec) and luminosity distance ($d_L$). Rather than a full three dimensional skymap, we then have a set of observed coordinates ($\rm RA_{obs}$, $\rm dec_{obs}$ and ${d_L}_{\rm obs}$), which are obtained from the gravitational wave event host galaxy coordinates by drawing from Gaussian distributions centered at the true event coordinates, ($\rm RA_{true}$, $\rm dec_{true}$ and ${d_L}_{\rm true}$)
\begin{equation}
    \label{eq: radecgaussian}
    \begin{split}
    &p(\rm RA_{\rm obs}, \rm dec_{\rm obs}) = \\
    &\dfrac{1}{2\pi \sigma_{\rm RA} \sigma_{\rm dec} }\exp\left(-\left(\dfrac{(\rm RA_{\rm obs}-\rm RA_{\rm true})^2}{2\sigma^2_{\rm RA}} + \dfrac{(\rm dec_{\rm obs}-\rm dec_{\rm true})^2}{2\sigma^2_{\rm dec}}\right)\right),
    \end{split}
\end{equation}
where $\sigma_{\rm RA}$ and $\sigma_{\rm dec}$ are fixed parameters and we have assumed $\rm RA$ and $\rm dec$ to be independent, and 
\begin{equation}
    p({d_L}_{\rm obs}) = \dfrac{1}{\sqrt{2\pi} \sigma_{d_L}} \exp\left(-\dfrac{({d_L}_{\rm obs}-{d_L}_{\rm true})^2}{2\sigma^2_{d_L}}\right)
    \label{eq: ObservedDl}
\end{equation}
 with 
$$\sigma_{d_L}=A{d_L}_{\rm true},$$
  where $A$ is the fractional distance uncertainty, a dimensionless constant.

It is also important to accurately model how an event is "detected": while real world gravitational wave events are considered detected when a threshold in signal-to-noise ratio or false alarm rate, computed using low-latency algorithms, are met \citep{latencyAubin_2021, latencyCANNON2021100680, latencyPhysRevD.95.042001, theligoscientificcollaboration2026gwtc50observationssecondfourth}, we instead employ a simplified model. Following \cite{Gair_2023}, we assume that an event is detected if its observed luminosity distance, ${d_L}_{\rm obs}$, is less than a certain luminosity distance threshold ${d_L}_{\rm thr}$. In this work, we will explore different detection thresholds covering the full range of the catalog, ${d_L}_{\rm thr} = 100,200,300$ and $400 \ \text{Mpc}$; furthermore, in Appendix \ref{Appendix edge} we explore the impact of different detection thresholds, in particular the case where the detection threshold is close to or even greater than the actual edge of the catalog.

\subsection{Bayesian Cosmological Analysis}

Once we have generated, given an injected value of $H_0$, an incomplete catalog and a set of $N_{\rm obs}$ gravitational wave events, we can proceed with the cosmological analysis. 
Following \cite{Gair_2023}, we write the posterior for the Hubble constant as:
\begin{equation}
    \label{eq: Final H0 Likelihood}
    p(H_0|\{X_{N_{\rm obs}}\}) \propto p(H_0) \prod_{i=1}^N \mathcal{L}(x_i|H_0)
\end{equation}
where $p(H_0)$ is our prior and  $X_{N_{\rm obs}}$ is the set of all data associated with the $N_{\rm obs}$ detections. The term $\mathcal{L}$ is the likelihood for a single event, which using Bayes' theorem can be written as 
\begin{equation}
    \label{eq: Single Event Likelihood}
    \mathcal{L}(x | H_0) = \dfrac{\int d\vec{\lambda} p(x|\vec{\lambda}, H_0) p_{\text{GW}}(\vec{\lambda})}{\int d\vec{\lambda} P_{\rm det}(\vec{\lambda}, H_0) p_{\text{GW}}(\vec{\lambda})} 
\end{equation}
% \MKc{In some cases $\lambda$ has a subscript and in some others no} \RBc{Corrected and took out the subset GW, kinda redundant}
where  $\vec{\lambda}$ is a set of parameters describing the event,  $p(x|\vec{\lambda}, H_0)$ is the likelihood of observing (including the detection of the data itself) data $x$ given an event with parameters $\vec{\lambda}$, $P_{\rm det}(x_i|\vec{\lambda}, H_0)$ is the probability of detection, necessary to account for selection effects, and $p_{\text{GW}}(\vec{\lambda})$ is the probability of having a gravitational wave event with parameters $\vec{\lambda}$, usually established as a prior dependent on our assumption about the formation channels of a GW event \citep{Gerosa_2021_formation}. In the rest of the paper, we will use the simplifying assumption that the only parameters relevant to cosmological analysis are the redshift z, the sky location $\Omega$ (expressed as $\rm RA$ and $\rm dec$) and the luminosity of the host galaxy, expressed as absolute magnitude $M$ and apparent magnitude $m$.

As already stated, we assume that every gravitational wave merger happens within a galaxy. Given this assumption, the probability $p_{\text{GW}}(\vec{\lambda})$ can be computed as:
\begin{equation}
    \label{eq: pGWterm}
    p_{\text{GW}}(\vec{\lambda}) = \dfrac{p_{\rm rate}(z)p_{\rm gal}(\vec{\lambda})}{\int p_{\rm rate}(z)p_{\rm gal}(\vec{\lambda})},
\end{equation}
where $p_{\rm gal}(\vec{\lambda})$ is the distribution of galaxies and $p_{\rm rate}(z)$ is the probability that a galaxy is the host of a gravitational wave merger (assuming that it depends exclusively on redshift).

If we further assume that the rate does not depend explicitly on redshift (assuming therefore that the mergers are distributed uniformly in comoving volume), we get 
$$p_{\text{GW}}(\vec{\lambda}) \sim p_{\rm gal}(\vec{\lambda}).$$
We are therefore able to compute the likelihood in \eqref{eq: Single Event Likelihood} if we know the distribution of galaxies in our Universe.

\subsubsection{Complete Catalog Case}
\label{subsub: complete catalog likelihood}
In the case of a complete galaxy catalog (meaning a survey that has detected every single galaxy in a certain area), we have full knowledge of the term $p_{\rm gal}$ and we can express it as a sum of delta functions centered at each galaxy's parameters:
$$p_{\rm gal} (\vec{\lambda}) \equiv p_{\rm cat} (\vec{\lambda}) = \sum_i^{N_{\rm gal}} \delta(\vec{\lambda}_i-\vec{\lambda})$$
Here we have assumed that all parameters of a galaxy ($z$, $\Omega$, $m$) are perfectly known without uncertainty. For a discussion of the impact of these uncertainties on the complete catalog case, we refer once again to \cite{Gair_2023}.

Equation \eqref{eq: Single Event Likelihood} then becomes:
\begin{equation}
    \label{eq: complete catalog likelihood}
    \mathcal{L}(x | H_0) = \dfrac{\sum_i^{N_{\rm gal}} p(x| \vec{\lambda_i}, H_0)}{\sum_i^{N_{\rm gal}}P_{\rm det}(\vec{\lambda_i}, H_0)}.
\end{equation}
 where, assuming that the likelihood depends exclusively on the galaxy location and not the galaxy magnitude (see section \ref{subsub: gw event generation} for comments regarding this assumption), $p(x|\vec{\lambda})$ is given by 
 \begin{equation}
     \label{eq: Numerator fullcat}
     %p(x|\vec{\lambda_i}, H_0) = p(d_L(z_i, H_0)|{d_L}_{\rm obs})p(\Omega_i|\Omega_{\rm obs}),
     p(x|\vec{\lambda_i}, H_0) = p({d_L}_{\rm obs}|d_L(z_i, H_0))p(\Omega_{\rm obs}|\Omega_i),
 \end{equation}
 where $\Omega_i$ are the angular (RA and dec) coordinates of galaxy $i$, $p({d_L}_{\rm obs} | d_L(z_i, H_0))$ and $p(\Omega_{\rm obs}|\Omega_i)$ are the distributions discussed in section \ref{subsub: event obs} %with the only caveat that in this case the "true" and "observed" subscripts for the luminosity distance are swapped, as we are now trying to infer the real values from the observed ones:
 \begin{equation}
     \label{eq: px_dl_direct}
     p({d_L}_{\rm obs} | {d_L}_{\rm true}) = \dfrac{1}{\sqrt{2\pi} A{d_L}_{\rm true}}\exp\left(-\dfrac{({d_L}_{\rm true}-{d_L}_{\rm obs})^2}{2(A{d_L}_{\rm true})^2}\right).
 \end{equation}
 We note that when viewed as a function of the unknown quantity, ${d_L}_{\rm true} $, or equivalently $H_0$, this is no longer a Gaussian, 
% Note that this is not a Gaussian anymore, 
as the denominator of the argument of the exponential contains the variable of interest itself, ${d_L}_{\rm true}$. The distributions for the angular parameters, on the other hand, are still Gaussian when viewed as functions of the true parameters, since the angular standard deviations are fixed and do not depend on the true parameters of the source.
 
The denominator can be easily derived, as the probability of detection $P_{\rm det}$ is defined as the integral of the likelihood over the range of detectable events. In our case, the only parameter that determines detection is ${d_L}_{\rm obs}$, therefore this term will depend exclusively on $z$ and $H_0$
 \begin{equation}
     P_{\rm det}(z,H_0) = \int_0^{{d_L}_{\rm thr}} p({d_L}_{\rm obs}|d_L(z,H_0)) d{d_L}_{\rm obs},
 \end{equation}
giving
\begin{equation}
    \label{eq: Pdet fullcat}
    P_{\rm det}(z,H_0) = \dfrac{1}{2}\left[1-\text{erf}\left(\dfrac{d_L(z, H_0)-{d_L}_{\rm thr}}{\sqrt{2}Ad_L(z, H_0)}\right)\right]
\end{equation}

Having defined all terms entering equation \eqref{eq: complete catalog likelihood}, we can now compute the Hubble constant posterior using equation \eqref{eq: Final H0 Likelihood}. Throughout this paper, we will always assume a flat prior on $H_0$ between 40 and 100 $\text{km}\ \text{s}^{-1} \ \text{Mpc}^{-1}$. Therefore, apart from a normalizing factor, the likelihood and the posterior are the same function. We will from here on out refer to the likelihood as $\mathcal{L}(H_0)$, and to the numerator and denominator of the likelihood as $p_x(H_0)$ and $p_D(H_0)$, where the subscript serve as a reminder that the numerator contains the data of the gravitational wave events $x$ while the denominator contains the probability of detection.

\subsubsection{Incomplete Catalog Case}
\label{subsub: Incomplete Catalog Bayesian}
When a galaxy catalog is not complete, we lack full knowledge of the term $p_{\rm gal}$, and we will rather have a fraction $f$ of the total number of galaxies in our catalog. We then have
$$p_{\rm gal}(\vec{\lambda}) = fp_{\rm cat}(\vec{\lambda}) + (1-f)p_{\rm miss}(\vec{\lambda}),$$
where $p_{\rm cat}$ is exactly the same as before, while $p_{\rm miss}$ describes the distribution of galaxies that are not present in the catalog. If the survey, as is always the case in real life and as we assume throughout this paper, is limited by the apparent luminosity (or equivalently, magnitude) of the galaxies, the term $p_{\rm miss}$ will depend on the redshift (galaxies farther away have a smaller chance of being detected by our survey), the apparent magnitude threshold $m_{\rm th}$ and two assumptions: the prior on the redshift distribution of the galaxies and the assumed distribution of magnitudes (the Schechter function). 

Equivalently, as is often done in the literature (see for example \cite{Gray_2020} for a detailed introduction), we can split the single event likelihood \eqref{eq: Single Event Likelihood} in two parts:
\begin{equation}
    \label{eq: Incomplete Catalog Likelihood}
    \mathcal{L}(H_0) = p_G(H_0) \mathcal{L}_{\rm in}(H_0) + p_{\bar{G}}(H_0) \mathcal{L}_{\rm out}(H_0)
\end{equation}
where $\mathcal{L}_{\rm in}$ is the likelihood assuming the GW source is inside the catalog (identical to the likelihood discussed in \ref{subsub: complete catalog likelihood}) and $\mathcal{L}_{\rm out}$ is the likelihood assuming the GW source is outside the catalog. 

The term $p_G$ is the probability that the host galaxy of the GW event is present in the catalog. This term can be calculated as a ratio of integrals over observable parameter space and the whole parameter space: 
\begin{equation}
    \label{eq: pG}
    \begin{split}
    &p_G(H_0) = \\
    &= \dfrac{\int dz \int dM  P_{\rm det}(z,H_0) \phi(M|H_0)p(z)\Theta(m_{\rm th}-m(z, M, H_0))}{\int dz \int dM  P_{\rm det}(z,H_0) \phi(M|H_0)p(z)} \\
    &= \dfrac{\int_0^{z(m_{\rm th}, M, H_0)} dz \int dM P_{\rm det}(z,H_0) \phi(M|H_0)p(z)}{\int dz \int dM P_{\rm det}(z,H_0) \phi(M|H_0)p(z)}
    \end{split}
\end{equation}
where $p(z)$ is the assumed prior on the redshift distribution of galaxies in the catalog and $\phi(M)$ is the Schechter function \citep{Schechter}
\begin{equation}
\Phi(M) = 0.4 \ln(10) \Phi^* \, 10^{0.4(\alpha+1)(M^*-M)} \exp\left(-10^{0.4(M^*-M)}\right),
\label{eq: schechter_mag}
\end{equation}
 in which $M^*$ is the characteristic galaxy magnitude, $\alpha$ is the slope of the Schechter function and $\Phi^*$ is a normalization constant. A Schechter function is usually defined by two additional parameters, ${M_{\rm sch}}_{\rm min}$ and ${M_{\rm sch}}_{\rm max}$, determining the minimum and maximum absolute magnitude of the galaxies. 

The term $p_{\bar{G}}$ is simply $1-p_G$, while the outside the catalog likelihood can be similarly computed, giving
\begin{equation}
    \label{eq: Outside Likelihood}
    \begin{split}
    &\mathcal{L}_{\rm out}(H_0) = \dfrac{{p_x}_{\bar{G}}(H_0)}{{p_D}_{\bar{G}}(H_0)} \\
    &=\dfrac{\int_{z(m_{\rm th}, M, H_0)}^\infty dz \int dM p({d_L}_{\rm obs}|d_L(z, H_0))\phi(M|H_0)p(z)}{\int_{z(m_{\rm th}, M, H_0)}^\infty dz \int dM P_{\rm det}(z, H_0)\phi(M|H_0)p(z)}.
    \end{split}
\end{equation}
where $p({d_L}_{\rm obs}|d_L(z, H_0))$ and $P_{\rm det}(z, H_0)$ are the same functions introduced in section \ref{subsub: complete catalog likelihood}

\subsection{Completing a galaxy catalog}
\label{sub: Catalog Completion}
Another approach one might take to address the incompleteness problem is to actually generate a complete catalog from an incomplete one. In order to do this, we have to add galaxies back to the catalog, remembering that the galaxies were cut not randomly but based on their apparent magnitude.

\subsubsection{Uniform Completion}
\label{subsub: Uniform Completion}
We estimate the fraction of galaxies at redshift $z$ that are still present in the catalog using the Schechter function, similar to the traditional incomplete analysis presented in section \ref{subsub: Incomplete Catalog Bayesian}:
\begin{equation}
\label{eq: Missing galaxies fraction}
    f(z)=\dfrac{\int_{M_{\rm min}(H_0)}^{M_{\rm th}(m_{\rm th}, z, H_0)} \phi(M) dM}{\int_{M_{\rm min}(H_0)}^{M_{\rm max}(H_0)} \phi(M) dM}
\end{equation}
where $M_{\rm min}$ and $M_{\rm max}$ are obtained from the fixed parameters of the Schechter function \eqref{eq: schechter_mag} through
\begin{equation}
    \label{eq: MtoMobs}
M = M_{\rm sch} + 5\log_{10}(H_0/100)
\end{equation}
and $M_{\rm th}$ is a threshold absolute magnitude, obtained from a reference apparent magnitude through:
\begin{equation}
\label{eq: M_mdl}
    M_{\rm th}(m_{\rm th}, z, H_0) = m_{\rm th}-5\log_{10}\left(\dfrac{d_L(z,H_0)}{1\ \text{Mpc}}\right)+25
\end{equation}
Note that equation \eqref{eq: Missing galaxies fraction} does not actually depend on $H_0$, as the limits of the integrals and the $M^*$ parameters that appears in \eqref{eq: schechter_mag} depends on $H_0$ through equation \eqref{eq: MtoMobs}, and the dependency can be therefore eliminated through a simple change of variables.
The number of missing galaxies at a redshift z can then be approximated by:
\begin{equation}
    \label{eq: missing gals z}
    {N_g}_{\rm miss}(z) = {N_g}_{\rm obs}(z) \cdot \frac{1-f(z)}{f(z)}.
\end{equation}
where ${N_g}_{\rm obs}(z)$ is the number of galaxies at redshift z present in the incomplete catalog. 
Using this redshift distribution, we can create new galaxies to repopulate the incomplete catalog, assigning random sky positions $\Omega$ and magnitudes $M$. We have now a "complete" galaxy catalog, that we can use to perform traditional Bayesian analysis as outlined in section \ref{subsub: complete catalog likelihood}.

In order to average out the inherent randomness of the process, we can perform this "re-population" and the relative cosmological analysis $N_{\rm repop}$ times and take the means of the $H_0$ posteriors \eqref{eq: Final H0 Likelihood}:
\begin{equation}
    \label{eq: AvaragingOutH_0Posteriors}
    p(H_0|N_{\rm repop}, {X_N}_{\rm obs}) = \dfrac{1}{N_{\rm repop}} \sum_{i=0}^{N_{\rm repop}} p_i(H_0|\{X_{N_{\rm obs}}\}).
\end{equation}
Performing this process is equivalent to the incomplete catalog bayesian analysis illustrated in \ref{subsub: Incomplete Catalog Bayesian}, as we show in section \ref{sec: Results}. Note that assigning random sky positions means assuming a uniform in comoving volume distribution for the galaxies, just like in equation \eqref{eq: pG} and \eqref{eq: Outside Likelihood} it is usually assumed a uniform in comoving volume distribution for the redshift prior $p(z)$.
If our original catalog is clustered, as is the case when using the MICE catalog, the clustering information is thus lost, resulting in a less precise estimate of $H_0$ and, although unlikely to be relevant, a potential bias, as the galaxy distribution used to generate the data does not match the assumptions in the data analysis.

\subsubsection{Clustered Completion}
\label{subsub: Clustered Completion}
We can, however, perform this completion process while accounting for clustering, making use of the 2-point correlation function \citep{peebles1980}. This function, which we will denote by $\xi (r)$, describes the probability of finding a galaxy in the volume element $dV_1$ and another galaxy in the volume element $dV_2$, 
\begin{equation}
    dP=n^2 dV_1 dV_2 [1+\xi (r)],
\end{equation}
with $r$ being the distance between the two volume elements and $n$ being the mean density of galaxies.

This can also be rewritten as the probability of finding a galaxy in the volume element $dV$ given the presence of a galaxy at a distance $r$:
\begin{equation}
    \label{eq: corrfunc sum introduction}
    dP=n dV [1+\xi(r)].
\end{equation}
$\xi(r)$ therefore represents the excess probability of finding two galaxies "close" to each other, with respect to a uniform in comoving volume distribution (which can be obtained by setting $\xi \equiv 0$). 
The correlation function is usually modeled as a power law
\begin{equation}
    \xi(r)=\left(\dfrac{r}{r_0}\right)^{-\gamma}
\end{equation}
where $\gamma$ is the slope of the power law and $r_0$ is a normalizing distance. This power law shape of $\xi(r)$ holds only in a limited range of distances, $[r_{\rm min}, r_{\rm max}]$. Outside this range, the correlation function moves away from a power law, and at distances significantly higher than $r_{\rm max}$ the correlation function becomes slightly negative, as the overall mean density of galaxies $n$ must be preserved \cite{peebles1980}. 

Typical ranges for the correlation function parameters are (\citet{longair2023clusteringbook}, \citet{peebles1980}, \citet{mo2010clusteringbook}):
\begin{equation}
    \label{eq: Clustering parameters}
    \begin{split}
        &\gamma \sim 1.7 - 1.8 \\
        &r_0 \sim 5 - 5.5 h^{-1}\ \rm{Mpc} \\
        &r_{\rm min} \sim 100h^{-1} \ \rm{kpc} \\ 
        &r_{\rm max} \sim 10-20h^{-1} \ \rm{Mpc} \\ 
    \end{split}
\end{equation}

Alternatively, given a complete galaxy catalog, the correlation function at distance r can be calculated by counting the neighbours inside a sphere of radius r, comparing it to the same quantity for a uniform distribution of galaxies. One such estimator for the correlation function is the Davis-Peebles (DP) estimator \citep{DavisPeebles}: 
\begin{equation}
    \label{eq: EmpiricalCorrelation}
    \xi (r) = \dfrac{\langle D,D \rangle (r)- \langle D,R \rangle (r)}{ \langle D,R \rangle (r)}
\end{equation}
where $\langle \ , \rangle(r)$ is the number of neighbors within a sphere of radius $r$, D represents the data distribution (a clustered catalog) and R is a uniform random distribution. Note that there are many ways to compute this empirical correlation function, with slightly different combinations of the terms in equation \eqref{eq: EmpiricalCorrelation} and with slight differences in their advantages and disadvantages (see \cite{Kerscher_2000} for a comparison of the different estimators). Throughout this paper, we will always assume a power law shape for the correlation function, with extremes of validity
\begin{equation}
    \begin{split}
        &r_{\rm min} \sim 100h \ \rm{kpc} \\ 
        &r_{\rm max} \sim 10 \ \rm{Mpc}. \\
    \end{split}
\end{equation}

Furthermore, for the majority of the results we will use parameters $\gamma$ and $r_0$ fit to the catalog correlation function of the full MICE catalog obtained by the DP estimator; in the results subsection \ref{subsub: correlation function results} we will briefly investigate the impact of different assumptions on the correlation function parameters.

In order to complete the catalog using this correlation function, we first divide our galaxy catalog redshift coverage into redshift shells and we count how many galaxies are present in each redshift shell. Just like in \eqref{eq: missing gals z}, we estimate the number of missing galaxies in each redshift shell:
\begin{equation}
    N_{\rm miss}(z_{\rm sh})=N_{\rm obs}(z_{\rm sh}) \cdot \frac{1-f(z_{\rm sh})}{f(z_{\rm sh})}
\end{equation}
where the subscript "sh" indicates that these quantities are relative to the redshift shell.

We then divide each redshift shell into pixels, using the python module \texttt{Healpy}~\citep{Healpy1, Healpy2} in order to have pixels of roughly the same area. The number of pixels in each redshift shell will depend on the redshift shell number itself according to
\begin{equation}
    N_{\rm pix}(z_{\rm sh})=12 (z_{\rm sh}+1)^2.
\end{equation}

We assign to each pixel a "clustering probability", determined by a sum of correlation functions (following equation \eqref{eq: corrfunc sum introduction}):
\begin{equation}
    \label{eq: ClusteringProbabilityPiel}
    p_{\rm clu}(\text{pix}_c)=1+\sum_{r<r_{\rm max}} \xi(r_{\text{pix}_c-\text{pix}_i}){N_g}_{\text{pix}_i}
\end{equation}
where $r_{\text{pix}_c-\text{pix}_i}$ is the distance between other pixels and the pixel in consideration, $N_{\text{pix}_i}$ is the number of galaxies contained in $\text{pix}_i$ and the sum is performed over all pixels within the maximum correlation distance $r_{\rm max}$. Once all the clustering probabilities are computed, we renormalize them, so that the sum of all clustering probabilities in a redshift shell is equal to 1

\begin{equation}
    \label{eq: renormalizing clust prob}
    \sum_{i=0}^{N_{\rm pix}} {p_{\rm clu}}_i = 1.
\end{equation}

Doing so, we are penalizing pixels that are far away from overdensities of galaxies and favoring pixels that are in the proximity of such overdensities.

Now that we have the number of missing galaxies per redshift shell and the clustering probabilities in each pixel, we can complete our galaxy catalog: we randomly choose a redshift shell that has at least 1 estimated missing galaxy. Among pixels in that redshift shell, we extract one randomly using as probability for each pixel the clustering probabilities computed in equation \eqref{eq: ClusteringProbabilityPiel} and add a single galaxy to the catalog in that pixel (the exact position is chosen randomly, uniformly in volume, within that pixel). We repeat this process, without recomputing the clustering probabilities, until we have added back in the catalog all the estimated missing galaxies in every redshift shell.

Once a catalog has been completed, we can now use it to perform a measurement of the Hubble constant, using the same approach as in the previous section, but this time assuming a complete catalog ($p_G=1, p_{\bar{G}}=0$), obtaining an $H_0$ posterior, and repeating this process $N_{\rm repop}$ time and averaging the posteriors as in equation 
\eqref{eq: AvaragingOutH_0Posteriors}.

\section{Results}
\label{sec: Results}

We now present the results: we first discuss the catalog completion process itself and afterwards the $H_0$ inference. Across this section, we will show plots for different detection thresholds and completeness fractions of the catalog, showing that our method works regardless of the details of the catalog and GW event generation.

\subsection{Catalog Completion}

\begin{figure}
    \centering
    \includegraphics[width=1\linewidth]{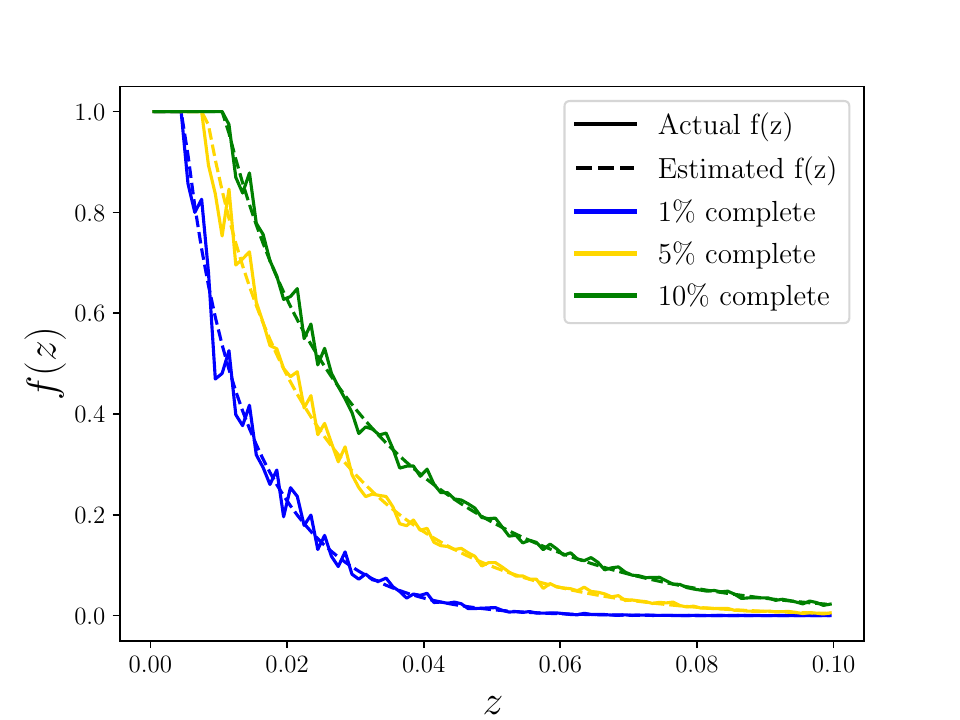}
    \caption{Actual completeness fraction (solid line)  versus estimated completeness (dashed line) through equation \eqref{eq: Missing galaxies fraction} as a function of redshift, for a 1\%, 5\% and 10\% overall complete MICE catalog. Note that, when we refer to the completeness of a catalog, we always refer to the global completeness of a catalog, as in total number of galaxies present in the incomplete catalog over total number of galaxies present in the complete catalog. The completeness of each redshift shell depends strongly on redshift, as can be seen above.}
    
    \label{fig:EstimatedvsActualIncompleteness}
\end{figure}

As explained in section \ref{sub: Catalog Completion}, our completion process starts with estimating the number of missing galaxies at each redshift shell. In figure \ref{fig:EstimatedvsActualIncompleteness} we show, as a function of redshift, the fraction of galaxies still present in the an incomplete (1\%, 5\% and 10\% overall completeness fraction) MICE catalog, compared with the completeness estimator introduced in equation \eqref{eq: Missing galaxies fraction}. We are able to accurately estimate the function $f(z)$, and through equation \eqref{eq: missing gals z} the number of missing galaxies in each redshift shell. By integrating $f(z)$ over redshift, we can also estimate directly the overall completeness fraction of the catalog and therefore the total number of missing galaxies. This will be useful for a sanity check (shown in figure \ref{fig: OurCompletion_ClustvsUni_sigmadl0.1}) when placing back galaxies in the catalog without clustering.

\begin{figure*}
	%\addtocounter{figure}{-1}
	\centering
	\includegraphics[width=0.45\textwidth]{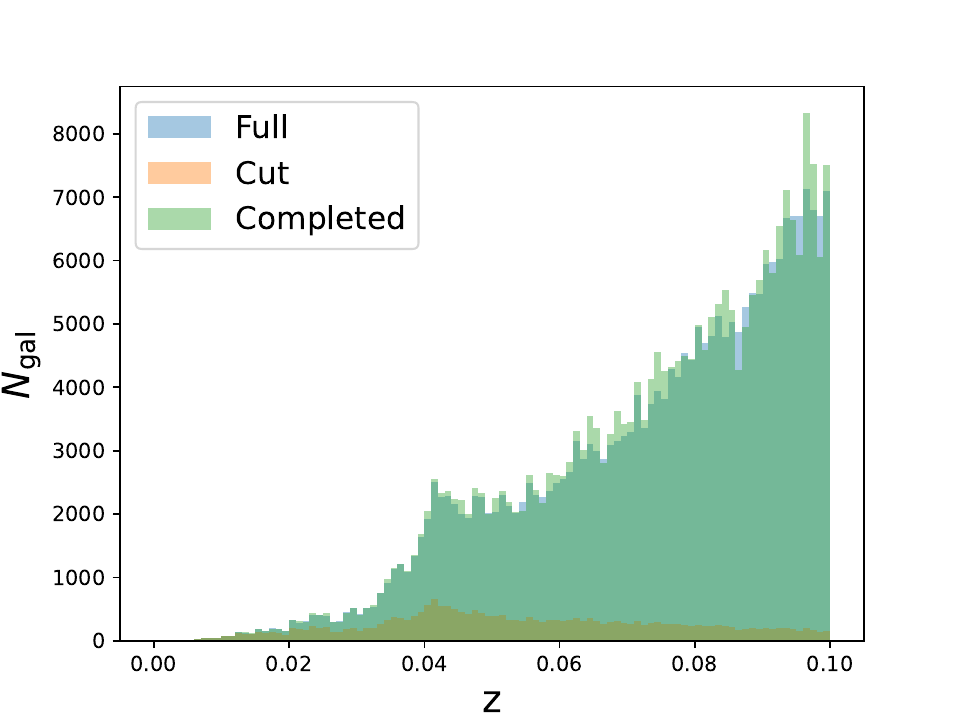}
	\hspace{1cm}
	\includegraphics[width=0.45\textwidth]{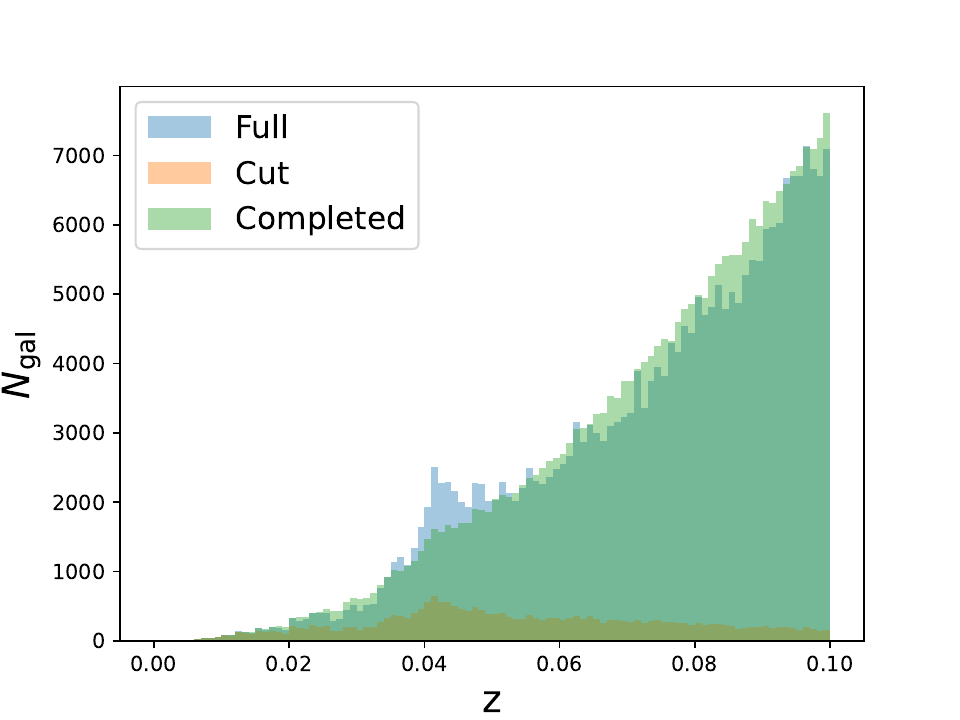}
    \vspace{1cm}
    \includegraphics[width=0.45\textwidth]{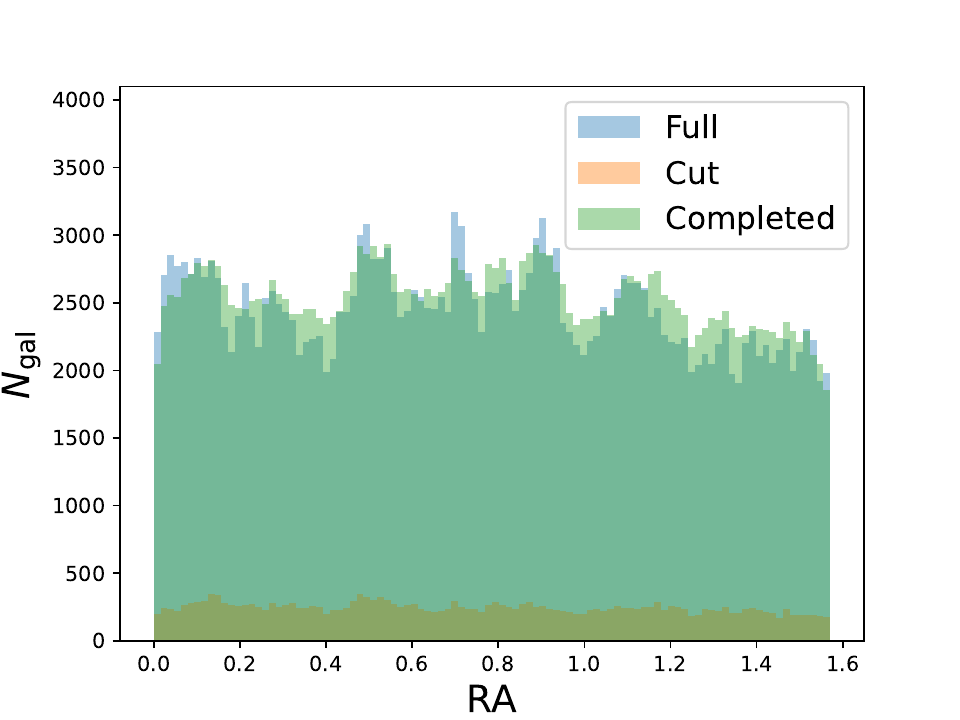}
	\hspace{1cm}
	\includegraphics[width=0.45\textwidth]{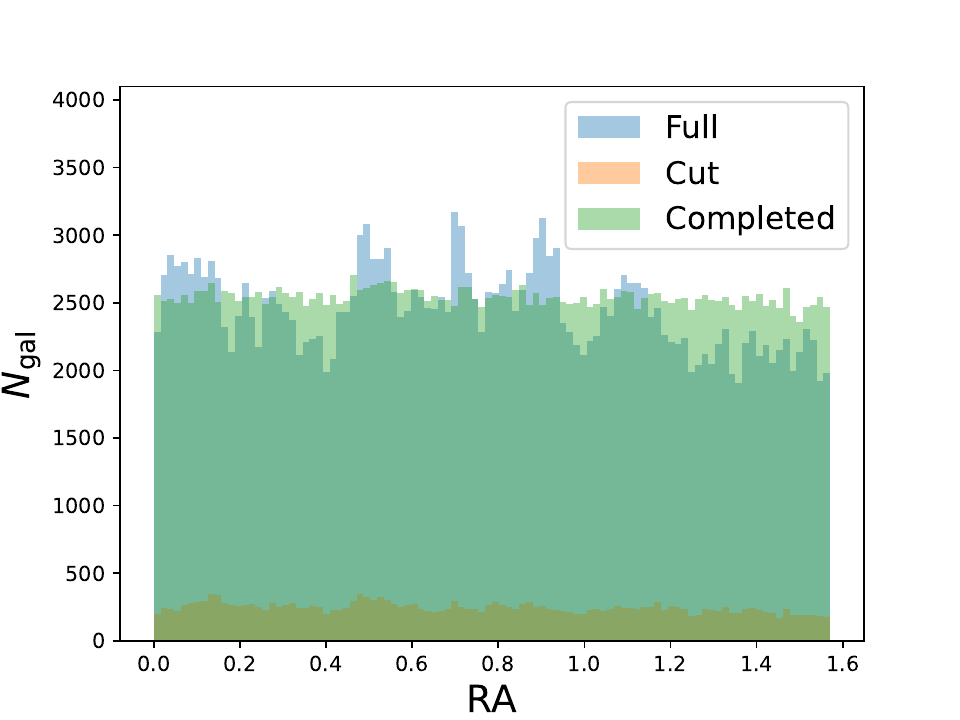}
    \vspace{1cm}
	\includegraphics[width=0.45\textwidth]{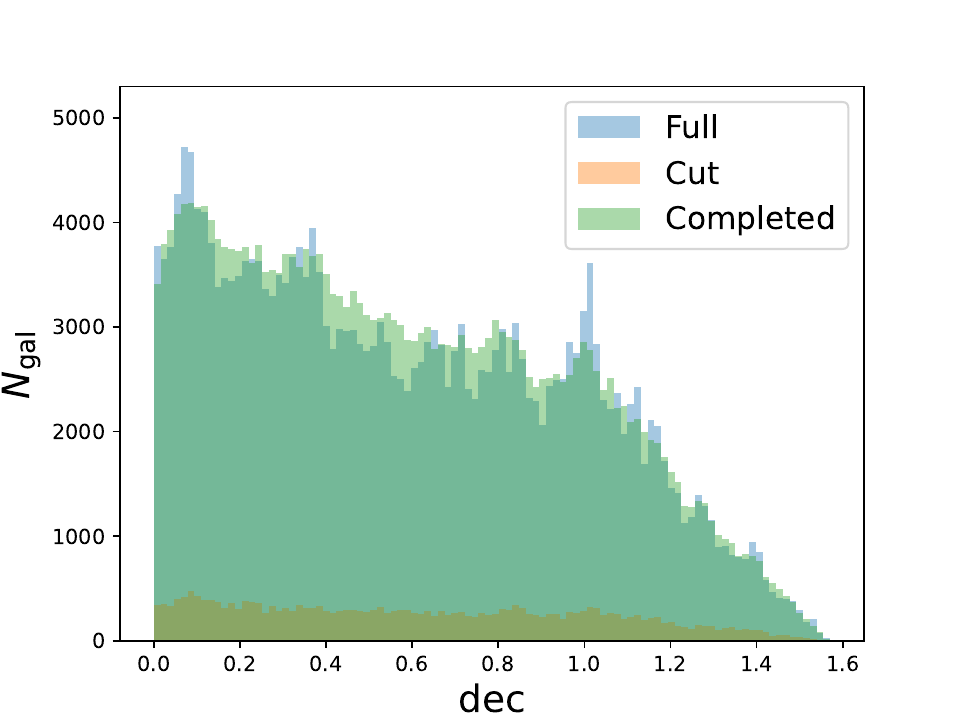}
	\hspace{1cm}
	\includegraphics[width=0.45\textwidth]{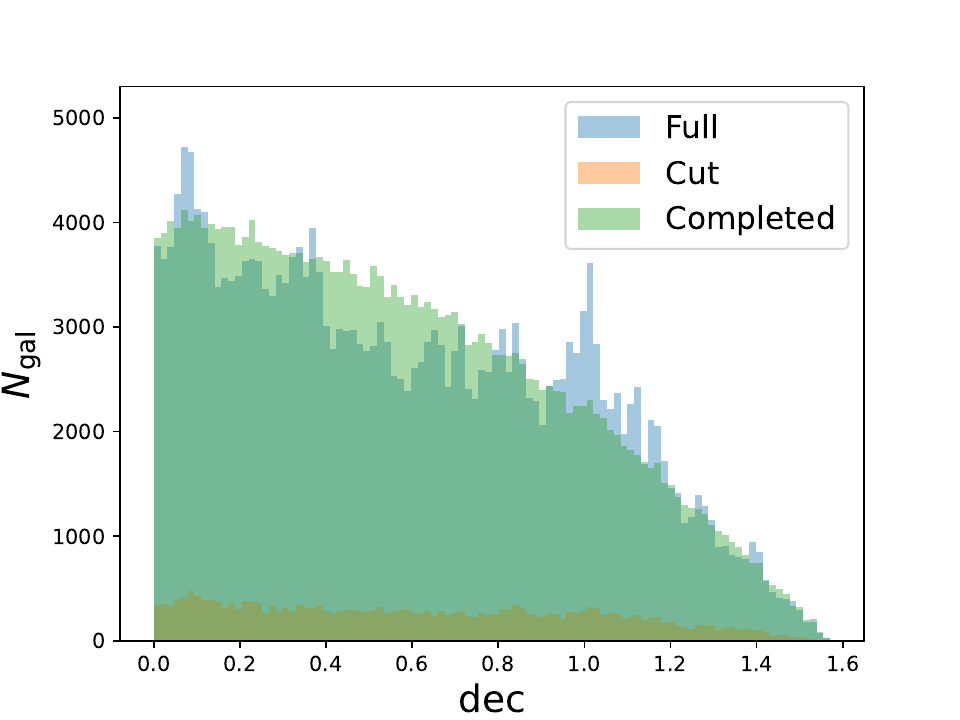}

	\caption{These histograms represent a comparison between the complete MICE catalog, the incomplete MICE catalog at a 10\% completeness and a re-completed catalog, constructed through the completion process presented in section \ref{sub: Catalog Completion}. In the left column, the completion was performed accounting for galaxy clustering, using a theoretical correlation function. On the right column, the same completion procedure was carried out uniformly in comoving volume. The first row shows a histogram for the redshift, the second row for the right ascension (RA), the last row for the declination (dec).}
	\label{fig: CompletionHistograms}
\end{figure*}

Given the clustering probabilities and the estimated missing galaxies at each redshift shell, we can reconstruct our galaxy catalog. In the panels of figure \ref{fig: CompletionHistograms}, we show a comparison of the distributions of the three coordinates ($z, {\rm RA}, {\rm dec}$) between the complete MICE catalog, the 10\% complete MICE catalog and our completed version.

As a comparison, we also show in the panels of figure \ref{fig: CompletionHistograms} the same histograms for a catalog completed according to the exact same procedure, but disregarding clustering (using an identically zero correlation function $\xi (r) = 0 $). With clustering we are able to reproduce much better various features of the galaxy distribution, which are completely lost when neglecting clustering.

\subsection{Cosmological Analysis}
In this section we will focus our attention on the Hubble constant posterior, comparing the uniform statistical method with our completion method. In particular, we will study the effects of:
\begin{itemize}
    \item Detection threshold
    \item Catalog completeness
    \item Number of events
    \item Uncertainty in distance and sky localization
    \item Assumptions about the correlation function
\end{itemize}

\subsubsection{Detection Threshold}
\label{subsub: detection threshold}

\begin{figure}
    \centering
    \includegraphics[width=1\linewidth]{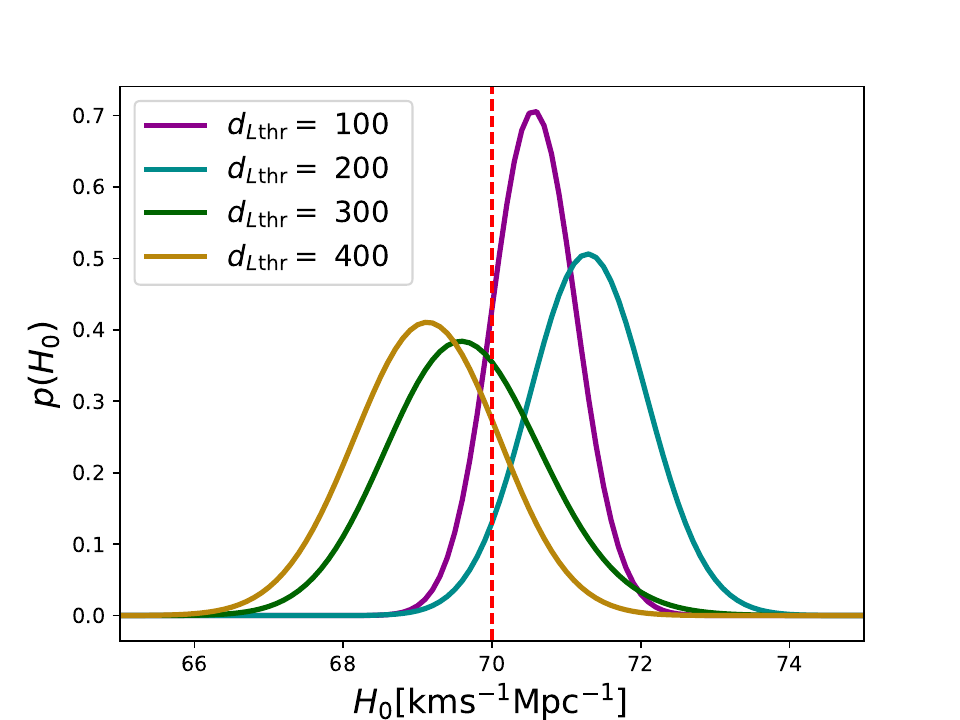}
    \caption{Likelihoods for 1000 gravitational wave events for the complete MICE catalog, generated assuming an error on sky position angles of $\sigma_{\Theta}=0.1$ and a fractional error on luminosity distance $A=0.1$, for different detection thresholds. Note that, given the different detection thresholds, each posterior is generated with a newly drawn batch of 1000 events. The dashed red line marks the $H_0$ input value, $H_0 = 70 \ \text{km}\ \text{s}^{-1} \ \text{Mpc}^{-1}$.}
    \label{fig: CompleteMICE}
\end{figure}

In figure \ref{fig: CompleteMICE} we show the $H_0$ posteriors obtained for the complete MICE catalog and 1000 gravitational wave events, with a fractional distance uncertainty $A=0.1$ and a Gaussian standard deviation in right ascension and declination $\sigma_{\rm RA}=\sigma_{\rm dec}=0.1$, corresponding to a 68\% sky localization contour of $\Delta_{\Omega} \sim 30 \ \text{deg}^2$ (see section \ref{sub: Data Generation} for details regarding these quantities) and for different detection thresholds. In this plot we can see how, as a general trend, the higher the detection threshold, the less precise the $H_0$ measurement. This is of course to be expected, as a higher detection threshold translates to a bigger error on the luminosity distance, and therefore to a higher number of galaxies that are possible hosts of the gravitational wave event. We note also an effect that will be even more relevant when dealing with an incomplete catalog: the described trend seems to stop at a detection threshold ${d_L}_{\text{th}} = 400 \ \text{Mpc}$, as this posterior is roughly the same width as the one for a detection threshold ${d_L}_{\text{th}} = 300 \ \text{Mpc}$. This is not a fluke, but arises because our galaxy catalog does not extend to infinity but only up to redshift 0.1, i.e., to a luminosity distance of about 440 Mpc, for $H_0 \sim 70 \ \text{km}\ \text{s}^{-1} \ \text{Mpc}^{-1}$, which is fairly close to this artificial edge of the catalog. This means that the luminosity distance posteriors for many events have significant coverage at distances past the edge of the catalog. If many events %luminosity distance posteriors 
cover the edge of the catalog, we gain information of the luminosity distance edge of the catalog, which combined with the redshift edge of the catalog gives a very precise measurement of $H_0$. For a more extensive discussion on the edge of the catalog effect, see appendix \ref{Appendix edge}.

\subsubsection{Catalog Completeness}
\label{subsub: catalog completeness}
\begin{figure}
    \centering
    \includegraphics[width=1\linewidth]{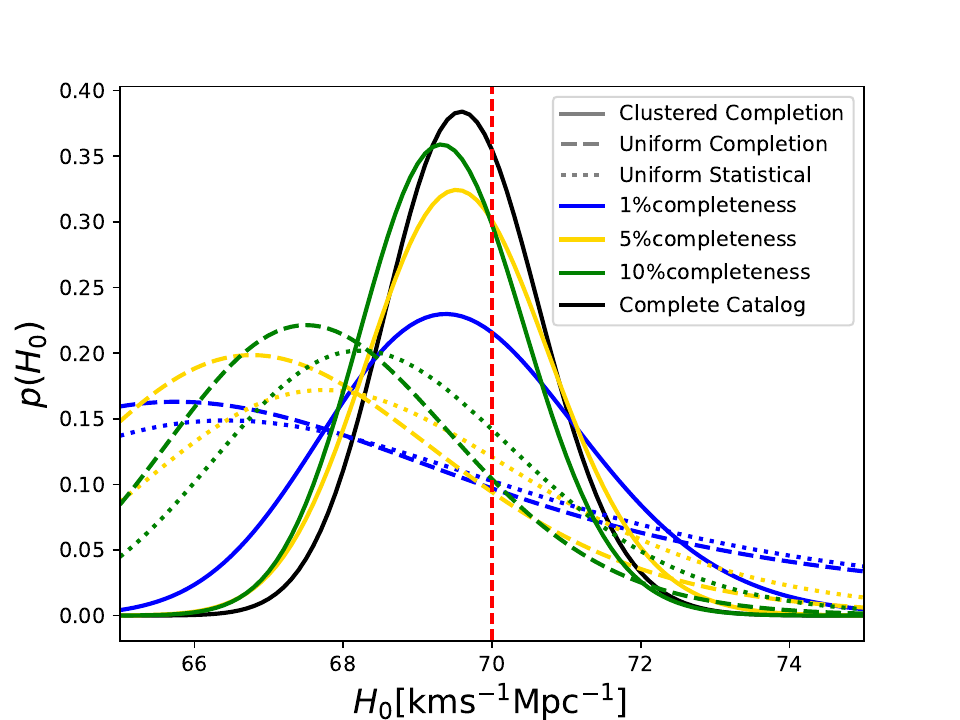}
    \caption{$H_0$ posteriors for different MICE catalog completeness fractions (1\%, 5\% and 10\%) and different methods of completion, for ${d_L}_{\text{thr}} = 300$ Mpc, $A=0.1$, $\Delta_\Omega\sim40 \ \text{deg}^2$, $N_{\rm events}=1000$. The dashed red line marks the $H_0$ input value, $H_0 = 70 \ \text{km}\ \text{s}^{-1} \ \text{Mpc}^{-1}$. The dotted lines represent the traditional uniform statistical method. The dashed and solid lines are instead obtained by completing the incomplete galaxy catalog 100 times, according to the procedure explained in Section~\ref{sub: Catalog Completion}, and averaging over the 100 realizations. The dashed lines correspond to the no-clustering (uniform) completion, while the solid lines correspond to the clustered (using the correlation function) completion. We show as reference the $H_0$ posterior for a complete catalog, obtained using the same gravitational wave events as the other ones. All posteriors are obtained assuming a flat $H_0$ prior.}
    \label{fig: OurCompletion_ClustvsUni_sigmadl0.1}
\end{figure}

We move now to investigate the most interesting and most relevant parameter, which is the incompleteness of the catalog. 
In figure \ref{fig: OurCompletion_ClustvsUni_sigmadl0.1} we compare Hubble constant posteriors for three different completeness fractions (1\%, 5\% and 10\%) and three different methods: the traditional statistical method, assuming a uniform in comoving volume distribution for the missing galaxies (see section \ref{subsub: Incomplete Catalog Bayesian} and appendix \ref{Appendix components} for details), our new clustered completion method (see \ref{sub: Catalog Completion}) and a uniform completion method, where we repopulate the catalog just like in the clustered completion method but distributing the galaxies uniformly, effectively using a correlation function which is identically zero. 
The "uniform completion" method should in practice be equivalent to the "uniform statistical" one, and we show it in figure \ref{fig: OurCompletion_ClustvsUni_sigmadl0.1} as a quick sanity check, to show that our process of adding back galaxies in the catalog does not intrinsically introduce any additional information or bias. Note that in this case we don't estimate the missing galaxies at each redshift shell, but we only estimate the number of overall missing galaxies, and distribute those galaxies uniformly in comoving volume: doing otherwise would still enforce some level of clustering in redshift, and would not be equivalent to the traditional statistical method.
We find that the posteriors for the uniform statistical method and uniform completion method are very close to each other, although not identical: while we expect them to be similar, we don't expect to be perfectly overlapping, as they are obtained with two different methods, one of which involves some inherent randomness.  The clustered completion method produces results which are both more precise and closer to the injected value, approaching the complete catalog posterior. We also note that, even in the case of a complete galaxy catalog, we don't expect the posterior on $H_0$ to be exactly centered on the injected value, as it depends on the specific distribution of gravitational wave events in the realised universe, which is inherently random.

\begin{figure}
    \centering
    \includegraphics[width=1\linewidth]{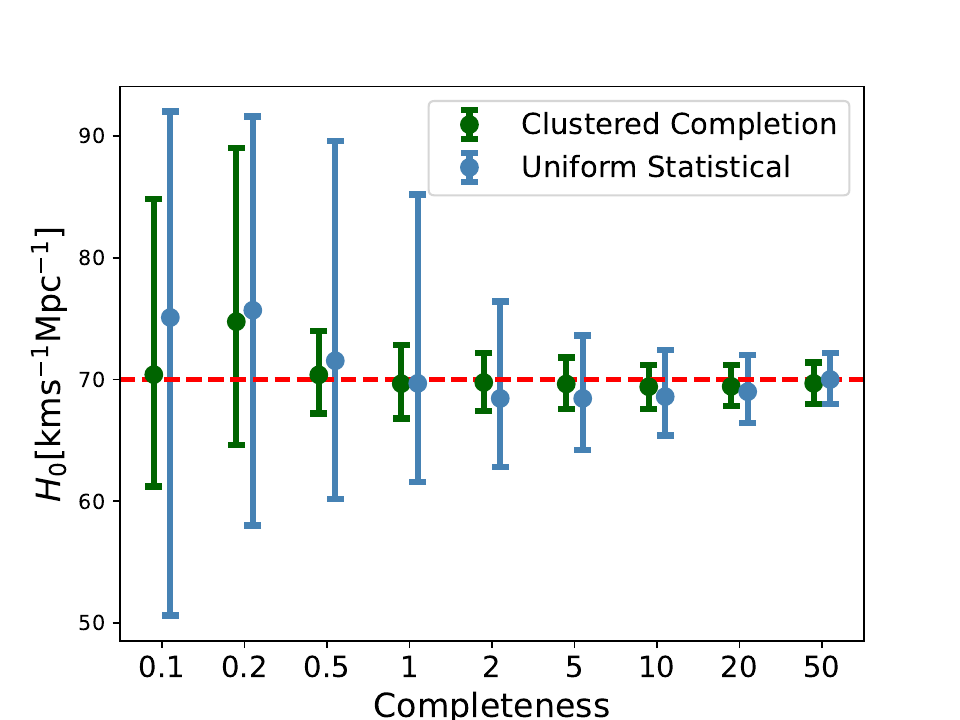}
    \caption{$H_0$ mean and 90\% credible intervals, for a fixed detection threshold of ${d_L}_\text{thr}=300$ Mpc and a range of overall catalog completeness fractions (as a percentage). The dashed red line marks the $H_0$ input value, $H_0 = 70 \ \text{km}\ \text{s}^{-1} \ \text{Mpc}^{-1}$. The blue symbols and error bars represent the traditional statistical method, assuming a uniform distribution of missing galaxies. The green symbols and error bars represent our clustered completion procedure, accounting for galaxy clustering. For completeness fractions between 0.5\% and 10\%, our method significantly outperforms the traditional one, obtaining measurements between 2 and 4 times as precise. At very high and very low completenesses our method becomes comparable to the traditional one.}
    \label{fig: ErrorBarPlot}
\end{figure}

In figure \ref{fig: ErrorBarPlot} we show the mean of the Hubble constant posteriors for a range of different catalog completenesses, together with their 90\% credible intervals. Our method performs drastically better than the traditional one for completenesses between 0.5\% and 10\%: this is the sweet spot where the catalog has a significant number of missing galaxies, but still retains enough structure to inform the placement, according to the correlation function, of the newly added missing galaxies. We have our best result for 0.5\% completeness fraction, where our method is more than 4 times as precise as the uniform statistical one, while for a 10\% completeness fraction the improvement is only a factor of $\sim2$. At high completeness fractions, the difference is minimal, as there is still so much information present in the incomplete catalog that simply being careful of accounting for the incompleteness at high redshift is enough to get an unbiased and precise measurement. On the other hand, at very low completeness fraction,  < 0.5\%, the difference is not as significant, as almost all the galaxies are missing from the catalog and therefore not enough structure is retained in the incomplete catalog to inform the placement of new galaxies with the clustered completion method.

\subsubsection{Number of events}
\begin{figure}
    \centering
    \includegraphics[width=1\linewidth]{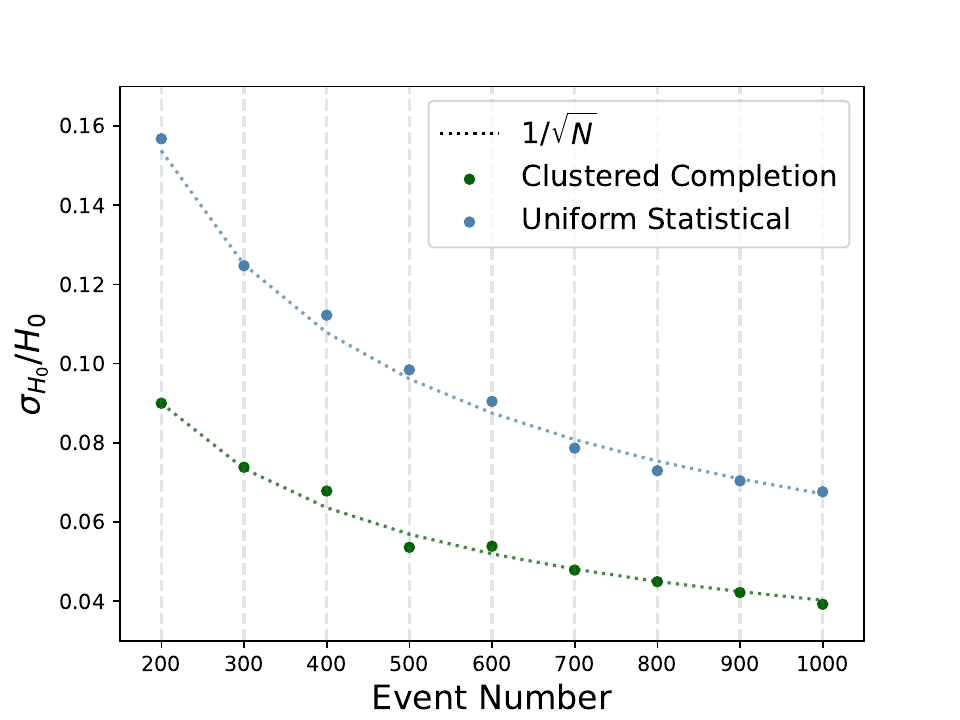}
    \caption{$H_0$ measurement precision, expressed as the width of the 90\% credible interval divided by the posterior mean, as a function of the number of events, for a 5\% complete catalog and a detection threshold ${d_L}_{\text{thr}}=200$ Mpc. The dark green dots represent the clustered completion method (together, in the dotted line, with the expected $1/\sqrt{N}$ behaviour, while the light blue one (and the corresponding dotted line) represent the traditional uniform statistical method. The same set of random events is used to generate the likelihoods for both the uniform statistical method and the clustered completion method.}
    \label{fig: sigmaH0overH0_vs_Events}
\end{figure}

Until now, we have shown $H_0$ posteriors obtained with a fixed number of gravitational wave events, $N_{\rm events}=1000$. In figure \ref{fig: sigmaH0overH0_vs_Events} we show the precision of the Hubble constant posterior, expressed as the width of the 90\% credible interval divided by the mean of the posterior, for a varying number of events, ranging from 100 to 1000. Firstly, we see that the expected behaviour of $1/\sqrt{N}$ is well recovered, which is a good sanity check that our process does not introduce biases. Furthermore, we can see how our method consistently outperforms the traditional statistical one, by a factor of almost two. Equivalently, we can see how the same precision on $H_0$ can be obtained by $\sim4$ times fewer events.

\subsubsection{Distance and angle uncertainties}
\label{sub: Distance and angle uncertainties}

\begin{figure}
    \centering
    \includegraphics[width=1\linewidth]{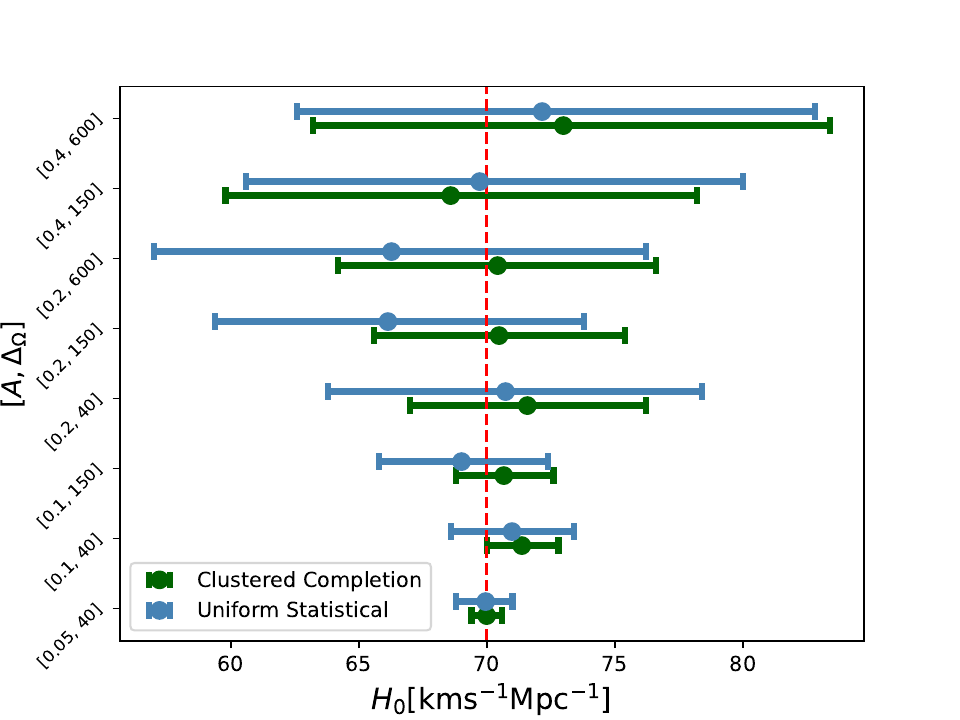}
    \caption{$H_0$ measurements, represented as 90\% credible intervals for various errors on distance and sky localization, for a 5\% complete catalog and a detection threshold $d_{L \rm thr}=200$ Mpc. The fractional error in distance is represented with the dimensionless parameter $A$, while the error in sky localization is represented as the 68\% contour in degrees squared. Note that the sky localization error is only an approximate indication, as the exact contour area will depend on the location of the event. Given the distance and angle uncertainty, a set of 1000 events is generated, and the same set of events is used to generate both the uniform and clustered likelihoods. The dashed red line marks the $H_0$ input value, $H_0 = 70 \ \text{km}\ \text{s}^{-1} \ \text{Mpc}^{-1}$. The blue lines represent the traditional Bayesian uniform method, while the green lines represent our clustered completion method.}
    \label{fig: varioussigmasdlradec}
\end{figure}

Until now we have always assumed a fixed error on luminosity distance and sky position of the gravitational wave events, $A=0.1$ and $\sigma_{\text{ra}}=\sigma_{\text{dec}}=0.1$, corresponding to a 68\% contour approximately equal to $30 \ \text{deg}^2$. These assumptions are reasonable for  binary neutron stars (BNS): GW170817 had very similar errors in both distance and sky localization \citep{GW170817discovery}. However, these numbers are extremely optimistic for binary black holes (BBH), which form the vast majority of events observed by the LVK collaboration \citep{theligoscientificcollaboration2026gwtc50observationssecondfourth}. In figure \ref{fig: varioussigmasdlradec} we show $H_0$ estimates (expressed as means and 90\% credible intervals) for different combinations of the dimensionless distance error parameter A and the $68\%$ sky localization area in $\text{deg}^2$, this time for a detection threshold of ${d_L}_{\rm thr}=200 \ \rm Mpc$ and an overall completeness fraction of 5\% (we changed these parameters simply to show that our method works well regardless of these choices). 

From this plot, we see two main behaviours:

\begin{itemize}
    \item The precision gain from using a clustered completion is less relevant at higher errors. For example, when the distance fractional error is 40\%, both the traditional method and our clustered completion produce almost identical credible intervals. This is not surprising, as the luminosity distance distribution is so wide that it covers almost uniformly the whole galaxy catalog, rendering our improved galaxy placement useless. However, starting from a 20\% fractional error, we can see a significant improvement with our clustered completion.

    \item The fractional error in distance, $A$, is more impactful than the sky localization error, $\Delta_\Omega$. Halving the fractional distance error improves significantly the results, while halving the error in right ascension/declination (therefore making the localization area 4 times as precise) improves the results only marginally. This is in agreement with \cite{MariosPaper}
\end{itemize}

\subsubsection{Assumptions about the correlation function}
\label{subsub: correlation function results}

All results shown until now are computed assuming a power law correlation function, whose parameters $r_0$ and $\gamma$ are fit from the catalog itself, using the DP estimator \eqref{eq: EmpiricalCorrelation}, to ensure that no biases are introduced by assuming a different correlation function from the one expressed by the MICE catalog. However, with real galaxy catalogs, we have at best an estimate of the correlation function, but no way of knowing the exact shape and parameters. One might then argue that we better retrieve the injected Hubble constant due to this additional information that we do not have in real life. Furthermore, different estimators can produce significantly different correlation functions from the same galaxy catalog, without any reason to choose an estimator over the other. We also note that throughout the paper, including this subsection, we always assume that the correlation function does not depend on redshift: this is a valid approximation at the low redshifts we are treating ($z<0.1$), but it is not valid anymore at higher redshifts \citep{Peacock_1997}. We leave an adaptation of this method to an evolving correlation function for future works.

\begin{figure}
    \centering
    \includegraphics[width=1\linewidth]{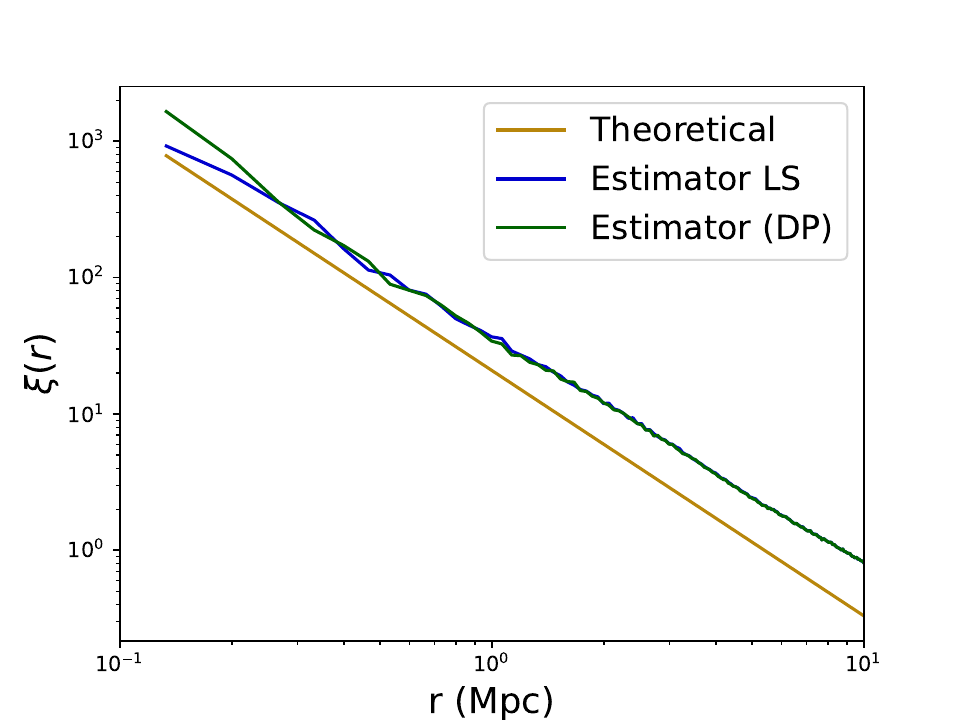}
    \caption{Loglog plot comparison between different power law correlation functions, keeping fixed the validity region between $r_{\rm min}=100 \ \rm kpc$ and $r_{\rm max}=10 \ \rm Mpc$. The gold line represents the theoretical correlation function, defined by the parameters of equation \eqref{Appendix eq: Clustering parameters}. The green line represents the correlation function measured directly from the full catalog through estimator \eqref{eq: EmpiricalCorrelation}. The blue line represents the correlation function measured directly from the full catalog via estimator \eqref{Estimator LS}.}
    \label{Appendix fig: correlation functions}
\end{figure}

Indeed, there are non negligible differences both between a theoretical correlation function and one fit from the catalog, and also between different estimators. In figure \ref{Appendix fig: correlation functions} we show, between the extremes $r_{\rm min}=100 \ \rm kpc$ and $r_{\rm max}=10\ \rm Mpc$, the comparison between the theoretical correlation function 
\begin{equation}
    \label{Appendix eq: theoretical correlation}
    \xi(r)=\left(\dfrac{r}{r_0}\right)^{-\gamma},
\end{equation}
with assumed parameters (taken from \cite{peebles1980})
\begin{equation}
    \label{Appendix eq: Clustering parameters}
    \begin{split}
        &\gamma = 1.8 \\
        &r_0 = 5.4\ \rm{Mpc}, \\ 
    \end{split}
\end{equation}
the correlation function fit from the catalog using the Davis-Peebles (DP) estimator \eqref{eq: EmpiricalCorrelation} \citep{DavisPeebles}, and the correlation function fit from the catalog using the Landy-Szalay (LS) estimator \citep{Landy-Szalay} 
\begin{equation}
\label{Estimator LS}
    \xi (r) = \dfrac{\langle D,D \rangle (r) - 2\langle D,R \rangle(r) +\langle R,R \rangle (r)}{ \langle R,R \rangle (r)} - 1.
\end{equation}

The difference between the different estimator is significant at small distances, but gets almost non existent at high distances. However, both estimators produce results that are fairly different from the theoretical correlation function, leading in particular to a stronger correlation function. One might expect that such differences, although small, in the assumed correlation function might produce significant differences in the resulting $H_0$ posterior, possibly even introducing biases.  

To check this, we compare 90\% credible intervals on $H_0$ obtained assuming a power law correlation function with three different sets of parameters: 
\begin{itemize}
    \item Theoretical parameters, taken from \cite{peebles1980}: $\gamma=1.8$, $r_0=5.4 \ \rm Mpc$;
    \item Parameters obtained fitting a power law to the DP estimator \eqref{eq: EmpiricalCorrelation}, obtaining $\gamma=2.15$, $r_0=4.14 \ \rm Mpc$. This is the correlation function used throughout the main body of this paper.
    \item Parameters obtained fitting a power law to the LS estimator \eqref{Estimator LS}, obtaining $\gamma=1.54$, $r_0=7.31 \ \rm Mpc$. 
\end{itemize}

\begin{figure}
    \centering
    \includegraphics[width=1\linewidth]{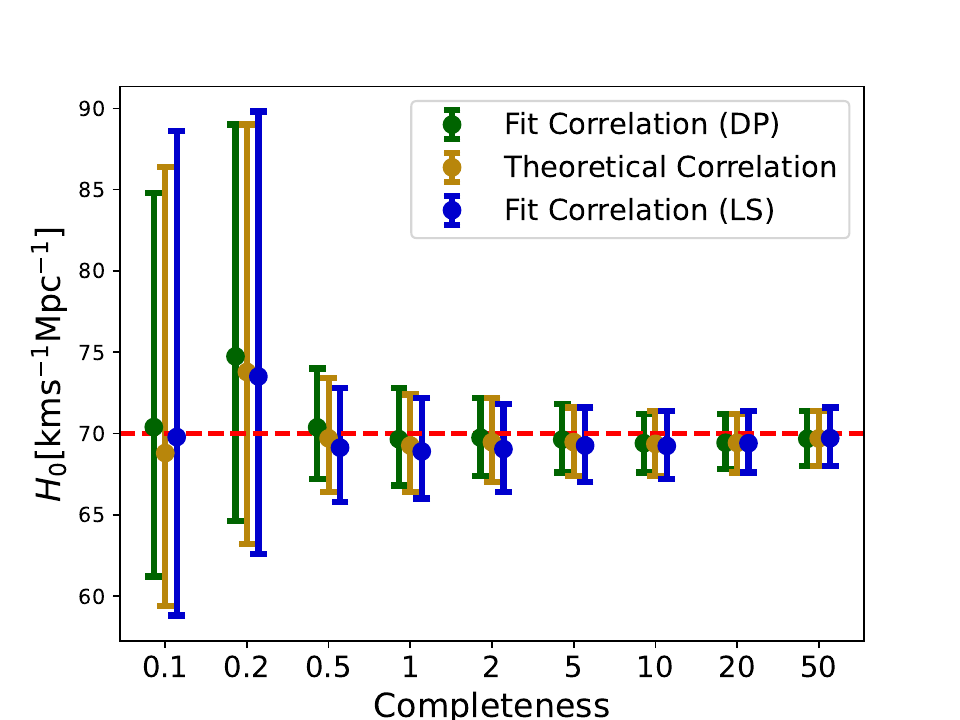}
    \caption{$H_0$ means and $90\%$ credible intervals for various completeness fractions and fixed parameters ${d_L}_{\rm thr} = 300 \ \text{Mpc}$, $A=0.1$, $\sigma_{\rm RA}=\sigma_{\rm dec}=0.1$. The dashed red line marks the $H_0$ input value, $H_0 = 70 \ \text{km}\ \text{s}^{-1} \ \text{Mpc}^{-1}$. All intervals are obtained assuming a power law, varying the slope $\gamma$ and the typical length $r_0$. The green bars use parameters obtained from the estimator \protect \eqref{eq: EmpiricalCorrelation}, which is the same correlation function used throughout the main body of this paper; the gold bars use theoretical parameters, taken from \protect \cite{peebles1980}; the blue bars use parameters obtained from the estimator \protect \eqref{Estimator LS}.}  \label{Appendix fig: correlation function comparison}
\end{figure}

For all three realizations we keep fixed the minimum and maximum range of the correlation function to the theoretical ones:
\begin{equation}
\begin{split}
    &r_{\rm min} = 100 \ \rm{kpc} \\ 
    &r_{\rm max} = 10 \ \rm{Mpc},
\end{split}
\end{equation}
 We show the comparison, for various completeness fractions, in figure \ref{Appendix fig: correlation function comparison}. The difference in the credible intervals is negligible, with all three different sets of parameters vastly improving over the uniform completion method. We can conclude that, as long as a reasonable correlation function is chosen, the exact shape and parameters don't have a significant impact on our method, and if any biases are introduced, they are well below the statistical uncertainty of the recovered $H_0$ posteriors. 

\subsection{Checking for biases}

As a final test of our method, we want to make sure our Hubble constant posteriors are not biased, or that a bias, if present, is well below the statistical uncertainty of the recovered posterior. In order to do this, we repeat the whole completion procedure (generating events, completing a galaxy catalog 100 times, obtaining 100 Hubble constant posteriors and avaraging them to get a final $H_0$ posterior) 25 times. For each of these different realizations, we extract a different injected value of $H_0$, ${H_0}_{\rm true}$, from our prior, which is flat between ${H_0}_{\rm min} = 40 \text{km}\ \text{s}^{-1} \ \text{Mpc}^{-1}$ and ${H_0}_{\rm max} = 100 \text{km}\ \text{s}^{-1} \ \text{Mpc}^{-1}$. Given the 25 posteriors and the associated injected values of $H_0$, we build a PP plot, with the posterior CDF at the injected value on the x-axis.

\begin{figure}
    \centering
    \includegraphics[width=1\linewidth]{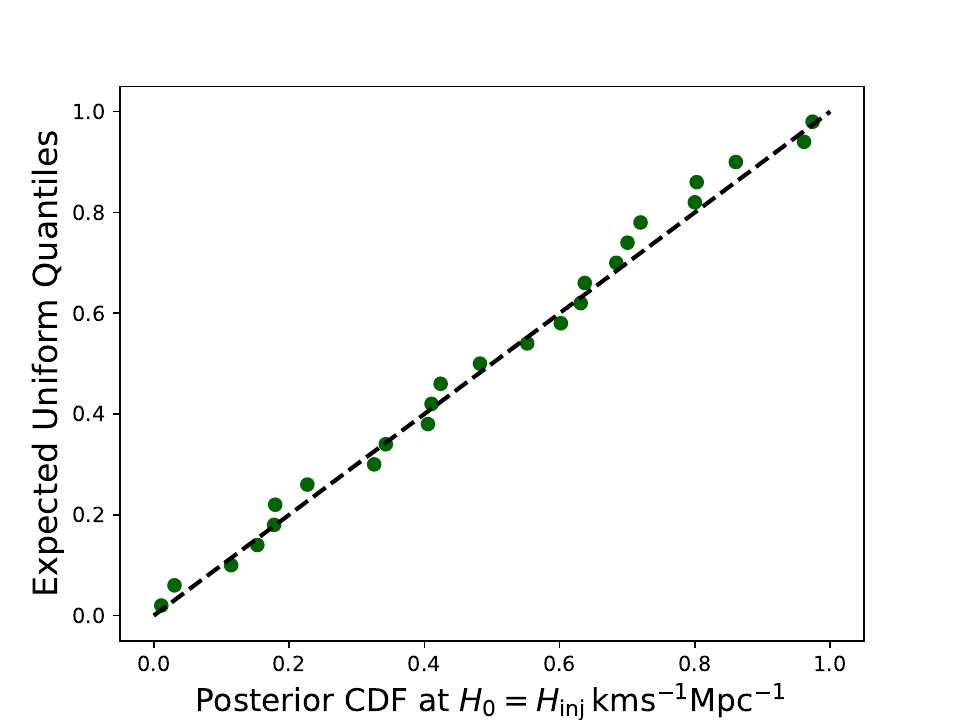}
    \caption{Probability-Probability (PP) plot for 25 realizations of our complete completion procedure, for a 10\% complete galaxy catalog, a detection threshold ${d_L}_{\rm thr}=300\ \space \rm Mpc$, and errors on distance and sky location respectively $A=0.1$ and $\sigma_{\rm RA}=\sigma_{\rm dec}=0.1$. 
    Each realization has a different injected value of ${H_0}_{\rm true}$, sampled from the flat prior we use throughout this paper. On the x-axis we show the posterior CDF value at the injected value, while on the y-axis we show the expected uniform quantiles: given that we expect a uniform distribution between 0 and 1, the dots are simply equally distanced in the y-axis. The PP plot does not show a significant bias, and the dots lie on the diagonal as expected.}
    \label{fig:placeholder}
\end{figure}

For a well calibrated inference procedure we expect the PP plot to show a diagonal, and this is indeed what we see. We can conclude that our procedure does not introduce any significant bias, neither in the width of the posterior nor in the location of the mean.

\section{Conclusions}
\label{sec: Conclusions}
In this paper, we have introduced a simple method to account for galaxy clustering in the estimation of the Hubble constant using GW dark sirens and incomplete galaxy catalogs. We have used a mock galaxy catalog (the MICE galaxy catalog) and mock gravitational wave events, employing a simplified model for their generation, detection and error estimation. Our method is based on a galaxy catalog completion, which means refilling an incomplete catalog with the missing galaxies, repeating this process several times and averaging over the final results. We generate incomplete galaxy catalogs, starting from the complete MICE catalog up to redshift 0.1, by imposing an apparent magnitude cut in order to obtain incomplete galaxy catalogs. We then performed cosmological analyses in various scenarios, comparing the traditional method to our clustered completion method.

We find that our method is, in the best case scenario of small errors in luminosity distance and sky location (comparable to current BNS errors) and a low completeness fraction (0.5 to 10\%) of the galaxy catalog, able to significantly outperform the traditional statistical method, which assumes a uniform in comoving volume distribution for the missing galaxies. The $H_0$ estimates are both closer in mean to the injected value of ${H_0}_\text{true} = 70 \ \text{km}\ \text{s}^{-1} \ \text{Mpc}^{-1}$ and more precise by a factor of between 1.5 and 4 in the very best case scenario (Figure \ref{fig: OurCompletion_ClustvsUni_sigmadl0.1}). At no point does our method introduce biases, nor does it perform worse than the uniform statistical one, but it does become comparable, or even indistinguishable, in two main cases:

\begin{itemize}
    \item When the completeness of the catalog is extremely low, <0.5\%. In this case, there are so few galaxies present in the incomplete catalog that it is impossible to place the missing galaxies enforcing clustering to the present ones (Figure \ref{fig: ErrorBarPlot}), and we resort to a completion procedure that is very close to a uniform one.
    \item When the uncertainties in luminosity distance (the dimensionless parameter $A$) and sky localization area (expressed in this paper as 68.3\% area) $\Delta_\Omega$ are too big ($>20\%$ and $150 \ \text{deg}^2$ respectively). This effect is actually dominated by the luminosity distance uncertainty, and is also fairly intuitive: with a fractional error of, for example, 40\%, the distance posterior assigns very similar probabilities to almost all areas of the catalog, therefore rendering exact placement of galaxies irrelevant
\end{itemize}

We have also showed that our method still works even if different correlation function parameters are used, proving that as long as reasonable assumptions are made about the correlation function, the exact assumptions matter only marginally. Finally, we showed through a PP plot that our method consistently recovers different injected values of the Hubble constant, while never introducing biases in either the mean or standard deviation of the posteriors.

Given the simplicity of our algorithm, the significant improvement over alternatives in reasonable scenarios, and its robustness, we believe that it can be a very useful tool in extracting as much information as possible from dark siren gravitational wave events. 
As natural extensions of this paper and application of this methods, we can look to studying the impact of introducing galaxy properties, such as the luminosity in certain bands, as weights in the likelihood or in the repopulation procedure (or both). It would also be extremely useful to add the option of a correlation function evolving in redshift, which should be fairly easy to introduce as the correlation probabilities are computed redshift shell by redshift shell. Furthermore, application to real galaxy catalogs and real gravitational wave events seems a natural follow up to test our method.

\section*{Acknowledgments}
We would like to thank Rachel Gray and Simone Mastrogiovanni for helpful discussions in very early stages of this work. SK acknowledges funding via STFC Small Grant ST/Y001133/1. JG acknowledges support from the Simons Foundation International via Grant No. SFI-MPS- BH-00012593-06. For the purpose of open access, the authors have applied a Creative Commons Attribution (CC BY) license to any Author Accepted Manuscript version arising from this submission.

\newpage

\bibliographystyle{mnras}
\bibliography{biblio}

@article{Gair_2023,
   title={The Hitchhiker’s Guide to the Galaxy Catalog Approach for Dark Siren Gravitational-wave Cosmology},
   volume={166},
   ISSN={1538-3881},
   url={http://dx.doi.org/10.3847/1538-3881/acca78},
   DOI={10.3847/1538-3881/acca78},
   number={1},
   journal={The Astronomical Journal},
   publisher={American Astronomical Society},
   author={Gair, Jonathan R. and Ghosh, Archisman and Gray, Rachel and Holz, Daniel E. and Mastrogiovanni, Simone and Mukherjee, Suvodip and Palmese, Antonella and Tamanini, Nicola and Baker, Tessa and Beirnaert, Freija and Bilicki, Maciej and Chen, Hsin-Yu and Dálya, Gergely and Ezquiaga, Jose Maria and Farr, Will M. and Fishbach, Maya and Garcia-Bellido, Juan and Ghosh, Tathagata and Huang, Hsiang-Yu and Karathanasis, Christos and Leyde, Konstantin and Hernandez, Ignacio Magaña and Noller, Johannes and Pierra, Gregoire and Raffai, Peter and Romano, Antonio Enea and Seglar-Arroyo, Monica and Steer, Danièle A. and Turski, Cezary and Vaccaro, Maria Paola and Vallejo-Peña, Sergio Andrés},
   year={2023},
   month=jun, pages={22} }

@article{VanWyngarden_et_al_2025_cat_completeness,
        title = "{How Low Can You Go: Constraining the Effects of Catalog Incompleteness on Dark Siren Cosmology}",
        author = {VanWyngarden, Madison and Fishbach, Maya and Vijaykumar, Aditya and Guerrero, Alexandra G. and Holz, Daniel E.},
      journal = {arXiv e-prints},
         year = 2025,
        month = nov,
          eid = {arXiv:2511.04786},
        pages = {arXiv:2511.04786},
          doi = {10.48550/arXiv.2511.04786},
archivePrefix = {arXiv},
       eprint = {2511.04786},
 primaryClass = {astro-ph.CO},
       adsurl = {https://ui.adsabs.harvard.edu/abs/2025arXiv251104786V}
}

@article{Gray_2020,
   title={Cosmological inference using gravitational wave standard sirens: A mock data analysis},
   volume={101},
   ISSN={2470-0029},
   url={http://dx.doi.org/10.1103/PhysRevD.101.122001},
   DOI={10.1103/physrevd.101.122001},
   number={12},
   journal={Physical Review D},
   publisher={American Physical Society (APS)},
   author={Gray, Rachel and Hernandez, Ignacio Magaña and Qi, Hong and Sur, Ankan and Brady, Patrick R. and Chen, Hsin-Yu and Farr, Will M. and Fishbach, Maya and Gair, Jonathan R. and Ghosh, Archisman and Holz, Daniel E. and Mastrogiovanni, Simone and Messenger, Christopher and Steer, Danièle A. and Veitch, John},
   year={2020},
   month=jun }

@ARTICLE{Schechter,
       author = {{Schechter}, P.},
        title = "{An analytic expression for the luminosity function for galaxies.}",
      journal = {\apj},
         year = 1976,
        month = jan,
       volume = {203},
        pages = {297-306},
          doi = {10.1086/154079},
       adsurl = {https://ui.adsabs.harvard.edu/abs/1976ApJ...203..297S}
}

@book{peebles1980,
  title={The Large-Scale Structure of the Universe},
  author={Peebles, Phillip James Edwin},
  year={1980},
  publisher={Princeton University Press},
  address={Princeton, NJ},
  series={Princeton Series in Physics},
  isbn={9780691082400}
}

@book{longair2023clusteringbook,
  title={Galaxy Formation},
  author={Longair, Malcolm S},
  year={2023},
  edition={3rd},
  publisher={Springer},
  series={Astronomy and Astrophysics Library},
  doi={10.1007/978-3-662-65891-8},
  isbn={978-3-662-65891-8}
}

@book{mo2010clusteringbook,
  title={Galaxy Formation and Evolution},
  author={Mo, Houjun and van den Bosch, Frank C and White, Simon},
  year={2010},
  publisher={Cambridge University Press},
  address={Cambridge},
  isbn={9780521857932}
}

@article{Kerscher_2000,
   title={A Comparison of Estimators for the Two-Point Correlation Function},
   volume={535},
   ISSN={0004-637X},
   url={http://dx.doi.org/10.1086/312702},
   DOI={10.1086/312702},
   number={1},
   journal={The Astrophysical Journal},
   publisher={American Astronomical Society},
   author={Kerscher, Martin and Szapudi, István and Szalay, Alexander S.},
   year={2000},
   month=may, pages={L13–L16} }

@article{Healpy1,
  doi = {10.21105/joss.01298},
  url = {https://doi.org/10.21105/joss.01298},
  year = {2019},
  month = mar,
  publisher = {The Open Journal},
  volume = {4},
  number = {35},
  pages = {1298},
  author = {Andrea Zonca and Leo Singer and Daniel Lenz and Martin Reinecke and Cyrille Rosset and Eric Hivon and Krzysztof Gorski},
  title = {healpy: equal area pixelization and spherical harmonics transforms for data on the sphere in Python},
  journal = {Journal of Open Source Software}
}

@article{Healpy2,
   author = {{G{\'o}rski}, K.~M. and {Hivon}, E. and {Banday}, A.~J. and 
	{Wandelt}, B.~D. and {Hansen}, F.~K. and {Reinecke}, M. and 
	{Bartelmann}, M.},
    title = "{HEALPix: A Framework for High-Resolution Discretization and Fast Analysis of Data Distributed on the Sphere}",
  journal = {\apj},
   eprint = {arXiv:astro-ph/0409513},
     year = 2005,
    month = apr,
   volume = 622,
    pages = {759-771},
      doi = {10.1086/427976},
   adsurl = {http://adsabs.harvard.edu/abs/2005ApJ...622..759G}
}

@article{MICECrocce_2015,
   title={The MICE Grand Challenge lightcone simulation – II. Halo and galaxy catalogues},
   volume={453},
   ISSN={1365-2966},
   url={http://dx.doi.org/10.1093/mnras/stv1708},
   DOI={10.1093/mnras/stv1708},
   number={2},
   journal={Monthly Notices of the Royal Astronomical Society},
   publisher={Oxford University Press (OUP)},
   author={Crocce, M. and Castander, F. J. and Gaztañaga, E. and Fosalba, P. and Carretero, J.},
   year={2015},
   month=aug, pages={1513–1530} }

@article{MICECarretero_2014,
   title={An algorithm to build mock galaxy catalogues using MICE simulations},
   volume={447},
   ISSN={0035-8711},
   url={http://dx.doi.org/10.1093/mnras/stu2402},
   DOI={10.1093/mnras/stu2402},
   number={1},
   journal={Monthly Notices of the Royal Astronomical Society},
   publisher={Oxford University Press (OUP)},
   author={Carretero, J. and Castander, F. J. and Gaztañaga, E. and Crocce, M. and Fosalba, P.},
   year={2014},
   month=dec, pages={646–670} }

@ARTICLE{Schutz1986,
       author = {{Schutz}, B.~F.},
        title = "{Determining the Hubble constant from gravitational wave observations}",
      journal = {\nat},
         year = 1986,
        month = sep,
       volume = {323},
       number = {6086},
        pages = {310-311},
          doi = {10.1038/323310a0},
       adsurl = {https://ui.adsabs.harvard.edu/abs/1986Natur.323..310S}
}

@ARTICLE{GW170817a,
       author = {{Abbott}, B.~P. and {Abbott}, R. and {Abbott}, T.~D. and {Acernese}, F. and {Ackley}, K. and {Adams}, C. and {Adams}, T. and {Addesso}, P. and {Adhikari}, R.~X. and {Adya}, V.~B. and {Affeldt}, C. and {Afrough}, M. and {Agarwal}, B. and {Agathos}, M. and {Agatsuma}, K. and {Aggarwal}, N. and {Aguiar}, O.~D. and {Aiello}, L. and {Ain}, A. and {Ajith}, P. and {Allen}, B. and {Allen}, G. and {Allocca}, A. and {Altin}, P.~A. and {Amato}, A. and {Ananyeva}, A. and {Anderson}, S.~B. and {Anderson}, W.~G. and {Angelova}, S.~V. and {Antier}, S. and {Appert}, S. and {Arai}, K. and {Araya}, M.~C. and {Areeda}, J.~S. and {Arnaud}, N. and {Arun}, K.~G. and {Ascenzi}, S. and {Ashton}, G. and {Ast}, M. and {Aston}, S.~M. and {Astone}, P. and {Atallah}, D.~V. and {Aufmuth}, P. and {Aulbert}, C. and {AultONeal}, K. and {Austin}, C. and {Avila-Alvarez}, A. and {Babak}, S. and {Bacon}, P. and {Bader}, M.~K.~M. and {Bae}, S. and {Bailes}, M. and {Baker}, P.~T. and {Baldaccini}, F. and {Ballardin}, G. and {Ballmer}, S.~W. and {Banagiri}, S. and {Barayoga}, J.~C. and {Barclay}, S.~E. and {Barish}, B.~C. and {Barker}, D. and {Barkett}, K. and {Barone}, F. and {Barr}, B. and {Barsotti}, L. and {Barsuglia}, M. and {Barta}, D. and {Barthelmy}, S.~D. and {Bartlett}, J. and {Bartos}, I. and {Bassiri}, R. and {Basti}, A. and {Batch}, J.~C. and {Bawaj}, M. and {Bayley}, J.~C. and {Bazzan}, M. and {B{\'e}csy}, B. and {Beer}, C. and {Bejger}, M. and {Belahcene}, I. and {Bell}, A.~S. and {Berger}, B.~K. and {Bergmann}, G. and {Bernuzzi}, S. and {Bero}, J.~J. and {Berry}, C.~P.~L. and {Bersanetti}, D. and {Bertolini}, A. and {Betzwieser}, J. and {Bhagwat}, S. and {Bhandare}, R. and {Bilenko}, I.~A. and {Billingsley}, G. and {Billman}, C.~R. and {Birch}, J. and {Birney}, R. and {Birnholtz}, O. and {Biscans}, S. and {Biscoveanu}, S. and {Bisht}, A. and {Bitossi}, M. and {Biwer}, C. and {Bizouard}, M.~A. and {Blackburn}, J.~K. and {Blackman}, J. and {Blair}, C.~D. and {Blair}, D.~G. and {Blair}, R.~M. and {Bloemen}, S. and {Bock}, O. and {Bode}, N. and {Boer}, M. and {Bogaert}, G. and {Bohe}, A. and {Bondu}, F. and {Bonilla}, E. and {Bonnand}, R. and {Boom}, B.~A. and {Bork}, R. and {Boschi}, V. and {Bose}, S. and {Bossie}, K. and {Bouffanais}, Y. and {Bozzi}, A. and {Bradaschia}, C. and {Brady}, P.~R. and {Branchesi}, M. and {Brau}, J.~E. and {Briant}, T. and {Brillet}, A. and {Brinkmann}, M. and {Brisson}, V. and {Brockill}, P. and {Broida}, J.~E. and {Brooks}, A.~F. and {Brown}, D.~A. and {Brown}, D.~D. and {Brunett}, S. and {Buchanan}, C.~C. and {Buikema}, A. and {Bulik}, T. and {Bulten}, H.~J. and {Buonanno}, A. and {Buskulic}, D. and {Buy}, C. and {Byer}, R.~L. and {Cabero}, M. and {Cadonati}, L. and {Cagnoli}, G. and {Cahillane}, C. and {Calder{\'o}n Bustillo}, J. and {Callister}, T.~A. and {Calloni}, E. and {Camp}, J.~B. and {Canepa}, M. and {Canizares}, P. and {Cannon}, K.~C. and {Cao}, H. and {Cao}, J. and {Capano}, C.~D. and {Capocasa}, E. and {Carbognani}, F. and {Caride}, S. and {Carney}, M.~F. and {Carullo}, G. and {Casanueva Diaz}, J. and {Casentini}, C. and {Caudill}, S. and {Cavagli{\`a}}, M. and {Cavalier}, F. and {Cavalieri}, R. and {Cella}, G. and {Cepeda}, C.~B. and {Cerd{\'a}-Dur{\'a}n}, P. and {Cerretani}, G. and {Cesarini}, E. and {Chamberlin}, S.~J. and {Chan}, M. and {Chao}, S. and {Charlton}, P. and {Chase}, E. and {Chassande-Mottin}, E. and {Chatterjee}, D. and {Chatziioannou}, K. and {Cheeseboro}, B.~D. and {Chen}, H.~Y. and {Chen}, X. and {Chen}, Y. and {Cheng}, H.-P. and {Chia}, H. and {Chincarini}, A. and {Chiummo}, A. and {Chmiel}, T. and {Cho}, H.~S. and {Cho}, M. and {Chow}, J.~H. and {Christensen}, N. and {Chu}, Q. and {Chua}, A.~J.~K. and {Chua}, S.},
        title = "{GW170817: Observation of Gravitational Waves from a Binary Neutron Star Inspiral}",
      journal = {\prl},
         year = 2017,
        month = oct,
       volume = {119},
       number = {16},
          eid = {161101},
        pages = {161101},
          doi = {10.1103/PhysRevLett.119.161101},
archivePrefix = {arXiv},
       eprint = {1710.05832},
 primaryClass = {gr-qc},
       adsurl = {https://ui.adsabs.harvard.edu/abs/2017PhRvL.119p1101A}
}

@ARTICLE{GW170817b,
   title={Multi-messenger Observations of a Binary Neutron Star Merger*},
   volume={848},
   ISSN={2041-8213},
   url={http://dx.doi.org/10.3847/2041-8213/aa91c9},
   DOI={10.3847/2041-8213/aa91c9},
   number={2},
   journal={The Astrophysical Journal Letters},
   publisher={American Astronomical Society},
   author={Abbott, B. P. and Abbott, R. and Abbott, T. D. and Acernese, F. and Ackley, K. and Adams, C. and Adams, T. and Addesso, P. and Adhikari, R. X. and Adya, V. B. and Affeldt, C. and Afrough, M. and Agarwal, B. and Agathos, M. and Agatsuma, K. and Aggarwal, N. and Aguiar, O. D. and Aiello, L. and Ain, A. and Ajith, P. and Allen, B. and Allen, G. and Allocca, A. and Altin, P. A. and Amato, A. and Ananyeva, A. and Anderson, S. B. and Anderson, W. G. and Angelova, S. V. and Antier, S. and Appert, S. and Arai, K. and Araya, M. C. and Areeda, J. S. and Arnaud, N. and Arun, K. G. and Ascenzi, S. and Ashton, G. and Ast, M. and Aston, S. M. and Astone, P. and Atallah, D. V. and Aufmuth, P. and Aulbert, C. and AultONeal, K. and Austin, C. and Avila-Alvarez, A. and Babak, S. and Bacon, P. and Bader, M. K. M. and Bae, S. and Baker, P. T. and Baldaccini, F. and Ballardin, G. and Ballmer, S. W. and Banagiri, S. and Barayoga, J. C. and Barclay, S. E. and Barish, B. C. and Barker, D. and Barkett, K. and Barone, F. and Barr, B. and Barsotti, L. and Barsuglia, M. and Barta, D. and Barthelmy, S. D. and Bartlett, J. and Bartos, I. and Bassiri, R. and Basti, A. and Batch, J. C. and Bawaj, M. and Bayley, J. C. and Bazzan, M. and Bécsy, B. and Beer, C. and Bejger, M. and Belahcene, I. and Bell, A. S. and Berger, B. K. and Bergmann, G. and Bero, J. J. and Berry, C. P. L. and Bersanetti, D. and Bertolini, A. and Betzwieser, J. and Bhagwat, S. and Bhandare, R. and Bilenko, I. A. and Billingsley, G. and Billman, C. R. and Birch, J. and Birney, I. A. and Birnholtz, O. and Biscans, S. and Biscoveanu, S. and Bisht, A. and Bitossi, M. and Biwer, C. and Bizouard, M. A. and Blackburn, J. K. and Blackman, J. and Blair, C. D. and Blair, D. G. and Blair, R. M. and Bloemen, S. and Bock, O. and Bode, N. and Boer, M. and Bogaert, G. and Bohe, A. and Bondu, F. and Bonilla, E. and Bonnand, R. and Boom, B. A. and Bork, R. and Boschi, V. and Bose, S. and Bossie, K. and Bouffanais, Y. and Bozzi, A. and Bradaschia, C. and Brady, P. R. and Branchesi, M. and Brau, J. E. and Briant, T. and Brillet, A. and Brinkmann, M. and Brisson, V. and Brockill, P. and Broida, J. E. and Brooks, A. F. and Brown, D. A. and Brown, D. D. and Brunett, S. and Buchanan, C. C. and Buikema, A. and Bulik, T. and Bulten, H. J. and Buonanno, A. and Buskulic, D. and Buy, C. and Byer, R. L. and Cabero, M. and Cadonati, L. and Cagnoli, G. and Cahillane, C. and Bustillo, J. Calderón and Callister, T. A. and Calloni, E. and Camp, J. B. and Canepa, M. and Canizares, P. and Cannon, K. C. and Cao, H. and Cao, J. and Capano, C. D. and Capocasa, E. and Carbognani, F. and Caride, S. and Carney, M. F. and Diaz, J. Casanueva and Casentini, C. and Caudill, S. and Cavaglià, M. and Cavalier, F. and Cavalieri, R. and Cella, G. and Cepeda, C. B. and Cerdá-Durán, P. and Cerretani, G. and Cesarini, E. and Chamberlin, S. J. and Chan, M. and Chao, S. and Charlton, P. and Chase, E. and Chassande-Mottin, E. and Chatterjee, D. and Chatziioannou, K. and Cheeseboro, B. D. and Chen, H. Y. and Chen, X. and Chen, Y. and Cheng, H.-P. and Chia, H. and Chincarini, A. and Chiummo, A. and Chmiel, T. and Cho, H. S. and Cho, M. and Chow, J. H. and Christensen, N. and Chu, Q. and Chua, A. J. K. and Chua, S. and Chung, A. K. W. and Chung, S. and Ciani, G. and Ciolfi, R. and Cirelli, C. E. and Cirone, A. and Clara, F. and Clark, J. A. and Clearwater, P. and Cleva, F. and Cocchieri, C. and Coccia, E. and Cohadon, P.-F. and Cohen, D. and Colla, A. and Collette, C. G. and Cominsky, L. R. and Jr., M. Constancio and Conti, L. and Cooper, S. J. and Corban, P. and Corbitt, T. R. and Cordero-Carrión, I. and Corley, K. R. and Cornish, N. and Corsi, A. and Cortese, S. and Costa, C. A. and Coughlin, M. W. and Coughlin, S. B. and Coulon, J.-P. and Countryman, S. T. and Couvares, P. and Covas, P. B. and Cowan, E. E. and Coward, D. M. and Cowart, M. J. and Coyne, D. C. and Coyne, R. and Creighton, J. D. E. and Creighton, T. D. and Cripe, J. and Crowder, S. G. and Cullen, T. J. and Cumming, A. and Cunningham, L. and Cuoco, E. and Canton, T. Dal and Dálya, G. and Danilishin, S. L. and D’Antonio, S. and Danzmann, K. and Dasgupta, A. and Da Silva Costa, C. F. and Dattilo, V. and Dave, I. and Davier, M. and Davis, D. and Daw, E. J. and Day, B. and De, S. and DeBra, D. and Degallaix, J. and Laurentis, M. De and Deléglise, S. and Pozzo, W. Del and Demos, N. and Denker, T. and Dent, T. and Pietri, R. De and Dergachev, V. and Rosa, R. De and DeRosa, R. T. and Rossi, C. De and DeSalvo, R. and Varona, O. de and Devenson, J. and Dhurandhar, S. and Díaz, M. C. and Fiore, L. Di and Giovanni, M. Di and Girolamo, T. Di and Lieto, A. Di and Pace, S. Di and Palma, I. Di and Renzo, F. Di and Doctor, Z. and Dolique, V. and Donovan, F. and Dooley, K. L. and Doravari, S. and Dorrington, I. and Douglas, R. and Álvarez, M. Dovale and Downes, T. P. and Drago, M. and Dreissigacker, C. and Driggers, J. C. and Du, Z. and Ducrot, M. and Dupej, P. and Dwyer, S. E. and Edo, T. B. and Edwards, M. C. and Effler, A. and Ehrens, P. and Eichholz, J. and Eikenberry, S. S. and Eisenstein, R. A. and Essick, R. C. and Estevez, D. and Etienne, Z. B. and Etzel, T. and Evans, M. and Evans, T. M. and Factourovich, M. and Fafone, V. and Fair, H. and Fairhurst, S. and Fan, X. and Farinon, S. and Farr, B. and Farr, W. M. and Fauchon-Jones, E. J. and Favata, M. and Fays, M. and Fee, C. and Fehrmann, H. and Feicht, J. and Fejer, M. M. and Fernandez-Galiana, A. and Ferrante, I. and Ferreira, E. C. and Ferrini, F. and Fidecaro, F. and Finstad, D. and Fiori, I. and Fiorucci, D. and Fishbach, M. and Fisher, R. P. and Fitz-Axen, M. and Flaminio, R. and Fletcher, M. and Fong, H. and Font, J. A. and Forsyth, P. W. F. and Forsyth, S. S. and Fournier, J.-D. and Frasca, S. and Frasconi, F. and Frei, Z. and Freise, A. and Frey, R. and Frey, V. and Fries, E. M. and Fritschel, P. and Frolov, V. V. and Fulda, P. and Fyffe, M. and Gabbard, H. and Gadre, B. U. and Gaebel, S. M. and Gair, J. R. and Gammaitoni, L. and Ganija, M. R. and Gaonkar, S. G. and Garcia-Quiros, C. and Garufi, F. and Gateley, B. and Gaudio, S. and Gaur, G. and Gayathri, V. and Gehrels, N. and Gemme, G. and Genin, E. and Gennai, A. and George, D. and George, J. and Gergely, L. and Germain, V. and Ghonge, S. and Ghosh, Abhirup and Ghosh, Archisman and Ghosh, S. and Giaime, J. A. and Giardina, K. D. and Giazotto, A. and Gill, K. and Glover, L. and Goetz, E. and Goetz, R. and Gomes, S. and Goncharov, B. and González, G. and Castro, J. M. Gonzalez and Gopakumar, A. and Gorodetsky, M. L. and Gossan, S. E. and Gosselin, M. and Gouaty, R. and Grado, A. and Graef, C. and Granata, M. and Grant, A. and Gras, S. and Gray, C. and Greco, G. and Green, A. C. and Gretarsson, E. M. and Griswold, B. and Groot, P. and Grote, H. and Grunewald, S. and Gruning, P. and Guidi, G. M. and Guo, X. and Gupta, A. and Gupta, M. K. and Gushwa, K. E. and Gustafson, E. K. and Gustafson, R. and Halim, O. and Hall, B. R. and Hall, E. D. and Hamilton, E. Z. and Hammond, G. and Haney, M. and Hanke, M. M. and Hanks, J. and Hanna, C. and Hannam, M. D. and Hannuksela, O. A. and Hanson, J. and Hardwick, T. and Harms, J. and Harry, G. M. and Harry, I. W. and Hart, M. J. and Haster, C.-J. and Haughian, K. and Healy, J. and Heidmann, A. and Heintze, M. C. and Heitmann, H. and Hello, P. and Hemming, G. and Hendry, M. and Heng, I. S. and Hennig, J. and Heptonstall, A. W. and Heurs, M. and Hild, S. and Hinderer, T. and Hoak, D. and Hofman, D. and Holt, K. and Holz, D. E. and Hopkins, P. and Horst, C. and Hough, J. and Houston, E. A. and Howell, E. J. and Hreibi, A. and Hu, Y. M. and Huerta, E. A. and Huet, D. and Hughey, B. and Husa, S. and Huttner, S. H. and Huynh-Dinh, T. and Indik, N. and Inta, R. and Intini, G. and Isa, H. N. and Isac, J.-M. and Isi, M. and Iyer, B. R. and Izumi, K. and Jacqmin, T. and Jani, K. and Jaranowski, P. and Jawahar, S. and Jiménez-Forteza, F. and Johnson, W. W. and Jones, D. I. and Jones, R. and Jonker, R. J. G. and Ju, L. and Junker, J. and Kalaghatgi, C. V. and Kalogera, V. and Kamai, B. and Kandhasamy, S. and Kang, G. and Kanner, J. B. and Kapadia, S. J. and Karki, S. and Karvinen, K. S. and Kasprzack, M. and Katolik, M. and Katsavounidis, E. and Katzman, W. and Kaufer, S. and Kawabe, K. and Kéfélian, F. and Keitel, D. and Kemball, A. J. and Kennedy, R. and Kent, C. and Key, J. S. and Khalili, F. Y. and Khan, I. and Khan, S. and Khan, Z. and Khazanov, E. A. and Kijbunchoo, N. and Kim, Chunglee and Kim, J. C. and Kim, K. and Kim, W. and Kim, W. S. and Kim, Y.-M. and Kimbrell, S. J. and King, E. J. and King, P. J. and Kinley-Hanlon, M. and Kirchhoff, R. and Kissel, J. S. and Kleybolte, L. and Klimenko, S. and Knowles, T. D. and Koch, P. and Koehlenbeck, S. M. and Koley, S. and Kondrashov, V. and Kontos, A. and Korobko, M. and Korth, W. Z. and Kowalska, I. and Kozak, D. B. and Krämer, C. and Kringel, V. and Krishnan, B. and Królak, A. and Kuehn, G. and Kumar, P. and Kumar, R. and Kumar, S. and Kuo, L. and Kutynia, A. and Kwang, S. and Lackey, B. D. and Lai, K. H. and Landry, M. and Lang, R. N. and Lange, J. and Lantz, B. and Lanza, R. K. and Larson, S. L. and Lartaux-Vollard, A. and Lasky, P. D. and Laxen, M. and Lazzarini, A. and Lazzaro, C. and Leaci, P. and Leavey, S. and Lee, C. H. and Lee, H. K. and Lee, H. M. and Lee, H. W. and Lee, K. and Lehmann, J. and Lenon, A. and Leonardi, M. and Leroy, N. and Letendre, N. and Levin, Y. and Li, T. G. F. and Linker, S. D. and Littenberg, T. B. and Liu, J. and Lo, R. K. L. and Lockerbie, N. A. and London, L. T. and Lord, J. E. and Lorenzini, M. and Loriette, V. and Lormand, M. and Losurdo, G. and Lough, J. D. and Lousto, C. O. and Lovelace, G. and Lück, H. and Lumaca, D. and Lundgren, A. P. and Lynch, R. and Ma, Y. and Macas, R. and Macfoy, S. and Machenschalk, B. and MacInnis, M. and Macleod, D. M. and Hernandez, I. Magaña and Magaña-Sandoval, F. and Zertuche, L. Magaña and Magee, R. M. and Majorana, E. and Maksimovic, I. and Man, N. and Mandic, V. and Mangano, V. and Mansell, G. L. and Manske, M. and Mantovani, M. and Marchesoni, F. and Marion, F. and Márka, S. and Márka, Z. and Markakis, C. and Markosyan, A. S. and Markowitz, A. and Maros, E. and Marquina, A. and Marsh, P. and Martelli, F. and Martellini, L. and Martin, I. W. and Martin, R. M. and Martynov, D. V. and Mason, K. and Massera, E. and Masserot, A. and Massinger, T. J. and Masso-Reid, M. and Mastrogiovanni, S. and Matas, A. and Matichard, F. and Matone, L. and Mavalvala, N. and Mazumder, N. and McCarthy, R. and McClelland, D. E. and McCormick, S. and McCuller, L. and McGuire, S. C. and McIntyre, G. and McIver, J. and McManus, D. J. and McNeill, L. and McRae, T. and McWilliams, S. T. and Meacher, D. and Meadors, G. D. and Mehmet, M. and Meidam, J. and Mejuto-Villa, E. and Melatos, A. and Mendell, G. and Mercer, R. A. and Merilh, E. L. and Merzougui, M. and Meshkov, S. and Messenger, C. and Messick, C. and Metzdorff, R. and Meyers, P. M. and Miao, H. and Michel, C. and Middleton, H. and Mikhailov, E. E. and Milano, L. and Miller, A. L. and Miller, B. B. and Miller, J. and Millhouse, M. and Milovich-Goff, M. C. and Minazzoli, O. and Minenkov, Y. and Ming, J. and Mishra, C. and Mitra, S. and Mitrofanov, V. P. and Mitselmakher, G. and Mittleman, R. and Moffa, D. and Moggi, A. and Mogushi, K. and Mohan, M. and Mohapatra, S. R. P. and Montani, M. and Moore, C. J. and Moraru, D. and Moreno, G. and Morriss, S. R. and Mours, B. and Mow-Lowry, C. M. and Mueller, G. and Muir, A. W. and Mukherjee, Arunava and Mukherjee, D. and Mukherjee, S. and Mukund, N. and Mullavey, A. and Munch, J. and Muñiz, E. A. and Muratore, M. and Murray, P. G. and Napier, K. and Nardecchia, I. and Naticchioni, L. and Nayak, R. K. and Neilson, J. and Nelemans, G. and Nelson, T. J. N. and Nery, M. and Neunzert, A. and Nevin, L. and Newport, J. M. and Newton, G. and Ng, K. K. Y. and Nguyen, P. and Nguyen, T. T. and Nichols, D. and Nielsen, A. B. and Nissanke, S. and Nitz, A. and Noack, A. and Nocera, F. and Nolting, D. and North, C. and Nuttall, L. K. and Oberling, J. and O’Dea, G. D. and Ogin, G. H. and Oh, J. J. and Oh, S. H. and Ohme, F. and Okada, M. A. and Oliver, M. and Oppermann, P. and Oram, Richard J. and O’Reilly, B. and Ormiston, R. and Ortega, L. F. and O’Shaughnessy, R. and Ossokine, S. and Ottaway, D. J. and Overmier, H. and Owen, B. J. and Pace, A. E. and Page, J. and Page, M. A. and Pai, A. and Pai, S. A. and Palamos, J. R. and Palashov, O. and Palomba, C. and Pal-Singh, A. and Pan, Howard and Pan, Huang-Wei and Pang, B. and Pang, P. T. H. and Pankow, C. and Pannarale, F. and Pant, B. C. and Paoletti, F. and Paoli, A. and Papa, M. A. and Parida, A. and Parker, W. and Pascucci, D. and Pasqualetti, A. and Passaquieti, R. and Passuello, D. and Patil, M. and Patricelli, B. and Pearlstone, B. L. and Pedraza, M. and Pedurand, R. and Pekowsky, L. and Pele, A. and Penn, S. and Perez, C. J. and Perreca, A. and Perri, L. M. and Pfeiffer, H. P. and Phelps, M. and Piccinni, O. J. and Pichot, M. and Piergiovanni, F. and Pierro, V. and Pillant, G. and Pinard, L. and Pinto, I. M. and Pirello, M. and Pitkin, M. and Poe, M. and Poggiani, R. and Popolizio, P. and Porter, E. K. and Post, A. and Powell, J. and Prasad, J. and Pratt, J. W. W. and Pratten, G. and Predoi, V. and Prestegard, T. and Price, L. R. and Prijatelj, M. and Principe, M. and Privitera, S. and Prodi, G. A. and Prokhorov, L. G. and Puncken, O. and Punturo, M. and Puppo, P. and Pürrer, M. and Qi, H. and Quetschke, V. and Quintero, E. A. and Quitzow-James, R. and Raab, F. J. and Rabeling, D. S. and Radkins, H. and Raffai, P. and Raja, S. and Rajan, C. and Rajbhandari, B. and Rakhmanov, M. and Ramirez, K. E. and Ramos-Buades, A. and Rapagnani, P. and Raymond, V. and Razzano, M. and Read, J. and Regimbau, T. and Rei, L. and Reid, S. and Reitze, D. H. and Ren, W. and Reyes, S. D. and Ricci, F. and Ricker, P. M. and Rieger, S. and Riles, K. and Rizzo, M. and Robertson, N. A. and Robie, R. and Robinet, F. and Rocchi, A. and Rolland, L. and Rollins, J. G. and Roma, V. J. and Romano, R. and Romel, C. L. and Romie, J. H. and Rosińska, D. and Ross, M. P. and Rowan, S. and Rüdiger, A. and Ruggi, P. and Rutins, G. and Ryan, K. and Sachdev, S. and Sadecki, T. and Sadeghian, L. and Sakellariadou, M. and Salconi, L. and Saleem, M. and Salemi, F. and Samajdar, A. and Sammut, L. and Sampson, L. M. and Sanchez, E. J. and Sanchez, L. E. and Sanchis-Gual, N. and Sandberg, V. and Sanders, J. R. and Sassolas, B. and Sathyaprakash, B. S. and Saulson, P. R. and Sauter, O. and Savage, R. L. and Sawadsky, A. and Schale, P. and Scheel, M. and Scheuer, J. and Schmidt, J. and Schmidt, P. and Schnabel, R. and Schofield, R. M. S. and Schönbeck, A. and Schreiber, E. and Schuette, D. and Schulte, B. W. and Schutz, B. F. and Schwalbe, S. G. and Scott, J. and Scott, S. M. and Seidel, E. and Sellers, D. and Sengupta, A. S. and Sentenac, D. and Sequino, V. and Sergeev, A. and Shaddock, D. A. and Shaffer, T. J. and Shah, A. A. and Shahriar, M. S. and Shaner, M. B. and Shao, L. and Shapiro, B. and Shawhan, P. and Sheperd, A. and Shoemaker, D. H. and Shoemaker, D. M. and Siellez, K. and Siemens, X. and Sieniawska, M. and Sigg, D. and Silva, A. D. and Singer, L. P. and Singh, A. and Singhal, A. and Sintes, A. M. and Slagmolen, B. J. J. and Smith, B. and Smith, J. R. and Smith, R. J. E. and Somala, S. and Son, E. J. and Sonnenberg, J. A. and Sorazu, B. and Sorrentino, F. and Souradeep, T. and Spencer, A. P. and Srivastava, A. K. and Staats, K. and Staley, A. and Steinke, M. and Steinlechner, J. and Steinlechner, S. and Steinmeyer, D. and Stevenson, S. P. and Stone, R. and Stops, D. J. and Strain, K. A. and Stratta, G. and Strigin, S. E. and Strunk, A. and Sturani, R. and Stuver, A. L. and Summerscales, T. Z. and Sun, L. and Sunil, S. and Suresh, J. and Sutton, P. J. and Swinkels, B. L. and Szczepańczyk, M. J. and Tacca, M. and Tait, S. C. and Talbot, C. and Talukder, D. and Tanner, D. B. and Tápai, M. and Taracchini, A. and Tasson, J. D. and Taylor, J. A. and Taylor, R. and Tewari, S. V. and Theeg, T. and Thies, F. and Thomas, E. G. and Thomas, M. and Thomas, P. and Thorne, K. A. and Thorne, K. S. and Thrane, E. and Tiwari, S. and Tiwari, V. and Tokmakov, K. V. and Toland, K. and Tonelli, M. and Tornasi, Z. and Torres-Forné, A. and Torrie, C. I. and Töyrä, D. and Travasso, F. and Traylor, G. and Trinastic, J. and Tringali, M. C. and Trozzo, L. and Tsang, K. W. and Tse, M. and Tso, R. and Tsukada, L. and Tsuna, D. and Tuyenbayev, D. and Ueno, K. and Ugolini, D. and Unnikrishnan, C. S. and Urban, A. L. and Usman, S. A. and Vahlbruch, H. and Vajente, G. and Valdes, G. and Bakel, N. van and Beuzekom, M. van and van den Brand, J. F. J. and Van Den Broeck, C. and Vander-Hyde, D. C. and van der Schaaf, L. and Heijningen, J. V. van and Veggel, A. A. van and Vardaro, M. and Varma, V. and Vass, S. and Vasúth, M. and Vecchio, A. and Vedovato, G. and Veitch, J. and Veitch, P. J. and Venkateswara, K. and Venugopalan, G. and Verkindt, D. and Vetrano, F. and Viceré, A. and Viets, A. D. and Vinciguerra, S. and Vine, D. J. and Vinet, J.-Y. and Vitale, S. and Vo, T. and Vocca, H. and Vorvick, C. and Vyatchanin, S. P. and Wade, A. R. and Wade, L. E. and Wade, M. and Walet, R. and Walker, M. and Wallace, L. and Walsh, S. and Wang, G. and Wang, H. and Wang, J. Z. and Wang, W. H. and Wang, Y. F. and Ward, R. L. and Warner, J. and Was, M. and Watchi, J. and Weaver, B. and Wei, L.-W. and Weinert, M. and Weinstein, A. J. and Weiss, R. and Wen, L. and Wessel, E. K. and Wessels, P. and Westerweck, J. and Westphal, T. and Wette, K. and Whelan, J. T. and Whitcomb, S. E. and Whiting, B. F. and Whittle, C. and Wilken, D. and Williams, D. and Williams, R. D. and Williamson, A. R. and Willis, J. L. and Willke, B. and Wimmer, M. H. and Winkler, W. and Wipf, C. C. and Wittel, H. and Woan, G. and Woehler, J. and Wofford, J. and Wong, K. W. K. and Worden, J. and Wright, J. L. and Wu, D. S. and Wysocki, D. M. and Xiao, S. and Yamamoto, H. and Yancey, C. C. and Yang, L. and Yap, M. J. and Yazback, M. and Yu, Hang and Yu, Haocun and Yvert, M. and Zadrożny, A. and Zanolin, M. and Zelenova, T. and Zendri, J.-P. and Zevin, M. and Zhang, L. and Zhang, M. and Zhang, T. and Zhang, Y.-H. and Zhao, C. and Zhou, M. and Zhou, Z. and Zhu, S. J. and Zhu, X. J. and Zimmerman, A. B. and Zucker, M. E. and Zweizig, J. and Wilson-Hodge, C. A. and Bissaldi, E. and Blackburn, L. and Briggs, M. S. and Burns, E. and Cleveland, W. H. and Connaughton, V. and Gibby, M. H. and Giles, M. M and Goldstein, A. and Hamburg, R. and Jenke, P. and Hui, C. M. and Kippen, R. M. and Kocevski, D. and McBreen, S. and Meegan, C. A. and Paciesas, W. S. and Poolakkil, S. and Preece, R. D. and Racusin, J. and Roberts, O. J. and Stanbro, M. and Veres, P. and von Kienlin, A. and Savchenko, V. and Ferrigno, C. and Kuulkers, E. and Bazzano, A. and Bozzo, E. and Brandt, S. and Chenevez, J. and Courvoisier, T. J.-L. and Diehl, R. and Domingo, A. and Hanlon, L. and Jourdain, E. and Laurent, P. and Lebrun, F. and Lutovinov, A. and Martin-Carrillo, A. and Mereghetti, S. and Natalucci, L. and Rodi, J. and Roques, J.-P. and Sunyaev, R. and Ubertini, P. and Aartsen, M. G. and Ackermann, M. and Adams, J. and Aguilar, J. A. and Ahlers, M. and Ahrens, M. and Samarai, I. Al and Altmann, D. and Andeen, K. and Anderson, T. and Ansseau, I. and Anton, G. and Argüelles, C. and Auffenberg, J. and Axani, S. and Bagherpour, H. and Bai, X. and Barron, J. P. and Barwick, S. W. and Baum, V. and Bay, R. and Beatty, J. J. and Tjus, J. Becker and Bernardini, E. and Besson, D. Z. and Binder, G. and Bindig, D. and Blaufuss, E. and Blot, S. and Bohm, C. and Börner, M. and Bos, F. and Bose, D. and Böser, S. and Botner, O. and Bourbeau, E. and Bourbeau, J. and Bradascio, F. and Braun, J. and Brayeur, L. and Brenzke, M. and Bretz, H.-P. and Bron, S. and Brostean-Kaiser, J. and Burgman, A. and Carver, T. and Casey, J. and Casier, M. and Cheung, E. and Chirkin, D. and Christov, A. and Clark, K. and Classen, L. and Coenders, S. and Collin, G. H. and Conrad, J. M. and Cowen, D. F. and Cross, R. and Day, M. and André, J. P. A. M. de and Clercq, C. De and DeLaunay, J. J. and Dembinski, H. and Ridder, S. De and Desiati, P. and Vries, K. D. de and Wasseige, G. de and With, M. de and DeYoung, T. and Díaz-Vélez, J. C. and Lorenzo, V. di and Dujmovic, H. and Dumm, J. P. and Dunkman, M. and Dvorak, E. and Eberhardt, B. and Ehrhardt, T. and Eichmann, B. and Eller, P. and Evenson, P. A. and Fahey, S. and Fazely, A. R. and Felde, J. and Filimonov, K. and Finley, C. and Flis, S. and Franckowiak, A. and Friedman, E. and Fuchs, T. and Gaisser, T. K. and Gallagher, J. and Gerhardt, L. and Ghorbani, K. and Giang, W. and Glauch, T. and Glüsenkamp, T. and Goldschmidt, A. and Gonzalez, J. G. and Grant, D. and Griffith, Z. and Haack, C. and Hallgren, A. and Halzen, F. and Hanson, K. and Hebecker, D. and Heereman, D. and Helbing, K. and Hellauer, R. and Hickford, S. and Hignight, J. and Hill, G. C. and Hoffman, K. D. and Hoffmann, R. and Hokanson-Fasig, B. and Hoshina, K. and Huang, F. and Huber, M. and Hultqvist, K. and Hünnefeld, M. and In, S. and Ishihara, A. and Jacobi, E. and Japaridze, G. S. and Jeong, M. and Jero, K. and Jones, B. J. P. and Kalaczynski, P. and Kang, W. and Kappes, A. and Karg, T. and Karle, A. and Kauer, M. and Keivani, A. and Kelley, J. L. and Kheirandish, A. and Kim, J. and Kim, M. and Kintscher, T. and Kiryluk, J. and Kittler, T. and Klein, S. R. and Kohnen, G. and Koirala, R. and Kolanoski, H. and Köpke, L. and Kopper, C. and Kopper, S. and Koschinsky, J. P. and Koskinen, D. J. and Kowalski, M. and Krings, K. and Kroll, M. and Krückl, G. and Kunnen, J. and Kunwar, S. and Kurahashi, N. and Kuwabara, T. and Kyriacou, A. and Labare, M. and Lanfranchi, J. L. and Larson, M. J. and Lauber, F. and Lesiak-Bzdak, M. and Leuermann, M. and Liu, Q. R. and Lu, L. and Lünemann, J. and Luszczak, W. and Madsen, J. and Maggi, G. and Mahn, K. B. M. and Mancina, S. and Maruyama, R. and Mase, K. and Maunu, R. and McNally, F. and Meagher, K. and Medici, M. and Meier, M. and Menne, T. and Merino, G. and Meures, T. and Miarecki, S. and Micallef, J. and Momenté, G. and Montaruli, T. and Moore, R. W. and Moulai, M. and Nahnhauer, R. and Nakarmi, P. and Naumann, U. and Neer, G. and Niederhausen, H. and Nowicki, S. C. and Nygren, D. R. and Pollmann, A. Obertacke and Olivas, A. and O’Murchadha, A. and Palczewski, T. and Pandya, H. and Pankova, D. V. and Peiffer, P. and Pepper, J. A. and Pérez de los Heros, C. and Pieloth, D. and Pinat, E. and Price, P. B. and Przybylski, G. T. and Raab, C. and Rädel, L. and Rameez, M. and Rawlins, K. and Rea, I. C. and Reimann, R. and Relethford, B. and Relich, M. and Resconi, E. and Rhode, W. and Richman, M. and Robertson, S. and Rongen, M. and Rott, C. and Ruhe, T. and Ryckbosch, D. and Rysewyk, D. and Sälzer, T. and Herrera, S. E. Sanchez and Sandrock, A. and Sandroos, J. and Santander, M. and Sarkar, S. and Sarkar, S. and Satalecka, K. and Schlunder, P. and Schmidt, T. and Schneider, A. and Schoenen, S. and Schöneberg, S. and Schumacher, L. and Seckel, D. and Seunarine, S. and Soedingrekso, J. and Soldin, D. and Song, M. and Spiczak, G. M. and Spiering, C. and Stachurska, J. and Stamatikos, M. and Stanev, T. and Stasik, A. and Stettner, J. and Steuer, A. and Stezelberger, T. and Stokstad, R. G. and Stössl, A. and Strotjohann, N. L. and Stuttard, T. and Sullivan, G. W. and Sutherland, M. and Taboada, I. and Tatar, J. and Tenholt, F. and Ter-Antonyan, S. and Terliuk, A. and Tešić, G. and Tilav, S. and Toale, P. A. and Tobin, M. N. and Toscano, S. and Tosi, D. and Tselengidou, M. and Tung, C. F. and Turcati, A. and Turley, C. F. and Ty, B. and Unger, E. and Usner, M. and Vandenbroucke, J. and Driessche, W. Van and Eijndhoven, N. van and Vanheule, S. and Santen, J. van and Vehring, M. and Vogel, E. and Vraeghe, M. and Walck, C. and Wallace, A. and Wallraff, M. and Wandler, F. D. and Wandkowsky, N. and Waza, A. and Weaver, C. and Weiss, M. J. and Wendt, C. and Werthebach, J. and Whelan, B. J. and Wiebe, K. and Wiebusch, C. H. and Wille, L. and Williams, D. R. and Wills, L. and Wolf, M. and Wood, T. R. and Woolsey, E. and Woschnagg, K. and Xu, D. L. and Xu, X. W. and Xu, Y. and Yanez, J. P. and Yodh, G. and Yoshida, S. and Yuan, T. and Zoll, M. and Balasubramanian, A. and Mate, S. and Bhalerao, V. and Bhattacharya, D. and Vibhute, A. and Dewangan, G. C. and Rao, A. R. and Vadawale, S. V. and Svinkin, D. S. and Hurley, K. and Aptekar, R. L. and Frederiks, D. D. and Golenetskii, S. V. and Kozlova, A. V. and Lysenko, A. L. and Oleynik, Ph. P. and Tsvetkova, A. E. and Ulanov, M. V. and Cline, T. and Li, T. P. and Xiong, S. L. and Zhang, S. N. and Lu, F. J. and Song, L. M. and Cao, X. L. and Chang, Z. and Chen, G. and Chen, L. and Chen, T. X. and Chen, Y. and Chen, Y. B. and Chen, Y. P. and Cui, W. and Cui, W. W. and Deng, J. K. and Dong, Y. W. and Du, Y. Y. and Fu, M. X. and Gao, G. H. and Gao, H. and Gao, M. and Ge, M. Y. and Gu, Y. D. and Guan, J. and Guo, C. C. and Han, D. W. and Hu, W. and Huang, Y. and Huo, J. and Jia, S. M. and Jiang, L. H. and Jiang, W. C. and Jin, J. and Jin, Y. J. and Li, B. and Li, C. K. and Li, G. and Li, M. S. and Li, W. and Li, X. and Li, X. B. and Li, X. F. and Li, Y. G. and Li, Z. J. and Li, Z. W. and Liang, X. H. and Liao, J. Y. and Liu, C. Z. and Liu, G. Q. and Liu, H. W. and Liu, S. Z. and Liu, X. J. and Liu, Y. and Liu, Y. N. and Lu, B. and Lu, X. F. and Luo, T. and Ma, X. and Meng, B. and Nang, Y. and Nie, J. Y. and Ou, G. and Qu, J. L. and Sai, N. and Sun, L. and Tan, Y. and Tao, L. and Tao, W. H. and Tuo, Y. L. and Wang, G. F. and Wang, H. Y. and Wang, J. and Wang, W. S. and Wang, Y. S. and Wen, X. Y. and Wu, B. B. and Wu, M. and Xiao, G. C. and Xu, H. and Xu, Y. P. and Yan, L. L. and Yang, J. W. and Yang, S. and Yang, Y. J. and Zhang, A. M. and Zhang, C. L. and Zhang, C. M. and Zhang, F. and Zhang, H. M. and Zhang, J. and Zhang, Q. and Zhang, S. and Zhang, T. and Zhang, W. and Zhang, W. C. and Zhang, W. Z. and Zhang, Y. and Zhang, Y. and Zhang, Y. F. and Zhang, Y. J. and Zhang, Z. and Zhang, Z. L. and Zhao, H. S. and Zhao, J. L. and Zhao, X. F. and Zheng, S. J. and Zhu, Y. and Zhu, Y. X. and Zou, C. L. and Albert, A. and André, M. and Anghinolfi, M. and Ardid, M. and Aubert, J.-J. and Aublin, J. and Avgitas, T. and Baret, B. and Barrios-Martí, J. and Basa, S. and Belhorma, B. and Bertin, V. and Biagi, S. and Bormuth, R. and Bourret, S. and Bouwhuis, M. C. and Brânzaş, H. and Bruijn, R. and Brunner, J. and Busto, J. and Capone, A. and Caramete, L. and Carr, J. and Celli, S. and Cherkaoui El Moursli, R. and Chiarusi, T. and Circella, M. and Coelho, J. A. B. and Coleiro, A. and Coniglione, R. and Costantini, H. and Coyle, P. and Creusot, A. and Díaz, A. F. and Deschamps, A. and Bonis, G. De and Distefano, C. and Palma, I. Di and Domi, A. and Donzaud, C. and Dornic, D. and Drouhin, D. and Eberl, T. and El Bojaddaini, I. and El Khayati, N. and Elsässer, D. and Enzenhöfer, A. and Ettahiri, A. and Fassi, F. and Felis, I. and Fusco, L. A. and Gay, P. and Giordano, V. and Glotin, H. and Grégoire, T. and Ruiz, R. Gracia and Graf, K. and Hallmann, S. and Haren, H. van and Heijboer, A. J. and Hello, Y. and Hernández-Rey, J. J. and Hössl, J. and Hofestädt, J. and Hugon, C. and Illuminati, G. and James, C. W. and Jong, M. de and Jongen, M. and Kadler, M. and Kalekin, O. and Katz, U. and Kiessling, D. and Kouchner, A. and Kreter, M. and Kreykenbohm, I. and Kulikovskiy, V. and Lachaud, C. and Lahmann, R. and Lefèvre, D. and Leonora, E. and Lotze, M. and Loucatos, S. and Marcelin, M. and Margiotta, A. and Marinelli, A. and Martínez-Mora, J. A. and Mele, R. and Melis, K. and Michael, T. and Migliozzi, P. and Moussa, A. and Navas, S. and Nezri, E. and Organokov, M. and Păvălaş, G. E. and Pellegrino, C. and Perrina, C. and Piattelli, P. and Popa, V. and Pradier, T. and Quinn, L. and Racca, C. and Riccobene, G. and Sánchez-Losa, A. and Saldaña, M. and Salvadori, I. and Samtleben, D. F. E. and Sanguineti, M. and Sapienza, P. and Sieger, C. and Spurio, M. and Stolarczyk, Th. and Taiuti, M. and Tayalati, Y. and Trovato, A. and Turpin, D. and Tönnis, C. and Vallage, B. and Elewyck, V. Van and Versari, F. and Vivolo, D. and Vizzoca, A. and Wilms, J. and Zornoza, J. D. and Zúñiga, J. and Beardmore, A. P. and Breeveld, A. A. and Burrows, D. N. and Cenko, S. B. and Cusumano, G. and D’Aì, A. and de Pasquale, M. and Emery, S. W. K. and Evans, P. A. and Giommi, P. and Gronwall, C. and Kennea, J. A. and Krimm, H. A. and Kuin, N. P. M. and Lien, A. and Marshall, F. E. and Melandri, A. and Nousek, J. A. and Oates, S. R. and Osborne, J. P. and Pagani, C. and Page, K. L. and Palmer, D. M. and Perri, M. and Siegel, M. H. and Sbarufatti, B. and Tagliaferri, G. and Tohuvavohu, A. and Tavani, M. and Verrecchia, F. and Bulgarelli, A. and Evangelista, Y. and Pacciani, L. and Feroci, M. and Pittori, C. and Giuliani, A. and Monte, E. Del and Donnarumma, I. and Argan, A. and Trois, A. and Ursi, A. and Cardillo, M. and Piano, G. and Longo, F. and Lucarelli, F. and Munar-Adrover, P. and Fuschino, F. and Labanti, C. and Marisaldi, M. and Minervini, G. and Fioretti, V. and Parmiggiani, N. and Gianotti, F. and Trifoglio, M. and Persio, G. Di and Antonelli, L. A. and Barbiellini, G. and Caraveo, P. and Cattaneo, P. W. and Costa, E. and Colafrancesco, S. and D’Amico, F. and Ferrari, A. and Morselli, A. and Paoletti, F. and Picozza, P. and Pilia, M. and Rappoldi, A. and Soffitta, P. and Vercellone, S. and Foley, R. J. and Coulter, D. A. and Kilpatrick, C. D. and Drout, M. R. and Piro, A. L. and Shappee, B. J. and Siebert, M. R. and Simon, J. D. and Ulloa, N. and Kasen, D. and Madore, B. F. and Murguia-Berthier, A. and Pan, Y.-C. and Prochaska, J. X. and Ramirez-Ruiz, E. and Rest, A. and Rojas-Bravo, C. and Berger, E. and Soares-Santos, M. and Annis, J. and Alexander, K. D. and Allam, S. and Balbinot, E. and Blanchard, P. and Brout, D. and Butler, R. E. and Chornock, R. and Cook, E. R. and Cowperthwaite, P. and Diehl, H. T. and Drlica-Wagner, A. and Drout, M. R. and Durret, F. and Eftekhari, T. and Finley, D. A. and Fong, W. and Frieman, J. A. and Fryer, C. L. and García-Bellido, J. and Gruendl, R. A. and Hartley, W. and Herner, K. and Kessler, R. and Lin, H. and Lopes, P. A. A. and Lourenço, A. C. C. and Margutti, R. and Marshall, J. L. and Matheson, T. and Medina, G. E. and Metzger, B. D. and Muñoz, R. R. and Muir, J. and Nicholl, M. and Nugent, P. and Palmese, A. and Paz-Chinchón, F. and Quataert, E. and Sako, M. and Sauseda, M. and Schlegel, D. J. and Scolnic, D. and Secco, L. F. and Smith, N. and Sobreira, F. and Villar, V. A. and Vivas, A. K. and Wester, W. and Williams, P. K. G. and Yanny, B. and Zenteno, A. and Zhang, Y. and Abbott, T. M. C. and Banerji, M. and Bechtol, K. and Benoit-Lévy, A. and Bertin, E. and Brooks, D. and Buckley-Geer, E. and Burke, D. L. and Capozzi, D. and Rosell, A. Carnero and Kind, M. Carrasco and Castander, F. J. and Crocce, M. and Cunha, C. E. and D’Andrea, C. B. and da Costa, L. N. and Davis, C. and DePoy, D. L. and Desai, S. and Dietrich, J. P. and Eifler, T. F. and Fernandez, E. and Flaugher, B. and Fosalba, P. and Gaztanaga, E. and Gerdes, D. W. and Giannantonio, T. and Goldstein, D. A. and Gruen, D. and Gschwend, J. and Gutierrez, G. and Honscheid, K. and James, D. J. and Jeltema, T. and Johnson, M. W. G. and Johnson, M. D. and Kent, S. and Krause, E. and Kron, R. and Kuehn, K. and Lahav, O. and Lima, M. and Maia, M. A. G. and March, M. and Martini, P. and McMahon, R. G. and Menanteau, F. and Miller, C. J. and Miquel, R. and Mohr, J. J. and Nichol, R. C. and Ogando, R. L. C. and Plazas, A. A. and Romer, A. K. and Roodman, A. and Rykoff, E. S. and Sanchez, E. and Scarpine, V. and Schindler, R. and Schubnell, M. and Sevilla-Noarbe, I. and Sheldon, E. and Smith, M. and Smith, R. C. and Stebbins, A. and Suchyta, E. and Swanson, M. E. C. and Tarle, G. and Thomas, R. C. and Troxel, M. A. and Tucker, D. L. and Vikram, V. and Walker, A. R. and Wechsler, R. H. and Weller, J. and Carlin, J. L. and Gill, M. S. S. and Li, T. S. and Marriner, J. and Neilsen, E. and Haislip, J. B. and Kouprianov, V. V. and Reichart, D. E. and Sand, D. J. and Tartaglia, L. and Valenti, S. and Yang, S. and Benetti, S. and Brocato, E. and Campana, S. and Cappellaro, E. and Covino, S. and D’Avanzo, P. and D’Elia, V. and Getman, F. and Ghirlanda, G. and Ghisellini, G. and Limatola, L. and Nicastro, L. and Palazzi, E. and Pian, E. and Piranomonte, S. and Possenti, A. and Rossi, A. and Salafia, O. S. and Tomasella, L. and Amati, L. and Antonelli, L. A. and Bernardini, M. G. and Bufano, F. and Capaccioli, M. and Casella, P. and Dadina, M. and Cesare, G. De and Paola, A. Di and Giuffrida, G. and Giunta, A. and Israel, G. L. and Lisi, M. and Maiorano, E. and Mapelli, M. and Masetti, N. and Pescalli, A. and Pulone, L. and Salvaterra, R. and Schipani, P. and Spera, M. and Stamerra, A. and Stella, L. and Testa, V. and Turatto, M. and Vergani, D. and Aresu, G. and Bachetti, M. and Buffa, F. and Burgay, M. and Buttu, M. and Caria, T. and Carretti, E. and Casasola, V. and Castangia, P. and Carboni, G. and Casu, S. and Concu, R. and Corongiu, A. and Deiana, G. L. and Egron, E. and Fara, A. and Gaudiomonte, F. and Gusai, V. and Ladu, A. and Loru, S. and Leurini, S. and Marongiu, L. and Melis, A. and Melis, G. and Migoni, Carlo and Milia, Sabrina and Navarrini, Alessandro and Orlati, A. and Ortu, P. and Palmas, S. and Pellizzoni, A. and Perrodin, D. and Pisanu, T. and Poppi, S. and Righini, S. and Saba, A. and Serra, G. and Serrau, M. and Stagni, M. and Surcis, G. and Vacca, V. and Vargiu, G. P. and Hunt, L. K. and Jin, Z. P. and Klose, S. and Kouveliotou, C. and Mazzali, P. A. and Møller, P. and Nava, L. and Piran, T. and Selsing, J. and Vergani, S. D. and Wiersema, K. and Toma, K. and Higgins, A. B. and Mundell, C. G. and di Serego Alighieri, S. and Gótz, D. and Gao, W. and Gomboc, A. and Kaper, L. and Kobayashi, S. and Kopac, D. and Mao, J. and Starling, R. L. C. and Steele, I. and van der Horst, A. J. and Acero, F. and Atwood, W. B. and Baldini, L. and Barbiellini, G. and Bastieri, D. and Berenji, B. and Bellazzini, R. and Bissaldi, E. and Blandford, R. D. and Bloom, E. D. and Bonino, R. and Bottacini, E. and Bregeon, J. and Buehler, R. and Buson, S. and Cameron, R. A. and Caputo, R. and Caraveo, P. A. and Cavazzuti, E. and Chekhtman, A. and Cheung, C. C. and Chiang, J. and Ciprini, S. and Cohen-Tanugi, J. and Cominsky, L. R. and Costantin, D. and Cuoco, A. and D’Ammando, F. and Palma, F. de and Digel, S. W. and Lalla, N. Di and Mauro, M. Di and Venere, L. Di and Dubois, R. and Fegan, S. J. and Focke, W. B. and Franckowiak, A. and Fukazawa, Y. and Funk, S. and Fusco, P. and Gargano, F. and Gasparrini, D. and Giglietto, N. and Giordano, F. and Giroletti, M. and Glanzman, T. and Green, D. and Grondin, M.-H. and Guillemot, L. and Guiriec, S. and Harding, A. K. and Horan, D. and Jóhannesson, G. and Kamae, T. and Kensei, S. and Kuss, M. and Mura, G. La and Latronico, L. and Lemoine-Goumard, M. and Longo, F. and Loparco, F. and Lovellette, M. N. and Lubrano, P. and Magill, J. D. and Maldera, S. and Manfreda, A. and Mazziotta, M. N. and McEnery, J. E. and Meyer, M. and Michelson, P. F. and Mirabal, N. and Monzani, M. E. and Moretti, E. and Morselli, A. and Moskalenko, I. V. and Negro, M. and Nuss, E. and Ojha, R. and Omodei, N. and Orienti, M. and Orlando, E. and Palatiello, M. and Paliya, V. S. and Paneque, D. and Pesce-Rollins, M. and Piron, F. and Porter, T. A. and Principe, G. and Rainò, S. and Rando, R. and Razzano, M. and Razzaque, S. and Reimer, A. and Reimer, O. and Reposeur, T. and Rochester, L. S. and Parkinson, P. M. Saz and Sgrò, C. and Siskind, E. J. and Spada, F. and Spandre, G. and Suson, D. J. and Takahashi, M. and Tanaka, Y. and Thayer, J. G. and Thayer, J. B. and Thompson, D. J. and Tibaldo, L. and Torres, D. F. and Torresi, E. and Troja, E. and Venters, T. M. and Vianello, G. and Zaharijas, G. and Allison, J. R. and Bannister, K. W. and Dobie, D. and Kaplan, D. L. and Lenc, E. and Lynch, C. and Murphy, T. and Sadler, E. M. and Hotan, A. and James, C. W. and Oslowski, S. and Raja, W. and Shannon, R. M. and Whiting, M. and Arcavi, I. and Howell, D. A. and McCully, C. and Hosseinzadeh, G. and Hiramatsu, D. and Poznanski, D. and Barnes, J. and Zaltzman, M. and Vasylyev, S. and Maoz, D. and Cooke, J. and Bailes, M. and Wolf, C. and Deller, A. T. and Lidman, C. and Wang, L. and Gendre, B. and Andreoni, I. and Ackley, K. and Pritchard, T. A. and Bessell, M. S. and Chang, S.-W. and Möller, A. and Onken, C. A. and Scalzo, R. A. and Ridden-Harper, R. and Sharp, R. G. and Tucker, B. E. and Farrell, T. J. and Elmer, E. and Johnston, S. and Krishnan, V. Venkatraman and Keane, E. F. and Green, J. A. and Jameson, A. and Hu, L. and Ma, B. and Sun, T. and Wu, X. and Wang, X. and Shang, Z. and Hu, Y. and Ashley, M. C. B. and Yuan, X. and Li, X. and Tao, C. and Zhu, Z. and Zhang, H. and Suntzeff, N. B. and Zhou, J. and Yang, J. and Orange, B. and Morris, D. and Cucchiara, A. and Giblin, T. and Klotz, A. and Staff, J. and Thierry, P. and Schmidt, B. P. and Tanvir, N. R. and Levan, A. J. and Cano, Z. and de Ugarte-Postigo, A. and González-Fernández, C. and Greiner, J. and Hjorth, J. and Irwin, M. and Krühler, T. and Mandel, I. and Milvang-Jensen, B. and O’Brien, P. and Rol, E. and Rosetti, S. and Rosswog, S. and Rowlinson, A. and Steeghs, D. T. H. and Thöne, C. C. and Ulaczyk, K. and Watson, D. and Bruun, S. H. and Cutter, R. and Figuera Jaimes, R. and Fujii, Y. I. and Fruchter, A. S. and Gompertz, B. and Jakobsson, P. and Hodosan, G. and Jèrgensen, U. G. and Kangas, T. and Kann, D. A. and Rabus, M. and Schrøder, S. L. and Stanway, E. R. and Wijers, R. A. M. J. and Lipunov, V. M. and Gorbovskoy, E. S. and Kornilov, V. G. and Tyurina, N. V. and Balanutsa, P. V. and Kuznetsov, A. S. and Vlasenko, D. M. and Podesta, R. C. and Lopez, C. and Podesta, F. and Levato, H. O. and Saffe, C. and Mallamaci, C. C. and Budnev, N. M. and Gress, O. A. and Kuvshinov, D. A. and Gorbunov, I. A. and Vladimirov, V. V. and Zimnukhov, D. S. and Gabovich, A. V. and Yurkov, V. V. and Sergienko, Yu. P. and Rebolo, R. and Serra-Ricart, M. and Tlatov, A. G. and Ishmuhametova, Yu. V. and Abe, F. and Aoki, K. and Aoki, W. and Asakura, Y. and Baar, S. and Barway, S. and Bond, I. A. and Doi, M. and Finet, F. and Fujiyoshi, T. and Furusawa, H. and Honda, S. and Itoh, R. and Kanda, N. and Kawabata, K. S. and Kawabata, M. and Kim, J. H. and Koshida, S. and Kuroda, D. and Lee, C.-H. and Liu, W. and Matsubayashi, K. and Miyazaki, S. and Morihana, K. and Morokuma, T. and Motohara, K. and Murata, K. L. and Nagai, H. and Nagashima, H. and Nagayama, T. and Nakaoka, T. and Nakata, F. and Ohsawa, R. and Ohshima, T. and Ohta, K. and Okita, H. and Saito, T. and Saito, Y. and Sako, S. and Sekiguchi, Y. and Sumi, T. and Tajitsu, A. and Takahashi, J. and Takayama, M. and Tamura, Y. and Tanaka, I. and Tanaka, M. and Terai, T. and Tominaga, N. and Tristram, P. J. and Uemura, M. and Utsumi, Y. and Yamaguchi, M. S. and Yasuda, N. and Yoshida, M. and Zenko, T. and Adams, S. M. and Anupama, G. C. and Bally, J. and Barway, S. and Bellm, E. and Blagorodnova, N. and Cannella, C. and Chandra, P. and Chatterjee, D. and Clarke, T. E. and Cobb, B. E. and Cook, D. O. and Copperwheat, C. and De, K. and Emery, S. W. K. and Feindt, U. and Foster, K. and Fox, O. D. and Frail, D. A. and Fremling, C. and Frohmaier, C. and Garcia, J. A. and Ghosh, S. and Giacintucci, S. and Goobar, A. and Gottlieb, O. and Grefenstette, B. W. and Hallinan, G. and Harrison, F. and Heida, M. and Helou, G. and Ho, A. Y. Q. and Horesh, A. and Hotokezaka, K. and Ip, W.-H. and Itoh, R. and Jacobs, Bob and Jencson, J. E. and Kasen, D. and Kasliwal, M. M. and Kassim, N. E. and Kim, H. and Kiran, B. S. and Kuin, N. P. M. and Kulkarni, S. R. and Kupfer, T. and Lau, R. M. and Madsen, K. and Mazzali, P. A. and Miller, A. A. and Miyasaka, H and Mooley, K. and Myers, S. T. and Nakar, E. and Ngeow, C.-C. and Nugent, P. and Ofek, E. O. and Palliyaguru, N. and Pavana, M. and Perley, D. A. and Peters, W. M. and Pike, S. and Piran, T. and Qi, H. and Quimby, R. M. and Rana, J. and Rosswog, S. and Rusu, F. and Sadler, E. M. and Sistine, A. Van and Sollerman, J. and Xu, Y. and Yan, L. and Yatsu, Y. and Yu, P.-C. and Zhang, C. and Zhao, W. and Chambers, K. C. and Huber, M. E. and Schultz, A. S. B. and Bulger, J. and Flewelling, H. and Magnier, E. A. and Lowe, T. B. and Wainscoat, R. J. and Waters, C. and Willman, M. and Ebisawa, K. and Hanyu, C. and Harita, S. and Hashimoto, T. and Hidaka, K. and Hori, T. and Ishikawa, M. and Isobe, N. and Iwakiri, W. and Kawai, H. and Kawai, N. and Kawamuro, T. and Kawase, T. and Kitaoka, Y. and Makishima, K. and Matsuoka, M. and Mihara, T. and Morita, T. and Morita, K. and Nakahira, S. and Nakajima, M. and Nakamura, Y. and Negoro, H. and Oda, S. and Sakamaki, A. and Sasaki, R. and Serino, M. and Shidatsu, M. and Shimomukai, R. and Sugawara, Y. and Sugita, S. and Sugizaki, M. and Tachibana, Y. and Takao, Y. and Tanimoto, A. and Tomida, H. and Tsuboi, Y. and Tsunemi, H. and Ueda, Y. and Ueno, S. and Yamada, S. and Yamaoka, K. and Yamauchi, M. and Yatabe, F. and Yoneyama, T. and Yoshii, T. and Coward, D. M. and Crisp, H. and Macpherson, D. and Andreoni, I. and Laugier, R. and Noysena, K. and Klotz, A. and Gendre, B. and Thierry, P. and Turpin, D. and Im, M. and Choi, C. and Kim, J. and Yoon, Y. and Lim, G. and Lee, S.-K. and Lee, C.-U. and Kim, S.-L. and Ko, S.-W. and Joe, J. and Kwon, M.-K. and Kim, P.-J. and Lim, S.-K. and Choi, J.-S. and Fynbo, J. P. U. and Malesani, D. and Xu, D. and Smartt, S. J. and Jerkstrand, A. and Kankare, E. and Sim, S. A. and Fraser, M. and Inserra, C. and Maguire, K. and Leloudas, G. and Magee, M. and Shingles, L. J. and Smith, K. W. and Young, D. R. and Kotak, R. and Gal-Yam, A. and Lyman, J. D. and Homan, D. S. and Agliozzo, C. and Anderson, J. P. and Angus, C. R. and Ashall, C. and Barbarino, C. and Bauer, F. E. and Berton, M. and Botticella, M. T. and Bulla, M. and Cannizzaro, G. and Cartier, R. and Cikota, A. and Clark, P. and De Cia, A. and Della Valle, M. and Dennefeld, M. and Dessart, L. and Dimitriadis, G. and Elias-Rosa, N. and Firth, R. E. and Flörs, A. and Frohmaier, C. and Galbany, L. and González-Gaitán, S. and Gromadzki, M. and Gutiérrez, C. P. and Hamanowicz, A. and Harmanen, J. and Heintz, K. E. and Hernandez, M.-S. and Hodgkin, S. T. and Hook, I. M. and Izzo, L. and James, P. A. and Jonker, P. G. and Kerzendorf, W. E. and Kostrzewa-Rutkowska, Z. and Kromer, M. and Kuncarayakti, H. and Lawrence, A. and Manulis, I. and Mattila, S. and McBrien, O. and Müller, A. and Nordin, J. and O’Neill, D. and Onori, F. and Palmerio, J. T. and Pastorello, A. and Patat, F. and Pignata, G. and Podsiadlowski, P. and Razza, A. and Reynolds, T. and Roy, R. and Ruiter, A. J. and Rybicki, K. A. and Salmon, L. and Pumo, M. L. and Prentice, S. J. and Seitenzahl, I. R. and Smith, M. and Sollerman, J. and Sullivan, M. and Szegedi, H. and Taddia, F. and Taubenberger, S. and Terreran, G. and Van Soelen, B. and Vos, J. and Walton, N. A. and Wright, D. E. and Wyrzykowski, Ł. and Yaron, O. and Chen, T.-W. and Krühler, T. and Schady, P. and Wiseman, P. and Greiner, J. and Rau, A. and Schweyer, T. and Klose, S. and Nicuesa Guelbenzu, A. and Palliyaguru, N. T. and Shara, M. M. and Williams, T. and Vaisanen, P. and Potter, S. B. and Colmenero, E. Romero and Crawford, S. and Buckley, D. A. H. and Mao, J. and Díaz, M. C. and Macri, L. M. and García Lambas, D. and Mendes de Oliveira, C. and Nilo Castellón, J. L. and Ribeiro, T. and Sánchez, B. and Schoenell, W. and Abramo, L. R. and Akras, S. and Alcaniz, J. S. and Artola, R. and Beroiz, M. and Bonoli, S. and Cabral, J. and Camuccio, R. and Chavushyan, V. and Coelho, P. and Colazo, C. and Costa-Duarte, M. V. and Cuevas Larenas, H. and Domínguez Romero, M. and Dultzin, D. and Fernández, D. and García, J. and Girardini, C. and Gonçalves, D. R. and Gonçalves, T. S. and Gurovich, S. and Jiménez-Teja, Y. and Kanaan, A. and Lares, M. and Lopes de Oliveira, R. and López-Cruz, O. and Melia, R. and Molino, A. and Padilla, N. and Peñuela, T. and Placco, V. M. and Quiñones, C. and Ramírez Rivera, A. and Renzi, V. and Riguccini, L. and Ríos-López, E. and Rodriguez, H. and Sampedro, L. and Schneiter, M. and Sodré, L. and Starck, M. and Torres-Flores, S. and Tornatore, M. and Zadrożny, A. and Castillo, M. and Castro-Tirado, A. J. and Tello, J. C. and Hu, Y.-D. and Zhang, B.-B. and Cunniffe, R. and Castellón, A. and Hiriart, D. and Caballero-García, M. D. and Jelínek, M. and Kubánek, P. and Pérez del Pulgar, C. and Park, I. H. and Jeong, S. and Castro Cerón, J. M. and Pandey, S. B. and Yock, P. C. and Querel, R. and Fan, Y. and Wang, C. and Beardsley, A and Brown, I. S. and Crosse, B. and Emrich, D. and Franzen, T. and Gaensler, B. M. and Horsley, L. and Johnston-Hollitt, M. and Kenney, D. and Morales, M. F. and Pallot, D. and Sokolowski, M. and Steele, K. and Tingay, S. J. and Trott, C. M. and Walker, M. and Wayth, R. and Williams, A. and Wu, C. and Yoshida, A. and Sakamoto, T. and Kawakubo, Y. and Yamaoka, K. and Takahashi, I. and Asaoka, Y. and Ozawa, S. and Torii, S. and Shimizu, Y. and Tamura, T. and Ishizaki, W. and Cherry, M. L. and Ricciarini, S. and Penacchioni, A. V. and Marrocchesi, P. S. and Pozanenko, A. S. and Volnova, A. A. and Mazaeva, E. D. and Minaev, P. Yu. and Krugov, M. A. and Kusakin, A. V. and Reva, I. V. and Moskvitin, A. S. and Rumyantsev, V. V. and Inasaridze, R. and Klunko, E. V. and Tungalag, N. and Schmalz, S. E. and Burhonov, O. and Abdalla, H. and Abramowski, A. and Aharonian, F. and Benkhali, F. Ait and Angüner, E. O. and Arakawa, M. and Arrieta, M. and Aubert, P. and Backes, M. and Balzer, A. and Barnard, M. and Becherini, Y. and Tjus, J. Becker and Berge, D. and Bernhard, S. and Bernlöhr, K. and Blackwell, R. and Böttcher, M. and Boisson, C. and Bolmont, J. and Bonnefoy, S. and Bordas, P. and Bregeon, J. and Brun, F. and Brun, P. and Bryan, M. and Büchele, M. and Bulik, T. and Capasso, M. and Caroff, S. and Carosi, A. and Casanova, S. and Cerruti, M. and Chakraborty, N. and Chaves, R. C. G. and Chen, A. and Chevalier, J. and Colafrancesco, S. and Condon, B. and Conrad, J. and Davids, I. D. and Decock, J. and Deil, C. and Devin, J. and deWilt, P. and Dirson, L. and Djannati-Ataï, A. and Donath, A. and O’C. Drury, L. and Dutson, K. and Dyks, J. and Edwards, T. and Egberts, K. and Emery, G. and Ernenwein, J.-P. and Eschbach, S. and Farnier, C. and Fegan, S. and Fernandes, M. V. and Fiasson, A. and Fontaine, G. and Funk, S. and Füssling, M. and Gabici, S. and Gallant, Y. A. and Garrigoux, T. and Gaté, F. and Giavitto, G. and Giebels, B. and Glawion, D. and Glicenstein, J. F. and Gottschall, D. and Grondin, M.-H. and Hahn, J. and Haupt, M. and Hawkes, J. and Heinzelmann, G. and Henri, G. and Hermann, G. and Hinton, J. A. and Hofmann, W. and Hoischen, C. and Holch, T. L. and Holler, M. and Horns, D. and Ivascenko, A. and Iwasaki, H. and Jacholkowska, A. and Jamrozy, M. and Jankowsky, D. and Jankowsky, F. and Jingo, M. and Jouvin, L. and Jung-Richardt, I. and Kastendieck, M. A. and Katarzyński, K. and Katsuragawa, M. and Kerszberg, D. and Khangulyan, D. and Khélifi, B. and King, J. and Klepser, S. and Klochkov, D. and Kluźniak, W. and Komin, Nu. and Kosack, K. and Krakau, S. and Kraus, M. and Krüger, P. P. and Laffon, H. and Lamanna, G. and Lau, J. and Lees, J.-P. and Lefaucheur, J. and Lemière, A. and Lemoine-Goumard, M. and Lenain, J.-P. and Leser, E. and Lohse, T. and Lorentz, M. and Liu, R. and Lypova, I. and Malyshev, D. and Marandon, V. and Marcowith, A. and Mariaud, C. and Marx, R. and Maurin, G. and Maxted, N. and Mayer, M. and Meintjes, P. J. and Meyer, M. and Mitchell, A. M. W. and Moderski, R. and Mohamed, M. and Mohrmann, L. and Morå, K. and Moulin, E. and Murach, T. and Nakashima, S. and Naurois, M. de and Ndiyavala, H. and Niederwanger, F. and Niemiec, J. and Oakes, L. and O’Brien, P. and Odaka, H. and Ohm, S. and Ostrowski, M. and Oya, I. and Padovani, M. and Panter, M. and Parsons, R. D. and Pekeur, N. W. and Pelletier, G. and Perennes, C. and Petrucci, P.-O. and Peyaud, B. and Piel, Q. and Pita, S. and Poireau, V. and Poon, H. and Prokhorov, D. and Prokoph, H. and Pühlhofer, G. and Punch, M. and Quirrenbach, A. and Raab, S. and Rauth, R. and Reimer, A. and Reimer, O. and Renaud, M. and de los Reyes, R. and Rieger, F. and Rinchiuso, L. and Romoli, C. and Rowell, G. and Rudak, B. and Rulten, C. B. and Sahakian, V. and Saito, S. and Sanchez, D. A. and Santangelo, A. and Sasaki, M. and Schlickeiser, R. and Schüssler, F. and Schulz, A. and Schwanke, U. and Schwemmer, S. and Seglar-Arroyo, M. and Settimo, M. and Seyffert, A. S. and Shafi, N. and Shilon, I. and Shiningayamwe, K. and Simoni, R. and Sol, H. and Spanier, F. and Spir-Jacob, M. and Stawarz, Ł. and Steenkamp, R. and Stegmann, C. and Steppa, C. and Sushch, I. and Takahashi, T. and Tavernet, J.-P. and Tavernier, T. and Taylor, A. M. and Terrier, R. and Tibaldo, L. and Tiziani, D. and Tluczykont, M. and Trichard, C. and Tsirou, M. and Tsuji, N. and Tuffs, R. and Uchiyama, Y. and van der Walt, D. J. and Eldik, C. van and Rensburg, C. van and Soelen, B. van and Vasileiadis, G. and Veh, J. and Venter, C. and Viana, A. and Vincent, P. and Vink, J. and Voisin, F. and Völk, H. J. and Vuillaume, T. and Wadiasingh, Z. and Wagner, S. J. and Wagner, P. and Wagner, R. M. and White, R. and Wierzcholska, A. and Willmann, P. and Wörnlein, A. and Wouters, D. and Yang, R. and Zaborov, D. and Zacharias, M. and Zanin, R. and Zdziarski, A. A. and Zech, A. and Zefi, F. and Ziegler, A. and Zorn, J. and Żywucka, N. and Fender, R. P. and Broderick, J. W. and Rowlinson, A. and Wijers, R. A. M. J. and Stewart, A. J. and ter Veen, S. and Shulevski, A. and Kavic, M. and Simonetti, J. H. and League, C. and Tsai, J. and Obenberger, K. S. and Nathaniel, K. and Taylor, G. B. and Dowell, J. D. and Liebling, S. L. and Estes, J. A. and Lippert, M. and Sharma, I. and Vincent, P. and Farella, B. and Abeysekara, A. U. and Albert, A. and Alfaro, R. and Alvarez, C. and Arceo, R. and Arteaga-Velázquez, J. C. and Avila Rojas, D. and Ayala Solares, H. A. and Barber, A. S. and Becerra Gonzalez, J. and Becerril, A. and Belmont-Moreno, E. and BenZvi, S. Y. and Berley, D. and Bernal, A. and Braun, J. and Brisbois, C. and Caballero-Mora, K. S. and Capistrán, T. and Carramiñana, A. and Casanova, S. and Castillo, M. and Cotti, U. and Cotzomi, J. and Coutiño de León, S. and De León, C. and De la Fuente, E. and Diaz Hernandez, R. and Dichiara, S. and Dingus, B. L. and DuVernois, M. A. and Díaz-Vélez, J. C. and Ellsworth, R. W. and Engel, K. and Enríquez-Rivera, O. and Fiorino, D. W. and Fleischhack, H. and Fraija, N. and García-González, J. A. and Garfias, F. and Gerhardt, M. and Gonzõlez Muñoz, A. and González, M. M. and Goodman, J. A. and Hampel-Arias, Z. and Harding, J. P. and Hernandez, S. and Hernandez-Almada, A. and Hona, B. and Hüntemeyer, P. and Iriarte, A. and Jardin-Blicq, A. and Joshi, V. and Kaufmann, S. and Kieda, D. and Lara, A. and Lauer, R. J. and Lennarz, D. and León Vargas, H. and Linnemann, J. T. and Longinotti, A. L. and Luis Raya, G. and Luna-García, R. and López-Coto, R. and Malone, K. and Marinelli, S. S. and Martinez, O. and Martinez-Castellanos, I. and Martínez-Castro, J. and Martínez-Huerta, H. and Matthews, J. A. and Miranda-Romagnoli, P. and Moreno, E. and Mostafá, M. and Nellen, L. and Newbold, M. and Nisa, M. U. and Noriega-Papaqui, R. and Pelayo, R. and Pretz, J. and Pérez-Pérez, E. G. and Ren, Z. and Rho, C. D. and Rivière, C. and Rosa-González, D. and Rosenberg, M. and Ruiz-Velasco, E. and Salazar, H. and Salesa Greus, F. and Sandoval, A. and Schneider, M. and Schoorlemmer, H. and Sinnis, G. and Smith, A. J. and Springer, R. W. and Surajbali, P. and Tibolla, O. and Tollefson, K. and Torres, I. and Ukwatta, T. N. and Weisgarber, T. and Westerhoff, S. and Wisher, I. G. and Wood, J. and Yapici, T. and Yodh, G. B. and Younk, P. W. and Zhou, H. and Álvarez, J. D. and Aab, A. and Abreu, P. and Aglietta, M. and Albuquerque, I. F. M. and Albury, J. M. and Allekotte, I. and Almela, A. and Alvarez Castillo, J. and Alvarez-Muñiz, J. and Anastasi, G. A. and Anchordoqui, L. and Andrada, B. and Andringa, S. and Aramo, C. and Arsene, N. and Asorey, H. and Assis, P. and Avila, G. and Badescu, A. M. and Balaceanu, A. and Barbato, F. and Barreira Luz, R. J. and Becker, K. H. and Bellido, J. A. and Berat, C. and Bertaina, M. E. and Bertou, X. and Biermann, P. L. and Biteau, J. and Blaess, S. G. and Blanco, A. and Blazek, J. and Bleve, C. and Boháčová, M. and Bonifazi, C. and Borodai, N. and Botti, A. M. and Brack, J. and Brancus, I. and Bretz, T. and Bridgeman, A. and Briechle, F. L. and Buchholz, P. and Bueno, A. and Buitink, S. and Buscemi, M. and Caballero-Mora, K. S. and Caccianiga, L. and Cancio, A. and Canfora, F. and Caruso, R. and Castellina, A. and Catalani, F. and Cataldi, G. and Cazon, L. and Chavez, A. G. and Chinellato, J. A. and Chudoba, J. and Clay, R. W. and Cobos Cerutti, A. C. and Colalillo, R. and Coleman, A. and Collica, L. and Coluccia, M. R. and Conceição, R. and Consolati, G. and Contreras, F. and Cooper, M. J. and Coutu, S. and Covault, C. E. and Cronin, J. and D’Amico, S. and Daniel, B. and Dasso, S. and Daumiller, K. and Dawson, B. R. and Day, J. A. and Almeida, R. M. de and Jong, S. J. de and Mauro, G. De and de Mello Neto, J. R. T. and Mitri, I. De and Oliveira, J. de and Souza, V. de and Debatin, J. and Deligny, O. and Díaz Castro, M. L. and Diogo, F. and Dobrigkeit, C. and D’Olivo, J. C. and Dorosti, Q. and Dos Anjos, R. C. and Dova, M. T. and Dundovic, A. and Ebr, J. and Engel, R. and Erdmann, M. and Erfani, M. and Escobar, C. O. and Espadanal, J. and Etchegoyen, A. and Falcke, H. and Farmer, J. and Farrar, G. and Fauth, A. C. and Fazzini, N. and Feldbusch, F. and Fenu, F. and Fick, B. and Figueira, J. M. and Filipčič, A. and Freire, M. M. and Fujii, T. and Fuster, A. and Gaïor, R. and García, B. and Gaté, F. and Gemmeke, H. and Gherghel-Lascu, A. and Ghia, P. L. and Giaccari, U. and Giammarchi, M. and Giller, M. and Głas, D. and Glaser, C. and Golup, G. and Gómez Berisso, M. and Gómez Vitale, P. F. and González, N. and Gorgi, A. and Gottowik, M. and Grillo, A. F. and Grubb, T. D. and Guarino, F. and Guedes, G. P. and Halliday, R. and Hampel, M. R. and Hansen, P. and Harari, D. and Harrison, T. A. and Harvey, V. M. and Haungs, A. and Hebbeker, T. and Heck, D. and Heimann, P. and Herve, A. E. and Hill, G. C. and Hojvat, C. and Holt, E. and Homola, P. and Hörandel, J. R. and Horvath, P. and Hrabovský, M. and Huege, T. and Hulsman, J. and Insolia, A. and Isar, P. G. and Jandt, I. and Johnsen, J. A. and Josebachuili, M. and Jurysek, J. and Kääpä, A. and Kampert, K. H. and Keilhauer, B. and Kemmerich, N. and Kemp, J. and Kieckhafer, R. M. and Klages, H. O. and Kleifges, M. and Kleinfeller, J. and Krause, R. and Krohm, N. and Kuempel, D. and Kukec Mezek, G. and Kunka, N. and Kuotb Awad, A. and Lago, B. L. and LaHurd, D. and Lang, R. G. and Lauscher, M. and Legumina, R. and Leigui de Oliveira, M. A. and Letessier-Selvon, A. and Lhenry-Yvon, I. and Link, K. and Lo Presti, D. and Lopes, L. and López, R. and López Casado, A. and Lorek, R. and Luce, Q. and Lucero, A. and Malacari, M. and Mallamaci, M. and Mandat, D. and Mantsch, P. and Mariazzi, A. G. and Maris, I. C. and Marsella, G. and Martello, D. and Martinez, H. and Martínez Bravo, O. and Masías Meza, J. J. and Mathes, H. J. and Mathys, S. and Matthews, J. and Matthiae, G. and Mayotte, E. and Mazur, P. O. and Medina, C. and Medina-Tanco, G. and Melo, D. and Menshikov, A. and Merenda, K.-D. and Michal, S. and Micheletti, M. I. and Middendorf, L. and Miramonti, L. and Mitrica, B. and Mockler, D. and Mollerach, S. and Montanet, F. and Morello, C. and Morlino, G. and Müller, A. L. and Müller, G. and Muller, M. A. and Müller, S. and Mussa, R. and Naranjo, I. and Nguyen, P. H. and Niculescu-Oglinzanu, M. and Niechciol, M. and Niemietz, L. and Niggemann, T. and Nitz, D. and Nosek, D. and Novotny, V. and Nožka, L. and Núñez, L. A. and Oikonomou, F. and Olinto, A. and Palatka, M. and Pallotta, J. and Papenbreer, P. and Parente, G. and Parra, A. and Paul, T. and Pech, M. and Pedreira, F. and Pȩkala, J. and Peña-Rodriguez, J. and Pereira, L. A. S. and Perlin, M. and Perrone, L. and Peters, C. and Petrera, S. and Phuntsok, J. and Pierog, T. and Pimenta, M. and Pirronello, V. and Platino, M. and Plum, M. and Poh, J. and Porowski, C. and Prado, R. R. and Privitera, P. and Prouza, M. and Quel, E. J. and Querchfeld, S. and Quinn, S. and Ramos-Pollan, R. and Rautenberg, J. and Ravignani, D. and Ridky, J. and Riehn, F. and Risse, M. and Ristori, P. and Rizi, V. and Rodrigues de Carvalho, W. and Rodriguez Fernandez, G. and Rodriguez Rojo, J. and Roncoroni, M. J. and Roth, M. and Roulet, E. and Rovero, A. C. and Ruehl, P. and Saffi, S. J. and Saftoiu, A. and Salamida, F. and Salazar, H. and Saleh, A. and Salina, G. and Sánchez, F. and Sanchez-Lucas, P. and Santos, E. M. and Santos, E. and Sarazin, F. and Sarmento, R. and Sarmiento-Cano, C. and Sato, R. and Schauer, M. and Scherini, V. and Schieler, H. and Schimp, M. and Schmidt, D. and Scholten, O. and Schovánek, P. and Schröder, F. G. and Schröder, S. and Schulz, A. and Schumacher, J. and Sciutto, S. J. and Segreto, A. and Shadkam, A. and Shellard, R. C. and Sigl, G. and Silli, G. and Šmída, R. and Snow, G. R. and Sommers, P. and Sonntag, S. and Soriano, J. F. and Squartini, R. and Stanca, D. and Stanič, S. and Stasielak, J. and Stassi, P. and Stolpovskiy, M. and Strafella, F. and Streich, A. and Suarez, F. and Suarez-Durán, M. and Sudholz, T. and Suomijärvi, T. and Supanitsky, A. D. and Šupík, J. and Swain, J. and Szadkowski, Z. and Taboada, A. and Taborda, O. A. and Timmermans, C. and Todero Peixoto, C. J. and Tomankova, L. and Tomé, B. and Torralba Elipe, G. and Travnicek, P. and Trini, M. and Tueros, M. and Ulrich, R. and Unger, M. and Urban, M. and Valdés Galicia, J. F. and Valiño, I. and Valore, L. and Aar, G. van and Bodegom, P. van and van den Berg, A. M. and Vliet, A. van and Varela, E. and Cárdenas, B. Vargas and Vázquez, R. A. and Veberič, D. and Ventura, C. and Vergara Quispe, I. D. and Verzi, V. and Vicha, J. and Villaseñor, L. and Vorobiov, S. and Wahlberg, H. and Wainberg, O. and Walz, D. and Watson, A. A. and Weber, M. and Weindl, A. and Wiedeński, M. and Wiencke, L. and Wilczyński, H. and Wirtz, M. and Wittkowski, D. and Wundheiler, B. and Yang, L. and Yushkov, A. and Zas, E. and Zavrtanik, D. and Zavrtanik, M. and Zepeda, A. and Zimmermann, B. and Ziolkowski, M. and Zong, Z. and Zuccarello, F. and Kim, S. and Schulze, S. and Bauer, F. E. and Corral-Santana, J. M. and de Gregorio-Monsalvo, I. and González-López, J. and Hartmann, D. H. and Ishwara-Chandra, C. H. and Martín, S. and Mehner, A. and Misra, K. and Michałowski, M. J. and Resmi, L. and Paragi, Z. and Agudo, I. and An, T. and Beswick, R. and Casadio, C. and Frey, S. and Jonker, P. and Kettenis, M. and Marcote, B. and Moldon, J. and Szomoru, A. and van Langevelde, H. J. and Yang, J. and Cwiek, A. and Cwiok, M. and Czyrkowski, H. and Dabrowski, R. and Kasprowicz, G. and Mankiewicz, L. and Nawrocki, K. and Opiela, R. and Piotrowski, L. W. and Wrochna, G. and Zaremba, M. and Żarnecki, A. F. and Haggard, D. and Nynka, M. and Ruan, J. J. and Bland, P. A. and Booler, T. and Devillepoix, H. A. R. and Gois, J. S. de and Hancock, P. J. and Howie, R. M. and Paxman, J. and Sansom, E. K. and Towner, M. C. and Tonry, J. and Coughlin, M. and Stubbs, C. W. and Denneau, L. and Heinze, A. and Stalder, B. and Weiland, H. and Eatough, R. P. and Kramer, M. and Kraus, A. and Troja, E. and Piro, L. and González, J. Becerra and Butler, N. R. and Fox, O. D. and Khandrika, H. G. and Kutyrev, A. and Lee, W. H. and Ricci, R. and Ryan Jr., R. E. and Sánchez-Ramírez, R. and Veilleux, S. and Watson, A. M. and Wieringa, M. H. and Burgess, J. M. and Eerten, H. van and Fontes, C. J. and Fryer, C. L. and Korobkin, O. and Wollaeger, R. T. and Camilo, F. and Foley, A. R. and Goedhart, S. and Makhathini, S. and Oozeer, N. and Smirnov, O. M. and Fender, R. P. and Woudt, P. A.},
   year={2017},
   month=oct, pages={L12} }

@article{Farr_2019,
   title={A Future Percent-level Measurement of the Hubble Expansion at Redshift 0.8 with Advanced LIGO},
   volume={883},
   ISSN={2041-8213},
   url={http://dx.doi.org/10.3847/2041-8213/ab4284},
   DOI={10.3847/2041-8213/ab4284},
   number={2},
   journal={The Astrophysical Journal Letters},
   publisher={American Astronomical Society},
   author={Farr, Will M. and Fishbach, Maya and Ye, Jiani and Holz, Daniel E.},
   year={2019},
   month=oct, pages={L42} }

@article{Suvodip_CrossCorrelation,
   title={Cross-correlating Dark Sirens and Galaxies: Constraints on H
               0 from GWTC-3 of LIGO–Virgo–KAGRA},
   volume={975},
   ISSN={1538-4357},
   url={http://dx.doi.org/10.3847/1538-4357/ad7d90},
   DOI={10.3847/1538-4357/ad7d90},
   number={2},
   journal={The Astrophysical Journal},
   publisher={American Astronomical Society},
   author={Mukherjee, Suvodip and Krolewski, Alex and Wandelt, Benjamin D. and Silk, Joseph},
   year={2024},
   month=nov, pages={189} }

@article{SuvodipSpectral,
    author = {Mukherjee, Suvodip},
    title = {The redshift dependence of black hole mass distribution: is it reliable for standard sirens cosmology?},
    journal = {Monthly Notices of the Royal Astronomical Society},
    volume = {515},
    number = {4},
    pages = {5495-5505},
    year = {2022},
    month = {10},
    issn = {0035-8711},
    doi = {10.1093/mnras/stac2152},
    url = {https://doi.org/10.1093/mnras/stac2152},
    eprint = {https://academic.oup.com/mnras/article-pdf/515/4/5495/45479584/stac2152.pdf},
}

@misc{leyde2022currentfutureconstraintscosmology,
      title={Current and future constraints on cosmology and modified gravitational wave friction from binary black holes}, 
      author={Konstantin Leyde and Simone Mastrogiovanni and Danièle A. Steer and Eric Chassande-Mottin and Christos Karathanasis},
      year={2022},
      eprint={2203.11680},
      archivePrefix={arXiv},
      primaryClass={gr-qc},
      url={https://arxiv.org/abs/2203.11680}, 
}

@article{EquiagaSpectralSiren,
  title = {Spectral Sirens: Cosmology from the Full Mass Distribution of Compact Binaries},
  author = {Ezquiaga, Jose Mar\'{\i}a and Holz, Daniel E.},
  journal = {Phys. Rev. Lett.},
  volume = {129},
  issue = {6},
  pages = {061102},
  numpages = {6},
  year = {2022},
  month = {Aug},
  publisher = {American Physical Society},
  doi = {10.1103/PhysRevLett.129.061102},
  url = {https://link.aps.org/doi/10.1103/PhysRevLett.129.061102}
}

@article{Karathanasis_2023,
   title={Binary black holes population and cosmology in new lights: signature of PISN mass and formation channel in GWTC-3},
   volume={523},
   ISSN={1365-2966},
   url={http://dx.doi.org/10.1093/mnras/stad1373},
   DOI={10.1093/mnras/stad1373},
   number={3},
   journal={Monthly Notices of the Royal Astronomical Society},
   publisher={Oxford University Press (OUP)},
   author={Karathanasis, Christos and Mukherjee, Suvodip and Mastrogiovanni, Simone},
   year={2023},
   month=june, pages={4539–4555} }

@article{Bera_2020,
   title={Incompleteness Matters Not: Inference of H0 from Binary Black Hole–Galaxy Cross-correlations},
   volume={902},
   ISSN={1538-4357},
   url={http://dx.doi.org/10.3847/1538-4357/abb4e0},
   DOI={10.3847/1538-4357/abb4e0},
   number={1},
   journal={The Astrophysical Journal},
   publisher={American Astronomical Society},
   author={Bera, Sayantani and Rana, Divya and More, Surhud and Bose, Sukanta},
   year={2020},
   month=oct, pages={79} }

@article{DiazCrossCorr,
    author = {Cigarrán Díaz, Cristina and Mukherjee, Suvodip},
    title = {Mapping the cosmic expansion history from LIGO-Virgo-KAGRA in synergy with DESI and SPHEREx},
    journal = {Monthly Notices of the Royal Astronomical Society},
    volume = {511},
    number = {2},
    pages = {2782-2795},
    year = {2022},
    month = {04},
    issn = {0035-8711},
    doi = {10.1093/mnras/stac208},
    url = {https://doi.org/10.1093/mnras/stac208},
    eprint = {https://academic.oup.com/mnras/article-pdf/511/2/2782/42537377/stac208.pdf},
}

@article{SuvodipCrossCorr2,
  title = {Accurate precision cosmology with redshift unknown gravitational wave sources},
  author = {Mukherjee, Suvodip and Wandelt, Benjamin D. and Nissanke, Samaya M. and Silvestri, Alessandra},
  journal = {Phys. Rev. D},
  volume = {103},
  issue = {4},
  pages = {043520},
  numpages = {15},
  year = {2021},
  month = {Feb},
  publisher = {American Physical Society},
  doi = {10.1103/PhysRevD.103.043520},
  url = {https://link.aps.org/doi/10.1103/PhysRevD.103.043520}
}

@article{DelPozzo,
  title = {Inference of cosmological parameters from gravitational waves: Applications to second generation interferometers},
  author = {Del Pozzo, Walter},
  journal = {Phys. Rev. D},
  volume = {86},
  issue = {4},
  pages = {043011},
  numpages = {13},
  year = {2012},
  month = {Aug},
  publisher = {American Physical Society},
  doi = {10.1103/PhysRevD.86.043011},
  url = {https://link.aps.org/doi/10.1103/PhysRevD.86.043011}
}

@article{Chen_2018,
   title={A two per cent Hubble constant measurement from standard sirens within five years},
   volume={562},
   ISSN={1476-4687},
   url={http://dx.doi.org/10.1038/s41586-018-0606-0},
   DOI={10.1038/s41586-018-0606-0},
   number={7728},
   journal={Nature},
   publisher={Springer Science and Business Media LLC},
   author={Chen, Hsin-Yu and Fishbach, Maya and Holz, Daniel E.},
   year={2018},
   month=oct, pages={545–547} }

@article{Fishbach_2019_GW170817,
   title={A Standard Siren Measurement of the Hubble Constant from GW170817 without the Electromagnetic Counterpart},
   volume={871},
   ISSN={2041-8213},
   url={http://dx.doi.org/10.3847/2041-8213/aaf96e},
   DOI={10.3847/2041-8213/aaf96e},
   number={1},
   journal={The Astrophysical Journal Letters},
   publisher={American Astronomical Society},
   author={Fishbach, M. and Gray, R. and Hernandez, I. Magaña and Qi, H. and Sur, A. and Acernese, F. and Aiello, L. and Allocca, A. and Aloy, M. A. and Amato, A. and Antier, S. and Arène, M. and Arnaud, N. and Ascenzi, S. and Astone, P. and Aubin, F. and Babak, S. and Bacon, P. and Badaracco, F. and Bader, M. K. M. and Baldaccini, F. and Ballardin, G. and Barone, F. and Barsuglia, M. and Barta, D. and Basti, A. and Bawaj, M. and Bazzan, M. and Bejger, M. and Belahcene, I. and Bernuzzi, S. and Bersanetti, D. and Bertolini, A. and Bitossi, M. and Bizouard, M. A. and Blair, C. D. and Bloemen, S. and Boer, M. and Bogaert, G. and Bondu, F. and Bonnand, R. and Boom, B. A. and Boschi, V. and Bouffanais, Y. and Bozzi, A. and Bradaschia, C. and Brady, P. R. and Branchesi, M. and Briant, T. and Brighenti, F. and Brillet, A. and Brisson, V. and Bulik, T. and Bulten, H. J. and Buskulic, D. and Buy, C. and Cagnoli, G. and Calloni, E. and Canepa, M. and Capocasa, E. and Carbognani, F. and Carullo, G. and Diaz, J. Casanueva and Casentini, C. and Caudill, S. and Cavalier, F. and Cavalieri, R. and Cella, G. and Cerdá-Durán, P. and Cerretani, G. and Cesarini, E. and Chaibi, O. and Chassande-Mottin, E. and Chatziioannou, K. and Chen, H. Y. and Chincarini, A. and Chiummo, A. and Christensen, N. and Chua, S. and Ciani, G. and Ciolfi, R. and Cipriano, F. and Cirone, A. and Cleva, F. and Coccia, E. and Cohadon, P.-F. and Cohen, D. and Conti, L. and Cordero-Carrión, I. and Cortese, S. and Coughlin, M. W. and Coulon, J.-P. and Croquette, M. and Cuoco, E. and Dálya, G. and D’Antonio, S. and Datrier, L. E. H. and Dattilo, V. and Davier, M. and Degallaix, J. and Laurentis, M. De and Deléglise, S. and Pozzo, W. Del and Denys, M. and Pietri, R. De and Rosa, R. De and Rossi, C. De and DeSalvo, R. and Dietrich, T. and Fiore, L. Di and Giovanni, M. Di and Girolamo, T. Di and Lieto, A. Di and Pace, S. Di and Palma, I. Di and Renzo, F. Di and Doctor, Z. and Drago, M. and Ducoin, J.-G. and Eisenmann, M. and Essick, R. C. and Estevez, D. and Fafone, V. and Farinon, S. and Farr, W. M. and Feng, F. and Ferrante, I. and Ferrini, F. and Fidecaro, F. and Fiori, I. and Fiorucci, D. and Flaminio, R. and Font, J. A. and Fournier, J.-D. and Frasca, S. and Frasconi, F. and Frey, V. and Gair, J. R. and Gammaitoni, L. and Garufi, F. and Gemme, G. and Genin, E. and Gennai, A. and George, D. and Germain, V. and Ghosh, A. and Giacomazzo, B. and Giazotto, A. and Giordano, G. and Castro, J. M. Gonzalez and Gosselin, M. and Gouaty, R. and Grado, A. and Granata, M. and Greco, G. and Groot, P. and Gruning, P. and Guidi, G. M. and Guo, Y. and Halim, O. and Harms, J. and Haster, C.-J. and Heidmann, A. and Heitmann, H. and Hello, P. and Hemming, G. and Hendry, M. and Hinderer, T. and Hoak, D. and Hofman, D. and Holz, D. E. and Hreibi, A. and Huet, D. and Idzkowski, B. and Iess, A. and Intini, G. and Isac, J.-M. and Jacqmin, T. and Jaranowski, P. and Jonker, R. J. G. and Katsanevas, S. and Katsavounidis, E. and Kéfélian, F. and Khan, I. and Koekoek, G. and Koley, S. and Kowalska, I. and Królak, A. and Kutynia, A. and Lange, J. and Lartaux-Vollard, A. and Lazzaro, C. and Leaci, P. and Letendre, N. and Li, T. G. F. and Linde, F. and Longo, A. and Lorenzini, M. and Loriette, V. and Losurdo, G. and Lumaca, D. and Macas, R. and Macquet, A. and Majorana, E. and Maksimovic, I. and Man, N. and Mantovani, M. and Marchesoni, F. and Markakis, C. and Marquina, A. and Martelli, F. and Massera, E. and Masserot, A. and Mastrogiovanni, S. and Meidam, J. and Mereni, L. and Merzougui, M. and Messenger, C. and Metzdorff, R. and Michel, C. and Milano, L. and Miller, A. and Minazzoli, O. and Minenkov, Y. and Montani, M. and Morisaki, S. and Mours, B. and Nagar, A. and Nardecchia, I. and Naticchioni, L. and Nelemans, G. and Nichols, D. and Nocera, F. and Obergaulinger, M. and Pagano, G. and Palomba, C. and Pannarale, F. and Paoletti, F. and Paoli, A. and Pasqualetti, A. and Passaquieti, R. and Passuello, D. and Patil, M. and Patricelli, B. and Pedurand, R. and Perreca, A. and Piccinni, O. J. and Pichot, M. and Piergiovanni, F. and Pillant, G. and Pinard, L. and Poggiani, R. and Popolizio, P. and Prodi, G. A. and Punturo, M. and Puppo, P. and Radulescu, N. and Raffai, P. and Rapagnani, P. and Raymond, V. and Razzano, M. and Regimbau, T. and Rei, L. and Ricci, F. and Rocchi, A. and Rolland, L. and Romanelli, M. and Romano, R. and Rosińska, D. and Ruggi, P. and Salconi, L. and Samajdar, A. and Sanchis-Gual, N. and Sassolas, B. and Schutz, B. F. and Sentenac, D. and Sequino, V. and Sieniawska, M. and Singh, N. and Singhal, A. and Sorrentino, F. and Stachie, C. and Steer, D. A. and Stratta, G. and Swinkels, B. L. and Tacca, M. and Tamanini, N. and Tiwari, S. and Tonelli, M. and Torres-Forné, A. and Travasso, F. and Tringali, M. C. and Trovato, A. and Trozzo, L. and Tsang, K. W. and Bakel, N. van and Beuzekom, M. van and Brand, J. F. J. van den and Broeck, C. Van Den and Schaaf, L. van der and Heijningen, J. V. van and Vardaro, M. and Vasúth, M. and Vedovato, G. and Veitch, J. and Verkindt, D. and Vetrano, F. and Viceré, A. and Vinet, J.-Y. and Vocca, H. and Walet, R. and Wang, G. and Wang, Y. F. and Was, M. and Williamson, A. R. and Yvert, M. and Zadrożny, A. and Zelenova, T. and Zendri, J.-P. and Zimmerman, A. B.},
   year={2019},
   month=jan, pages={L13} }

@article{Gray_Pixelatedgwcosmo,
    author = {Gray, R and Messenger, C and Veitch, J},
    title = {A pixelated approach to galaxy catalogue incompleteness: improving the dark siren measurement of the Hubble constant},
    journal = {Monthly Notices of the Royal Astronomical Society},
    volume = {512},
    number = {1},
    pages = {1127-1140},
    year = {2022},
    month = {05},
    issn = {0035-8711},
    doi = {10.1093/mnras/stac366},
    url = {https://doi.org/10.1093/mnras/stac366},
    eprint = {https://academic.oup.com/mnras/article-pdf/512/1/1127/45303118/stac366.pdf},
}

@article{Leandro,
  title = {Measuring the Hubble constant with black sirens},
  author = {Leandro, Hebertt and Marra, Valerio and Sturani, Riccardo},
  journal = {Phys. Rev. D},
  volume = {105},
  issue = {2},
  pages = {023523},
  numpages = {12},
  year = {2022},
  month = {Jan},
  publisher = {American Physical Society},
  doi = {10.1103/PhysRevD.105.023523},
  url = {https://link.aps.org/doi/10.1103/PhysRevD.105.023523}
}

@article{GLADE+,
   title={GLADE+ : an extended galaxy catalogue for multimessenger searches with advanced gravitational-wave detectors},
   volume={514},
   ISSN={1365-2966},
   url={http://dx.doi.org/10.1093/mnras/stac1443},
   DOI={10.1093/mnras/stac1443},
   number={1},
   journal={Monthly Notices of the Royal Astronomical Society},
   publisher={Oxford University Press (OUP)},
   author={Dálya, G and Díaz, R and Bouchet, F R and Frei, Z and Jasche, J and Lavaux, G and Macas, R and Mukherjee, S and Pálfi, M and de Souza, R S and Wandelt, B D and Bilicki, M and Raffai, P},
   year={2022},
   month=may, pages={1403–1411} }

@ARTICLE{Dalang_Fiorini_Baker_2024,
       author = {{Dalang}, Charles and {Fiorini}, Bartolomeo and {Baker}, Tessa},
        title = "{Large scale structure prior knowledge in the dark siren method}",
      journal = {arXiv e-prints},
         year = 2024,
        month = oct,
          eid = {arXiv:2410.03275},
        pages = {arXiv:2410.03275},
          doi = {10.48550/arXiv.2410.03275},
archivePrefix = {arXiv},
       eprint = {2410.03275},
 primaryClass = {astro-ph.CO},
       adsurl = {https://ui.adsabs.harvard.edu/abs/2024arXiv241003275D}
}

@ARTICLE{Cosmic_cartography_2025,
       author = {{Leyde}, Konstantin and {Baker}, Tessa and {Enzi}, Wolfgang},
        title = "{Cosmic Cartography II: completing galaxy catalogs for gravitational-wave cosmology}",
      journal = {arXiv e-prints},
         year = 2025,
        month = jul,
          eid = {arXiv:2507.12171},
        pages = {arXiv:2507.12171},
          doi = {10.48550/arXiv.2507.12171},
archivePrefix = {arXiv},
       eprint = {2507.12171},
 primaryClass = {astro-ph.CO},
       adsurl = {https://ui.adsabs.harvard.edu/abs/2025arXiv250712171L}
}

@misc{MariosPaper,
      title={Clustering effects on the Dark Siren determination of $H_0$: A simulation study}, 
      author={Marios Kalomenopoulos and Riccardo Barbieri and Sadegh Khochfar and Jonathan Gair and Robert J. McGibbon},
      year={2025},
      eprint={2511.12334},
      archivePrefix={arXiv},
      primaryClass={astro-ph.CO},
      url={https://arxiv.org/abs/2511.12334}, 
}

@article{Planck,
   title={<i>Planck</i>
                    2018 results: VI. Cosmological parameters},
   volume={641},
   ISSN={1432-0746},
   url={http://dx.doi.org/10.1051/0004-6361/201833910},
   DOI={10.1051/0004-6361/201833910},
   journal={Astronomy &amp; Astrophysics},
   publisher={EDP Sciences},
   author={Aghanim, N. and Akrami, Y. and Ashdown, M. and Aumont, J. and Baccigalupi, C. and Ballardini, M. and Banday, A. J. and Barreiro, R. B. and Bartolo, N. and Basak, S. and Battye, R. and Benabed, K. and Bernard, J.-P. and Bersanelli, M. and Bielewicz, P. and Bock, J. J. and Bond, J. R. and Borrill, J. and Bouchet, F. R. and Boulanger, F. and Bucher, M. and Burigana, C. and Butler, R. C. and Calabrese, E. and Cardoso, J.-F. and Carron, J. and Challinor, A. and Chiang, H. C. and Chluba, J. and Colombo, L. P. L. and Combet, C. and Contreras, D. and Crill, B. P. and Cuttaia, F. and de Bernardis, P. and de Zotti, G. and Delabrouille, J. and Delouis, J.-M. and Di Valentino, E. and Diego, J. M. and Doré, O. and Douspis, M. and Ducout, A. and Dupac, X. and Dusini, S. and Efstathiou, G. and Elsner, F. and Enßlin, T. A. and Eriksen, H. K. and Fantaye, Y. and Farhang, M. and Fergusson, J. and Fernandez-Cobos, R. and Finelli, F. and Forastieri, F. and Frailis, M. and Fraisse, A. A. and Franceschi, E. and Frolov, A. and Galeotta, S. and Galli, S. and Ganga, K. and Génova-Santos, R. T. and Gerbino, M. and Ghosh, T. and González-Nuevo, J. and Górski, K. M. and Gratton, S. and Gruppuso, A. and Gudmundsson, J. E. and Hamann, J. and Handley, W. and Hansen, F. K. and Herranz, D. and Hildebrandt, S. R. and Hivon, E. and Huang, Z. and Jaffe, A. H. and Jones, W. C. and Karakci, A. and Keihänen, E. and Keskitalo, R. and Kiiveri, K. and Kim, J. and Kisner, T. S. and Knox, L. and Krachmalnicoff, N. and Kunz, M. and Kurki-Suonio, H. and Lagache, G. and Lamarre, J.-M. and Lasenby, A. and Lattanzi, M. and Lawrence, C. R. and Le Jeune, M. and Lemos, P. and Lesgourgues, J. and Levrier, F. and Lewis, A. and Liguori, M. and Lilje, P. B. and Lilley, M. and Lindholm, V. and López-Caniego, M. and Lubin, P. M. and Ma, Y.-Z. and Macías-Pérez, J. F. and Maggio, G. and Maino, D. and Mandolesi, N. and Mangilli, A. and Marcos-Caballero, A. and Maris, M. and Martin, P. G. and Martinelli, M. and Martínez-González, E. and Matarrese, S. and Mauri, N. and McEwen, J. D. and Meinhold, P. R. and Melchiorri, A. and Mennella, A. and Migliaccio, M. and Millea, M. and Mitra, S. and Miville-Deschênes, M.-A. and Molinari, D. and Montier, L. and Morgante, G. and Moss, A. and Natoli, P. and Nørgaard-Nielsen, H. U. and Pagano, L. and Paoletti, D. and Partridge, B. and Patanchon, G. and Peiris, H. V. and Perrotta, F. and Pettorino, V. and Piacentini, F. and Polastri, L. and Polenta, G. and Puget, J.-L. and Rachen, J. P. and Reinecke, M. and Remazeilles, M. and Renzi, A. and Rocha, G. and Rosset, C. and Roudier, G. and Rubiño-Martín, J. A. and Ruiz-Granados, B. and Salvati, L. and Sandri, M. and Savelainen, M. and Scott, D. and Shellard, E. P. S. and Sirignano, C. and Sirri, G. and Spencer, L. D. and Sunyaev, R. and Suur-Uski, A.-S. and Tauber, J. A. and Tavagnacco, D. and Tenti, M. and Toffolatti, L. and Tomasi, M. and Trombetti, T. and Valenziano, L. and Valiviita, J. and Van Tent, B. and Vibert, L. and Vielva, P. and Villa, F. and Vittorio, N. and Wandelt, B. D. and Wehus, I. K. and White, M. and White, S. D. M. and Zacchei, A. and Zonca, A.},
   year={2020},
   month=Sept, pages={A6} }

@article{SHOES,
   title={Large Magellanic Cloud Cepheid Standards Provide a 1% Foundation for the Determination of the Hubble Constant and Stronger Evidence for Physics beyond ΛCDM},
   volume={876},
   ISSN={1538-4357},
   url={http://dx.doi.org/10.3847/1538-4357/ab1422},
   DOI={10.3847/1538-4357/ab1422},
   number={1},
   journal={The Astrophysical Journal},
   publisher={American Astronomical Society},
   author={Riess, Adam G. and Casertano, Stefano and Yuan, Wenlong and Macri, Lucas M. and Scolnic, Dan},
   year={2019},
   month=May, pages={85} }

\appendix

\section{Understanding the components of the Bayesian likelihood}
\label{Appendix components}
In this appendix, we show the role of each component of the Bayesian likelihood, hopefully providing the reader with a useful intuition for each term.

In the case of a complete catalog, the likelihood is simply:
\begin{equation}
    \label{Appendix eq: full cat likelihood}
    \mathcal{L}(H_0)=\dfrac{p_x(D_{\text{GW}} | H_0, D_{\text{cat}})}{p_D(H_0|{d_L}_\text{thr}, D_{\text{cat}})},
\end{equation}
where we have made explicit the dependencies of each term: $D_{\rm GW}$ represents the gravitational wave data and $D_{\rm cat}$ represents the galaxy catalog data, and ${d_L}_\text{thr}$ represents the gravitational wave event luminosity distance detection threshold, which is our only criteria of detection.

The term $p_x$ is the only one containing information of the individual GW event, and is expressed as:
\begin{equation}
    \label{Appendix eq: pxH0}
    p_x(D_{\text{GW}} | H_0, D_{\text{cat}}) = \sum_i^{N_{\text{gal}}} p({d_L}_{\text{GW}} | d_L(z_i, H_0)) p(\Omega_i | \Omega_\text{GW}) p_{\rm s}(z_i) {p_W}_i
\end{equation}
where $p(\Omega_i | \Omega_\text{GW})$ is a 2-dimensional Gaussian centered on the GW event observed sky location $\Omega_\text{GW} = (\text{RA}_{\text{GW}}, \text{dec}_\text{GW})$, while $p({d_L}_{\text{GW}} | d_L(z_i, H_0))$ is given by
 \begin{equation}
     \label{Appendix eq: px_dl_direct}
     p({d_L}_{\rm GW}|{d_L}) = \dfrac{1}{\sqrt{2\pi} A{d_L}}\exp\left(-\dfrac{({d_L}-{d_L}_{\rm GW})^2}{2(A{d_L})^2}\right).
 \end{equation}
 which, when regarded as a function of the unknown $d_L$, is not a Gaussian, as the variable $d_L$ appears also in the denominator of the normalization factor in front of the exponential and in the denominator of the argument of the exponential itself. We show the shape of the term $p({d_L}_{\rm GW}|{d_L})$ as a function of $d_L$ in figure \ref{Appendix fig: pdldirect}.

 \begin{figure}
     \centering
     \includegraphics[width=1\linewidth]{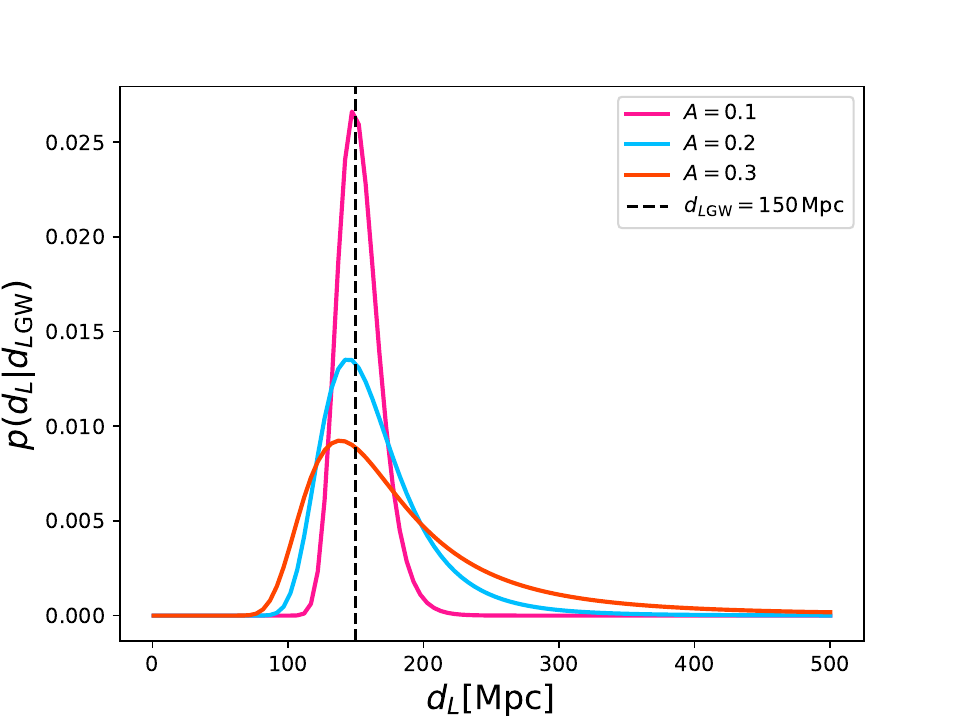}
     \caption{Likelihood of the true source distance given an observed event distance ${d_L}_{\rm GW}$, for different values of the fractional distance error $A$. The function is always peaked at the observed distance ${d_L}_{\rm GW}$. Naturally, the higher the error $A$, the wider the posterior. It is important to notice that this is not a Gaussian, as can be seen by the long tail at high distances.}
     \label{Appendix fig: pdldirect}
 \end{figure}

 The terms $p_{\rm s}(z)$ represents a weighting that allows the %probability that the 
 merger rate to depend on redshift, while we grouped in $p_{W_i}$ all possible weights that can be assigned to a galaxy depending on its properties that indicate if %to make 
 it is more (or less) likely to be a host of a gravitational wave event, most commonly its luminosity in certain bands. Across this work, we assume no dependency on redshift and no additional galaxy weights, so both these terms are set to one.

 The term $p_D(H_0|{d_L}_\text{thr})$ is the probability of detection of a gravitational wave event. This term does not depend on the individual gravitational wave event, but only on the detection thresholds of our "detector", which in this work we simplify to be a fixed (observed) luminosity distance detection threshold ${d_L}_\text{thr}$. It can be computed as:
 \begin{equation}
     \label{Appendix eq: pD}
     p_D(H_0|{d_L}_\text{thr}) = \sum_i^{N_\text{gal}} P_{\rm det}(z_i,H_0) p_s(z_i) {p_W}_i,
 \end{equation}
where
\begin{equation}
    \label{eq: Pdet fullcat}
    P_{\rm det}(z,H_0) = \dfrac{1}{2}\left[1-\text{erf}\left(\dfrac{d_L(z, H_0)-{d_L}_{\rm thr}}{\sqrt{2}Ad_L(z, H_0)}\right)\right].
\end{equation}

We show the shape of the probability of detection function for a fixed fractional error $A$ and for a fixed detection threshold ${d_L}_\text{thr}$ in the two plots of figure \ref{Appendix fig: Pdet}, while figure \ref{Appendix fig: pDG} shows the shape of the function $p_D(H_0)$ for various detection thresholds.

\begin{figure*}
	%\addtocounter{figure}{-1}
	\centering
	\includegraphics[width=0.45\textwidth]{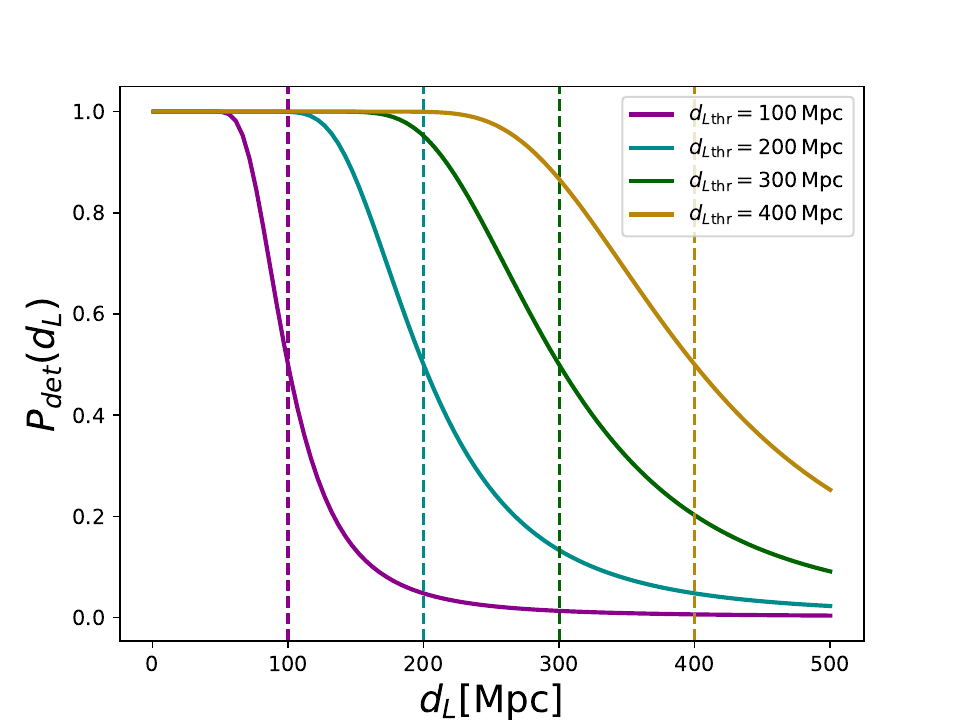}
	\hspace{1cm}
	\includegraphics[width=0.45\textwidth]{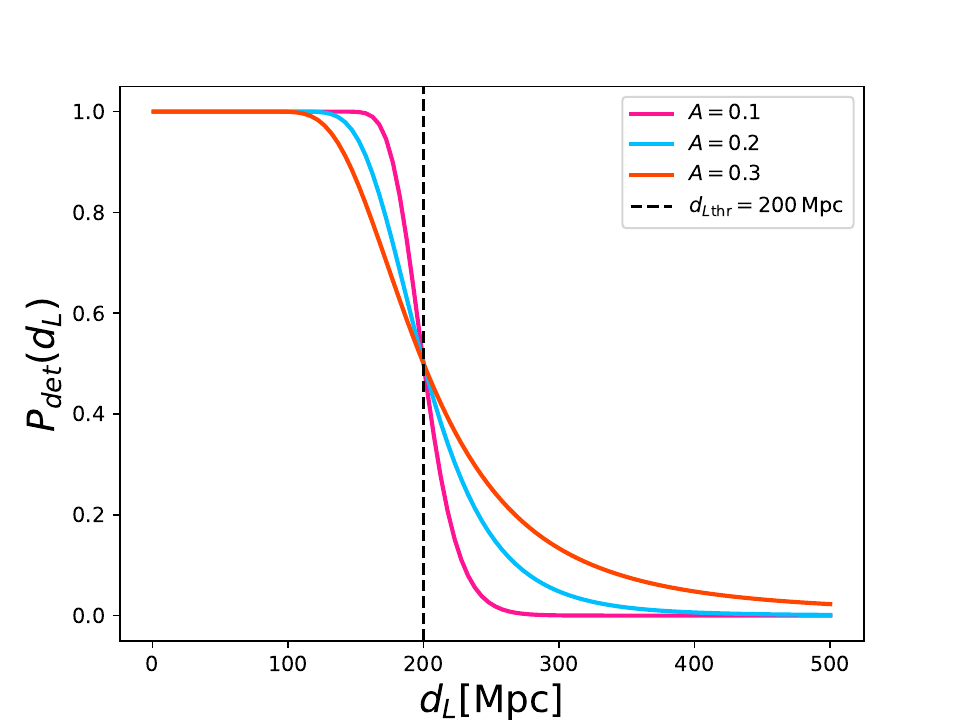}
	\caption{These plots show the shape of the function $P_{\rm det}(d_L)$ , the probability that an event happening at a certain true luminosity distance is detected, as a function of the true luminosity distance. On the left side, we show $P_{\rm det}(d_L)$ for a fixed fractional error in the luminosity distance $A=10 \%$ and various detection thresholds. Naturally, the higher the detection threshold, the higher the probability of detection.
    On the right side, we keep a fixed distance detection threshold of ${d_L}_{\rm thr} = 200$ Mpc and we vary the fractional error A. The lower the fractional error $A$, the steeper is the probability of detection. For a fractional error $A \xrightarrow{} 0$, we would have a perfect Heaviside step function.
    In both cases, the probability of detection reaches exactly 0.5 at the detection threshold: this is expected, as it is equally likely that a source at that distance gets perturbed to a higher distance (putting it outside the detection range) with respect to a lower distance (putting it inside the detection range).}
	\label{Appendix fig: Pdet}
\end{figure*}

\begin{figure}
    \centering
    \includegraphics[width=1\linewidth]{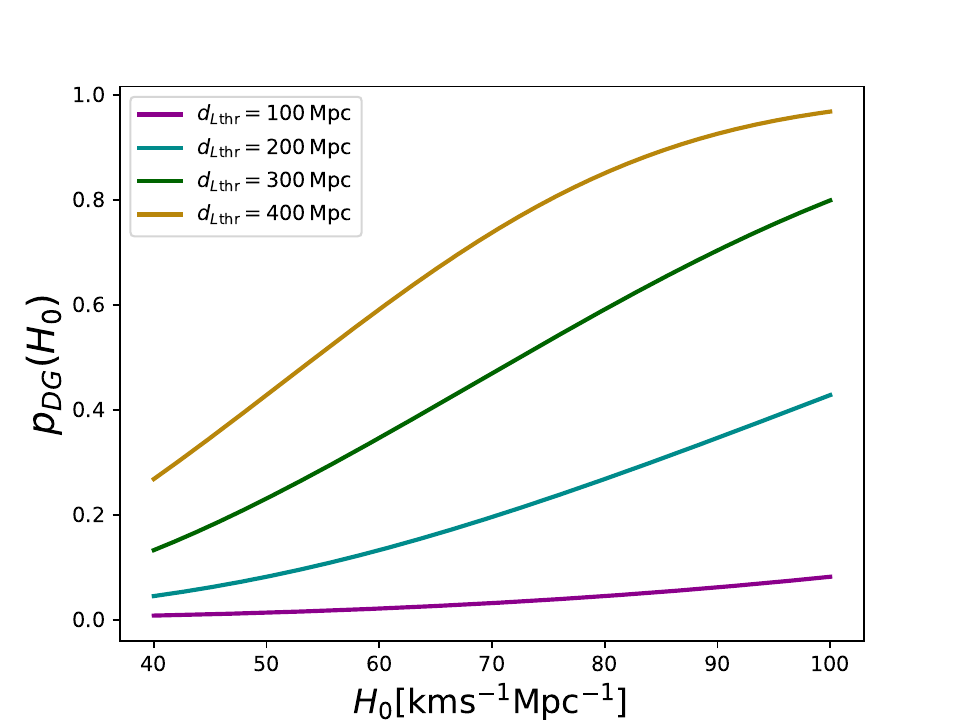}
    \caption{We show the function $p_D(H_0)$, given by equation \ref{Appendix eq: pD}, for different values of the detection threshold ${d_L}_{\rm thr}$. This function expresses the probability of detection for a source inside the catalog. Naturally, the higher the detection threshold, the higher the probability of detection. Furthermore, the higher $H_0$, the lower the distance for a fixed redshift, therefore the higher the probability of detection (as our only criteria of detection is an observed luminosity distance smaller than a certain threshold).}
    \label{Appendix fig: pDG}
\end{figure}

This likelihood is used across this work both in the case of a complete galaxy catalog and in the case of a reconstructed (according to the procedure presented in \ref{sub: Catalog Completion}) galaxy catalog. For the sake of completeness, we report here also an explanation of the terms entering the usual statistical method Bayesian likelihood for an incomplete galaxy catalog.

In the case of an incomplete galaxy catalog, the likelihood becomes:
\begin{equation}
    \label{Appendix eq: incomplete cat likelihood}
    \begin{split}    
    &\mathcal{L}(H_0)=p_G(H_0|m_{\text{th}}, {d_L}_{\text{thr}}) \dfrac{{p_x}_G(D_{\text{GW}} | H_0, D_{\text{cat}})}{{p_D}_{G}(H_0|{d_L}_{\text{thr}}, D_{\text{cat}})} + \\ 
    & \hspace{2cm}p_{\bar{G}} (H_0|m_{\text{th}}, {d_L}_{\text{thr}}) \dfrac{{p_x}_{\bar{G}}(D_{\text{GW}} | H_0, m_{\text{th}})}{{p_D}_{\bar{G}}(H_0|{d_L}_{\text{thr}}, m_{\text{th}})},
    \end{split}
\end{equation}
where once again we made manifest the dependencies of each term. 

The numerator and denominator of the first term are exactly the same as before, however they are now multiplied by the term $p_G(H_0)$, the probability that the host of the observed gravitational wave event is a galaxy contained in the catalog. This term is calculated as
\begin{equation}
    \label{Appendix eq: pG}
    \begin{split}
    &p_G(H_0|m_{\text{th}}, {d_L}_\text{thr}) = \\
    &= \dfrac{\int_0^{z(m_{\rm th}, M, H_0)} dz \int dM P_{\rm det}(z,H_0) \phi(M|H_0)p(z)}{\int dz \int dM P_{\rm det}(z,H_0) \phi(M|H_0)p(z)},
    \end{split}
\end{equation}
where $\phi(M)$ is an assumed Schechter magnitude function 
\begin{equation}
\Phi(M) = 0.4 \ln(10) \Phi^* \, 10^{0.4(\alpha+1)(M^*-M)} \exp\left(-10^{0.4(M^*-M)}\right),
\end{equation}
defined by 4 parameters $M^*_{\rm sch}$, $\alpha$ and the two limits of the distribution ${M_{\rm sch}}_{\rm min}$ and ${M_{\rm sch}}_{\rm max}$. The dependency on $H_0$ enters through 

$M^* = M^*_{\rm sch} + 5\text{log}_{10}(H_0/100).$

As reported in the main text, we have assumed r-band Schechter parameters for the Schecter function.

$p(z)$ is an assumed redshift prior. As is commonly done in the literature, we have assumed throughout this work a uniform in comoving volume redshift prior for this term. The limit of the integral $z(m_{\rm th}, M, H_0)$ is computed by first obtaining $d_L(m_{\rm th}, M)$ inverting the formula

$$M_{\rm th}(m_{\rm th}, d_L) = m_{\rm th}-5\log_{10}(d_L)+25$$

and then inserting into (and inverting)

$$d_L(z, H_0) = \dfrac{c}{H_0} (1+z) \int_0^z \dfrac{H_0}{H(z')} dz.$$

In figure \ref{Appendix fig: pG} we show the shape of this function for various %dldet 
luminosity distance detection thresholds and completenesses. 

\begin{figure*}
	%\addtocounter{figure}{-1}
	\centering
	\includegraphics[width=0.45\textwidth]{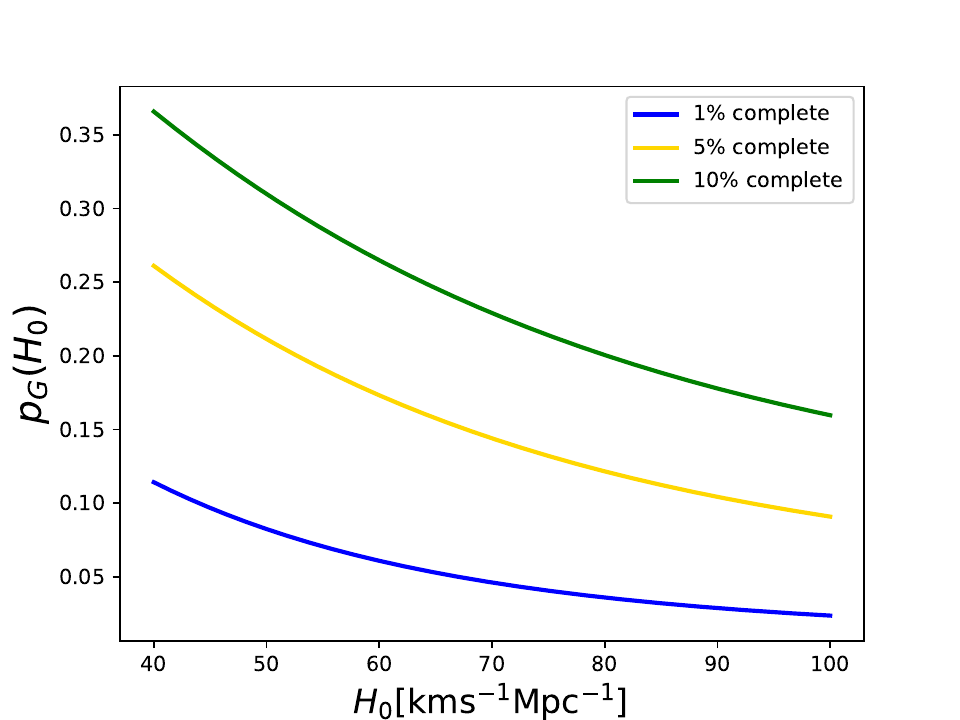}
	\hspace{1cm}
	\includegraphics[width=0.45\textwidth]{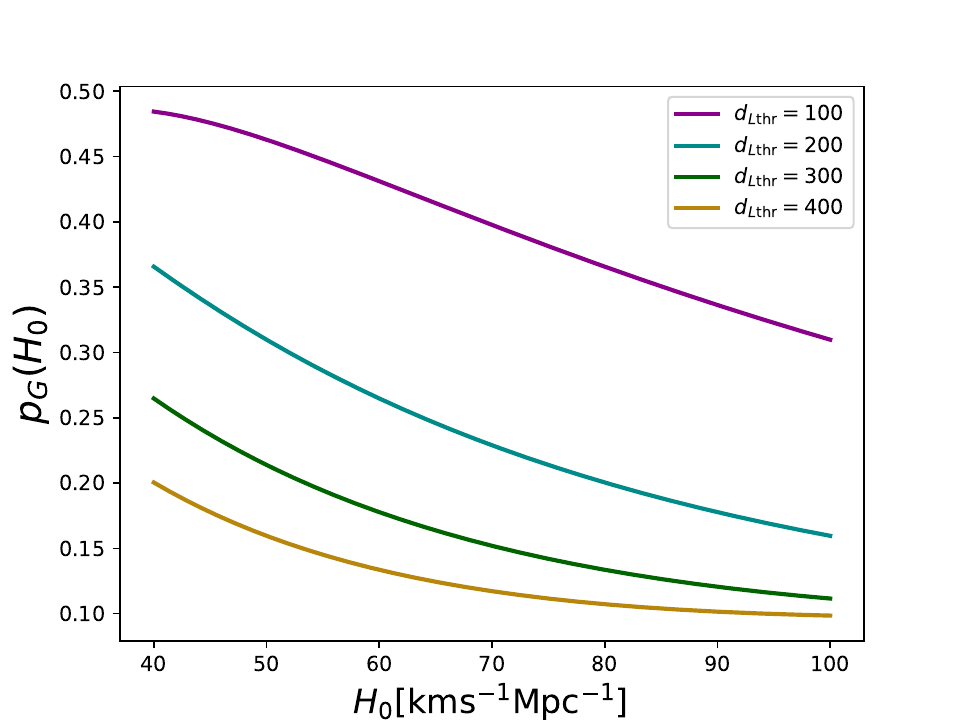}
	\caption{These plots show the shape of the function $p_G(H_0)$, the probability of the host being inside the catalog in the uniform statistical Bayesian method. On the left side, we show $p_G$ for a fixed detection threshold of ${d_L}_{\rm thr} = 200 \ \rm Mpc$, varying the completeness of the catalog. Naturally, the more complete the catalog, the higher the probability that the host is contained in the galaxy catalog.
    On the right side, we keep fixed the completeness fraction of the catalog to 10\%, and we show the impact of different detection thresholds. The higher the detection threshold, the further into the catalog we can "look" with the detected gravitational wave events; therefore, we explore higher redshifts areas which are less complete, giving a lower $p_G$. 
    In both cases, the function is strictly decreasing: the higher $H_0$, the higher the redshift for a given luminosity distance, therefore the lower the probability that the host is inside the galaxy catalog.}
	\label{Appendix fig: pG}
\end{figure*}

The term $p_G(H_0)$, together with the same likelihood as before, forms the inside the catalog half of the likelihood.

The second term in the likelihood \ref{Appendix eq: incomplete cat likelihood} has equivalently three components: $p_{\bar{G}}$, ${p_x}_{\bar{G}}$ and ${p_D}_{\bar{G}}$.

$p_{\bar{G}}$ is the probability that the galaxy host of the gravitational wave event is not contained in the catalog: this is of course given simply by $1-p_G$.

The term ${p_x}_{\bar{G}}(H_0| D_{\text{GW}}, m_{\text{th}})$ is the term containing information about the individual gravitational wave event, just like the term ${p_x}_G$. However, this term assumes that the host is not inside the catalog, the catalog information entering this term only through the apparent magnitude threshold $m_{th}$. This term is computed as:

\begin{equation}
    \label{Appendix eq: pxnG}
    \begin{split}
        &{p_x}_{\bar{G}}(H_0| D_{\text{GW}}, m_{\text{th}})= \\ &\int_{z(m_{\rm th}, M, H_0)}^\infty dz \int dM p(d_L(z, H_0)|{d_L}_{\rm GW})\phi(M|H_0)p(z).
        \dfrac{}{}
    \end{split}
\end{equation}

The term ${p_D}_{\bar{G}}$ is similarly obtained:
\begin{equation}
    \label{Appendix eq: pDnG}
    \begin{split}
    &{p_D}_{\bar{G}}(H_0|{d_L}_{\text{thr}}(D_{\text{GW}}), m_{\text{th}}) = \\
    &\int_{z(m_{\rm th}, M, H_0)}^\infty dz \int dM P_{\rm det}(z, H_0)\phi(M|H_0)p(z).
    \end{split}
\end{equation}

We show the impact of different completenesses and detection thresholds on this term in figure \ref{Appendix fig: pDnG}.

\begin{figure*}
	%\addtocounter{figure}{-1}
	\centering
	\includegraphics[width=0.45\textwidth]{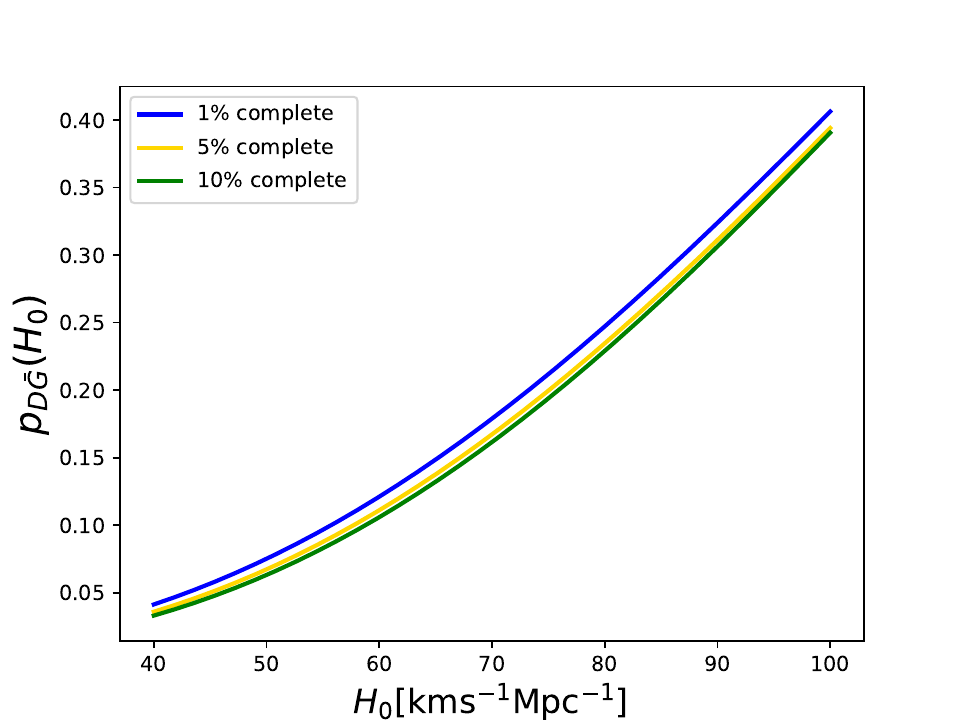}
	\hspace{1cm}
	\includegraphics[width=0.45\textwidth]{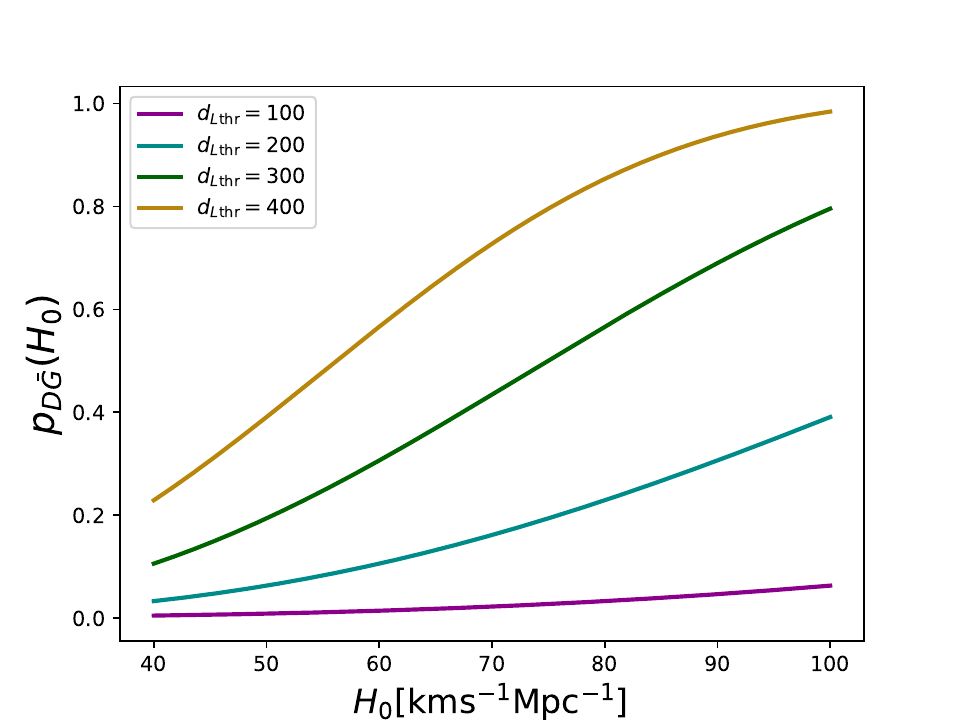}
	\caption{These plots show the shape of the function ${p_D}_{\bar{G}}(H_0)$, the probability of detection given a host outside the galaxy catalog. On the left side, we show ${p_D}_{\bar{G}}$ for a fixed detection threshold of ${d_L}_{\rm thr} = 200 \ \rm Mpc$, varying the completeness of the catalog. The less complete a galaxy catalog, the closer is the average galaxy not contained in the catalog, therefore the higher the probability of detection.
    On the right side, we keep fixed the completeness of the catalog to 10\%, and we show the impact of different detection thresholds. For low detection thresholds, the probability of detection is very small, as a source outside the catalog is most likely located at high redshift/distances, which would need to be perturbed strongly to be detected at low ${d_L}_{\rm thr}$. 
    In both cases, the function is strictly increasing: the higher $H_0$, the lower the distance for a fixed redshift, therefore the higher the probability of being detected (as our only criteria of detection is an observed luminosity distance smaller than a certain threshold).}
	\label{Appendix fig: pDnG}
\end{figure*}

\section{Edge of the catalog effect}
\label{Appendix edge}
In figure \ref{fig: CompleteMICE} we show how the posterior on $H_0$ for a complete galaxy catalog is impacted by the detection threshold ${d_L}_{\rm thr}$. The general trend is that, the higher the detection threshold, the wider the posterior. This is intuitive, as the higher the detection threshold, the higher the number of candidate host galaxies for each event. However, this trend breaks down when the detection threshold approaches the maximum range of the catalog (for the injected $H_0$ value):
$${d_L}_{\rm thr} \sim {d_L}_{\rm edge} = d_L(z_{\rm max}, {H_0}_{\rm true}).$$
We also note that it is necessary to provide the information of the redshift edge of the catalog to both the completion method and the statistical method: if such information is not provided, the expected but missing galaxies at high redshift can bias the $H_0$ posterior. This problem applies only to simulated galaxy catalogs, as real galaxy catalogs don't have any artificial edge. 
In particular, this effect occurs when the distance posterior of each event, given by equation \ref{Appendix eq: px_dl_direct}, has significant coverage past the edge of the catalog ${d_L}_{\rm edge}$. When this happens, we are able to get a "measurement" of the luminosity distance edge of the catalog, which, when combined with the redshift edge of the catalog, can lead to a much tighter posterior. Notably, if this effect is strong enough (either by having a very high detection threshold or a sufficiently large number of events), the precision of the $H_0$ posterior (measured as the width of the posterior) does not depend anymore on the completeness of the catalog, as (almost) all the information is obtained by combining the measured luminosity distance edge of the catalog with the measured redshift edge of the catalog.

In figure \ref{Appendix fig: highdldets} we show the $H_0$ credible intervals for various detection thresholds, all higher than the luminosity distance edge of the catalog, for three cases: using a 5\% complete catalog and the traditional statistical method, a 5\% complete catalog and our clustered completion method, and a complete catalog. The $H_0$ measurements for both incomplete methods are extremely close to the complete catalog, and they become almost indistinguishable at the higher detection thresholds. In these cases, all the information comes from the edge of the catalog "measurement", the incompleteness being basically irrelevant. Furthermore, the 90\% credible interval on $H_0$ becomes tighter the higher the detection threshold, as the edge of the catalog is more accurately measured.

\begin{figure}
    \centering
    \includegraphics[width=1.0\linewidth]{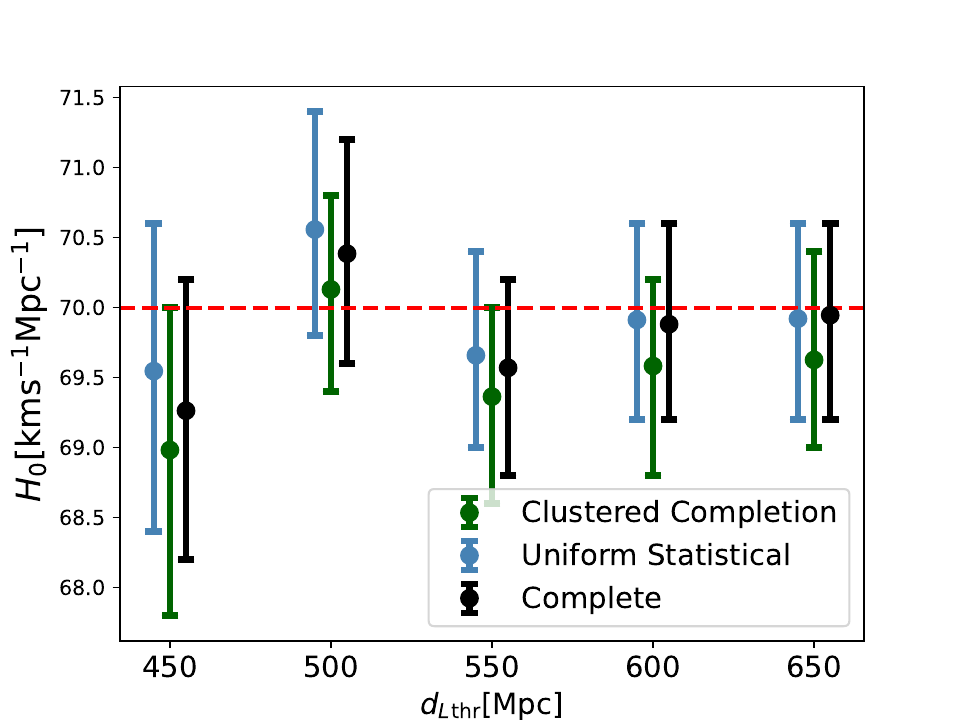}
    \caption{$H_0$ mean and 90\% credible intervals for a range of detection thresholds, all higher than the luminosity distance edge of the catalog ${d_L}_{\rm edge} = d_L(z_{\rm max}, {H_0}_{\rm true})\sim 430 \ \rm Mpc$. The dashed red line marks the $H_0$ input value, $H_0 = 70 \ \text{km}\ \text{s}^{-1} \ \text{Mpc}^{-1}$. The light blue represents the traditional uniform statistical method, the dark green represents the clustered completion introduced in this paper; both of these are computed using a 5\% complete galaxy catalog. The black points represent the complete catalog. }
    \label{Appendix fig: highdldets}
\end{figure}

\section{Octave of the sky}
\label{Appendix octave}

The MICE galaxy catalog is a simulation that covers only 1/8th of the sky. Therefore the catalog used in this paper has not only a maximum redshift (0.1), but also a minimum and maximum right ascension (0<RA<$\pi/2$) and a minimum and maximum declination (0<dec<$\pi/2$). When using the pixelation provided by \texttt{healpy}, we will then have some "border" pixels who are partially within the catalog and partially outside the catalog. If we calculate normally the "clustering probability" provided by equation \eqref{eq: ClusteringProbabilityPiel}, we will place missing galaxies outside the catalog, where we know there shouldn't be any. If we limit the placing to only inside the catalog (therefore "cutting" the pixel), we will place the same number of galaxies in a smaller volume, creating an artificial overdensity. To avoid this, we also introduced a volume factor to equation \eqref{eq: ClusteringProbabilityPiel}, which then becomes:
\begin{equation}
    \label{Appendix eq: ClusteringProbabilityPiel updated}
    p_{\rm clu}(\text{pix}_{c})=V(\text{pix}_{c})\left(1+\sum_{r<r_{\rm max}} \xi(r_{\text{pix}_c-\text{pix}_i}){N_g}_{\text{pix}_i}\right).
\end{equation}

This correction can also be used if one wants to use pixels of different volumes throughout the catalog, and not just at the edges.

We also note that the "edge of the catalog" effect, explored in appendix \ref{Appendix edge}, where a significant fraction of an event distance posterior lies outside the limits of the catalog, is also present at the angular edges of the catalog. However, in this case the edge of the catalog does not provide a redshift/distance measurement and therefore does not directly impact the cosmological analysis. The events lying close to the edge of the catalog will simply be slightly more informative (given the fewer candidate galaxies). Furthermore, given that we keep fixed the RA and dec uncertainty (usually to 0.1), we can estimate that the fraction of events lying within 1 sigma from the angular edge of the catalog is roughly 20\% (estimated by simply measuring the volume within $0.1<\rm RA < \pi/2 -0.1$ and $0.1<\rm dec <\pi/2 -0.1$), which does not significantly impact the overall likelihood.

\end{document}